\PassOptionsToPackage{usenames,dvipsnames}{xcolor}
\documentclass{lmcs}
\pdfoutput=1

\usepackage{lastpage}
\lmcsdoi{19}{4}{5}
\lmcsheading{}{\pageref{LastPage}}{}{}%
{Dec.~15,~2021}{Oct.~23,~2023}{}

\pdfoutput=1
\usepackage[T1]{fontenc}
\usepackage[utf8]{inputenc}
\usepackage[scaled=0.85]{beramono}
\usepackage{amssymb}
\usepackage{iris}
\usepackage[
  breaklinks=true,
  colorlinks=true,
  citecolor=blue,
  linkcolor=blue,
  urlcolor=blue,
]{hyperref}
\usepackage[final]{listings}

\newcommand{\ocamlcommentstyle}{\color{blue}}
\newcommand{\ocamldoccommentstyle}{\color{BlueGreen}}

\lstdefinelanguage{ocaml}[Objective]{Caml}{
  deletekeywords={closed,ref},
  morekeywords={initializer,effect,perform,continue},
  flexiblecolumns=false,
  showstringspaces=false,
  framesep=5pt,
  commentstyle=\ocamlcommentstyle,
  basicstyle=\tt\small,
  numberstyle=\footnotesize,
  mathescape=true,
  escapeinside={(*@}{*)},
  rangeprefix=(*\ ,
  rangesuffix=\ *),
  morecomment=[s][\ocamldoccommentstyle]{(**}{*)},
}

\def\oc|#1|{\text{\lstinline[language=ocaml,basicstyle=\tt,flexiblecolumns=true]|#1|}}

\def\qoc|#1|{``\oc|#1|''}

\lstdefinelanguage{none}{
  identifierstyle=
}
\usepackage{locallabel}
\let\softlink\hyperlink
\let\softtarget\hypertarget
\renewcommand{\LlabelTypeset}[1]{(\textrm{#1})}
\renewcommand{\LrefTypeset}[1]{\textrm{#1}}
\usepackage{mathpartir}
\renewcommand{\TirNameStyle}[1]{\hypertarget{#1}{\text{\rm\textsc{\small #1}}}}
\let\oldfootnote\footnote
\renewcommand{\footnote}[1]{\oldfootnote{\,#1}}
\usepackage{moreverb}
\usepackage{stmaryrd}
\usepackage{xspace}
\usepackage{caption}
\usepackage{subcaption}
\usepackage{hhline}
\usetikzlibrary{positioning}

\newcommand{\sref}[1]{\S\ref{#1}}
\newcommand{\fref}[1]{Figure~\ref{#1}}
\newcommand{\lemmaref}[1]{Lemma~\ref{#1}}
\newcommand{\defref}[1]{Definition~\ref{#1}}
\newcommand{\goalref}[1]{Goal~\ref{#1}}
\newcommand{\hypref}[1]{Hypothesis~\ref{#1}}
\newcommand{\lineref}[1]{line~\ref{line:#1}}
\newcommand{\linerange}[2]{lines~\ref{line:#1}--\ref{line:#2}}
\newcommand{\linesand}[2]{lines~\ref{line:#1} and~\ref{line:#2}}

\renewcommand{\eqdef}{\;\triangleq\;}

\def\etal.{et al.}

\newcommand{\SL}{Separation Logic\xspace}

\newcommand{\mocaml}{Multicore OCaml\xspace}
\newcommand{\hazel}{Hazel\xspace}
\newcommand{\lc}{$\lambda$-calculus\xspace}
\newcommand{\VLAD}{\textsc{VLAD}\xspace}

\newcommand{\vakar}{V{\'{a}}k{\'{a}}r\xspace}

\newcommand{\ad}{automatic differentiation\xspace}
\newcommand{\Ad}{Automatic differentiation\xspace}
\newcommand{\AD}{AD\xspace}
\newcommand{\rmAD}{reverse-mode \AD}
\newcommand{\rmad}{reverse-mode automatic differentiation\xspace}

\newcommand{\pd}{partial derivative\xspace}

\colorlet{fp}{green!10!red!90!}

\colorlet{paulo}{orange}

\newcommand{\kw}[1]{\text{\sf #1}}

\newcommand{\App}[2]{#1\;#2}
\newcommand{\AppTwo}[3]{#1\;#2\;#3}

\makeatletter
  \newcommand{\Read@op}{\TextOrMath{\kw{!}}{\mathop{!}}}
  \newcommand{\Write@op}{\TextOrMath{\kw{:=}}{\mathrel{\coloneqq}}}
  \newcommand{\Read}[1]{\Read@op #1}
  \newcommand{\Write}[2]{#1 \Write@op #2}
\makeatother

\newcommand{\lang}{\text{\textit{HH}}\xspace}

\newcommand\sepimp{\mathrel{-\mkern-12mu-\mkern-10mu*}}

\newcommand{\isep}{\,\mathrel{*}\,}
\newcommand{\ISEP}{\scalebox{1.7}{\raisebox{-0.3ex}{$\ast$}}\!}

\renewcommand\vsWand{%
  \equiv
  \kern-1.5ex
  \smash{\raisebox{-0.25mm}{\scalerel*{\vphantom-\ast}{\strut}}}
  \kern-0.2ex
}

\newcommand{\ev}{e} 
\newcommand{\E}{E} 
\newcommand{\El}{E_\locl}
\newcommand{\Er}{E_\locr}
\newcommand{\F}{F}
\newcommand{\T}{T}
\newcommand{\loc}{\ensuremath{\ell}}

\newcommand{\kont}{k} 

\newcommand{\prot}{\Psi}

\newcommand{\efill}[2]{#1[#2]} 

\renewcommand{\pred}{\Phi}

\newcommand{\ar}{\rightarrow}

\newcommand{\pair}[2]{\left(#1,#2\right)}

\newcommand{\wpname}{\mathit{wp}}
\renewcommand{\wp}[3]{\wpname_{#1}\;#2\;\{#3\}}
\newcommand{\wpnomask}[2]{\wpname\;#1\;\{#2\}}

\newcommand{\RULE}[1]{\hyperlink{#1}{\rm\textsc{#1}}\xspace}

\newcommand{\codelabel}[1]{%
  \ocamldoccommentstyle
  (*\;\,%
  \hyperlink{text:#1}{%
    \ocamldoccommentstyle
    \hypertarget{codelabel:#1}{#1:}%
  }%
  \;\,*)%
}
\newcommand{\uniquecoderef}[1]{%
  \hyperlink{codelabel:#1}{%
    \ocamldoccommentstyle
    \normalfont
    \hypertarget{text:#1}{\texttt{#1}}%
  }%
}
\newcommand{\extracoderef}[1]{%
  \hyperlink{codelabel:#1}{%
    \ocamldoccommentstyle
    \normalfont
    \texttt{#1}%
  }%
}

\newcommand{\pluseq}{\mathrel{+\kern-1mm=}}

\newcommand{\protbot}{\ensuremath{\bot}}
\newcommand{\SEND}{\,{!}\,}
\newcommand{\RECV}{{?}\,}

\newcommand{\x}{x}
\newcommand{\y}{y}

\newcommand{\protocol}[1]{\langle#1\rangle}
\newcommand{\condition}[1]{\{#1\}}

\newcommand{\ewpname}{\mathit{ewp}}

\newcommand{\ewpnomask}[3]{\ewpname\;#1\;\protocol{#2}\condition{#3}}

\newcommand{\ewpVerticalAlt}[5]{
  \begin{array}[t]{@{}l@{}}
  \ewpname_{#1}\;#2\; \protocol{#3} \{#4.\crqb{#5\}}
  \end{array}
}

\newcommand{\handler}[2]{(\,#1\,\mid\,#2\,)}
\newcommand{\shape}[2]{\protocol{#1}\condition{#2}}
\newcommand{\shapespaced}[2]{\protocol{#1}\;\condition{#2}}

\newcommand{\nameIsDeepHandler}{\textit{deep-handler}}
\newcommand{\isDeepHandler}[7]{%
  \nameIsDeepHandler_{#1}\;\shape{#4}{#6}\;\handler{#2}{#3}\;\shape{#5}{#7}
}
\newcommand{\isDeepHandlerVertical}[6]{%
  \begin{array}[t]{@{}r@{\;}l}
  \nameIsDeepHandler & \shapespaced{#3}{#5} \\
                     & \handler{#1}{#2} \\
                     & \shapespaced{#4}{#6}
  \end{array}
}
\newcommand{\jointeeff}[2]{#1\;\&\;#2}

\newcommand{\boite}[1]{\begin{array}{@{}l@{}}#1\end{array}}
\newcommand{\crqqb}[1]{\\\kern1.5em\boite{#1}}
\newcommand{\crqb}[1]{\\\quad\boite{#1}}
\newcommand{\crb}[1]{\\\boite{#1}}

\newcommand{\nameReasoningTryWithDeep}{Try-With-Deep}

\newcommand{\nameReasoningComp}{Composition}

\newcommand{\ReasoningTryWithDeep}{\RULE\nameReasoningTryWithDeep}

\newcommand{\listconcat}{;}
\newcommand{\listnil}{[]}
\newcommand{\listsingleton}[1]{#1}
\newcommand{\listcons}[2]{#1;#2}
\newcommand{\listsnoc}[2]{#1\listconcat\listsingleton{#2}}
\newcommand{\tlist}[1]{\mathit{list}\;#1}

\newcommand{\post}{\phi} 

\lstdefinelanguage{lang}[]{ocaml}{
  deletekeywords={try},
  alsoletter=-,
  morekeywords={shallow-try,deep-try},
}
\def\hh|#1|{\text{\lstinline[language=lang,basicstyle=\tt,flexiblecolumns=true]|#1|}}

\newcommand{\tyExp}{\oc|exp|}
\newcommand{\tnumdual}{\oc|t|\xspace}

\newcommand{\diff}{\oc|diff|\xspace}
\newcommand{\monomial}{\oc|monomial|\xspace}

\newcommand{\tx}{\oc|x|\xspace}

\newcommand{\tMul}{\oc|Mul|\xspace}
\newcommand{\tI}{\oc|I|\xspace}

\newcommand{\tVar}{\oc|Var|\xspace}

\newcommand{\num}{r}
\newcommand{\numb}{s}
\newcommand{\numm}{m}
\newcommand{\numd}{d}
\newcommand{\Num}{\mathcal{R}}

\newcommand{\valnum}{n}
\newcommand{\dual}[1]{#1^2}
\newcommand{\tov}[1]{#1}
\newcommand{\tod}[1]{\dot{#1}}

\newcommand{\opVar}{\mathit{op}}

\newcommand{\DExpr}[1]{\mathit{Exp}_{#1}}

\newcommand{\Node}[3]{#2\mathbin{#1}#3}
\newcommand{\Leaf}[1]{\mathbf{Leaf}\;#1}
\newcommand{\Zero}{\mathbf{0}}
\newcommand{\One}{\mathbf{1}}
\newcommand{\Add}{\text{\tt\bfseries +}}
\newcommand{\Mul}{\text{\tt\bfseries *}}

\newcommand{\X}{X}
\newcommand{\SX}{{\{\X\}}}
\newcommand{\DExprSX}{{\DExpr\SX}}

\newcommand{\envLeaf}{\mathit{Leaf}}
\newcommand{\envX}[1]{\lambda\X.#1} 
\newcommand{\parenvX}[1]{\envX{#1}}
\newcommand{\basicenv}{\mathit{\eta}} 
\newcommand{\Equiv}[3]{#2 \equiv_{#1} #3}
\newcommand{\ExprEquivname}{\equiv_{\DExprSX}}

\newcommand{\el}{E_1}
\newcommand{\er}{E_2}

\newcommand{\iSet}{\mathcal{I}}
\newcommand{\jSet}{\mathcal{J}}
\newcommand{\env}{\varrho}
\newcommand{\envB}{\vartheta}

\newcommand{\locx}{x}
\newcommand{\locy}{y}
\newcommand{\locl}{a}
\newcommand{\locr}{b}
\newcommand{\locu}{u}

\newcommand{\binding}{B}
\newcommand{\K}{K} 
\newcommand{\Kl}{K_1}
\newcommand{\Kr}{K_2}
\newcommand{\bind}[4]{\mathsf{let\;}#1=#3\mathop{#2}#4}
\newcommand{\kwin}{\;\mathsf{in}\;}
\newcommand{\typicalbinding}{\bind\locu\opVar\locl\locr}
\newcommand{\defsname}{\mathit{defs}}
\newcommand{\defs}[1]{\defsname(#1)}
\newcommand{\dfieldname}{\oc|d|}
\newcommand{\dfield}{\dfieldname~field\xspace}
\newcommand{\dfields}{\dfieldname~fields\xspace}

\newcommand{\vfieldname}{\oc|v|}
\newcommand{\vfield}{\vfieldname~field\xspace}
\newcommand{\vfields}{\vfieldname~field\xspace}

\newcommand{\Bindname}{\mathit{bind}}
\newcommand{\emapname}{\mathit{map}}
\newcommand{\evalname}{\mathit{eval}}
\newcommand{\varsname}{\mathit{leaves}}
\newcommand{\interpname}{\interp\cdot\cdot}

\newcommand{\evalmapname}{\llbracket\cdot\rrbracket_{(\cdot)}}

\newcommand{\Bind}[2]{\Bindname\;#1\;#2}

\newcommand{\emap}[2]{\emapname\;#1\;#2}
\newcommand{\eval}[1]{\evalname\;#1}
\newcommand{\evalmap}[2]{\llbracket #1 \rrbracket_{#2}}
\newcommand{\pp}[2]{\partial{#1}/\partial{#2}}
\newcommand{\Diff}[3]{\partial{#2} / \partial{#3} \; (#1)}
\renewcommand{\Diff}[3]{\evalmap{\pp{#2}{#3}}{#1}}
\newcommand{\vars}[1]{\varsname(#1)}
\newcommand{\interp}[2]{#1[#2]}

\newcommand{\evalbasic}[1]{\llparenthesis#1\rrparenthesis}
\newcommand{\Diffbasic}[2]{\evalbasic{\pp{#1}{#2}}}
\newcommand{\evalext}[2]{\evalbasic{#1}_{#2}}
\newcommand{\Diffext}[3]{\evalext{\pp{#2}{#3}}{#1}}

\newcommand{\metalet}[3]{\mathit{let}\;#1:=#2\;\mathit{in}\;#3}

\newcommand{\metaIfThenElse}[3]
{\mathit{if}\;#1\;\mathit{then}\;#2\;\mathit{else}\;#3}

\newcommand{\typeArrow}[2]{{#1} \rightarrow {#2}}

\newcommand{\typeSet}  [1]{\mathcal{P}(#1)}

\newcommand{\typeVal}     {\mathit{Val}}
\newcommand{\typeLoc}     {\mathit{Loc}}

\newcommand{\funOverwrite}[3]{#1[#2\,:=\,#3]}
\newcommand{\envExtension}[2]{#1\{#2\}}

\newcommand{\deriv}[1]{#1'}

\newcommand{\implementsname}{\mathit{isNum}}
\newcommand{\implements}[2]{#1\;\implementsname\;#2}

\newcommand{\isExpname}{\mathit{isExp}}
\newcommand{\isExp}[2]{#1\;\isExpname\;#2}

\newcommand{\numSpecname}{\mathit{isDict}}
\newcommand{\numSpec}[7]{\numSpecname\,(#5,#6,#7,#1,#2,#3,#4)}

\newcommand{\zero}{\oc|zero|\xspace}
\newcommand{\one}{\oc|one|\xspace}
\newcommand{\add}{\oc|add|\xspace}
\newcommand{\mul}{\oc|mul|\xspace}
\newcommand{\tn}{\oc|n|\xspace}

\newcommand{\zeroP}{\oc|zero'|\xspace}
\newcommand{\oneP}{\oc|one'|\xspace}
\newcommand{\addP}{\oc|add'|\xspace}
\newcommand{\mulP}{\oc|mul'|\xspace}

\newcommand{\zeroR}{\oc|zero|_\Num\xspace}
\newcommand{\oneR}{\oc|one|_\Num\xspace}
\newcommand{\addR}{\oc|add|_\Num\xspace}
\newcommand{\mulR}{\oc|mul|_\Num\xspace}
\newcommand{\nR}{\tn_\Num\xspace}

\newcommand{\implementsnameR}{\mathit{isNum}_\Num}
\newcommand{\implementsR}[2]{#1\;\implementsnameR\;#2}
\newcommand{\protR}{\prot_\Num}

\newcommand{\numStruct}{\oc|dict|\xspace}
\newcommand{\toStruct}[4]{\{#1;\,#2;\,#3;\,#4\}}
\newcommand{\explicitNumStruct}{\toStruct\zero\one\add\mul}
\newcommand{\explicitNumStructR}{\toStruct\zeroR\oneR\addR\mulR}

\newcommand{\dotEval}[1]{#1\oc|.eval|} 

\newcommand{\ForwardInvname}{\mathit{ForwardInv}}
\newcommand{ \ForwardInv}[1]{\ForwardInvname\;#1}

\newcommand{\BackwardInvname}{\mathit{BackwardInv}}
\newcommand{\BackwardInvBodyname}{\mathit{BackwardInvBody}}
\newcommand{\BackwardInv}[1]{\BackwardInvname\;#1}
\newcommand{\BackwardInvBody}[3]{\BackwardInvBodyname\;#1\;#2\;#3}

\newcommand{\representsname}{\mathit{isSubExpRaw}}
\newcommand{\represents}[2]{#1\;\representsname\;#2}

\newcommand{\representsmodname}{\mathit{isSubExp}}
\newcommand{\representsmod}[2]{#1\;\representsmodname\,\;#2}

\newcommand{\isVarname}{\mathit{isVar}}
\newcommand{\isVar}[3]{#1\;\isVarname\,(#2,\;#3)}

\newcommand{\isHandlername}{\mathit{isHandler}}
\newcommand{\isHandler}[1]{\isHandlername\,#1}

\newcommand{\isContname}{\mathit{isCont}}
\newcommand{\isCont}[1]{\isContname\,#1}

\newcommand{\namecurrCtx}{\mathit{isContext}}
\newcommand{\currCtx}[1]{\namecurrCtx\,#1}

\newcommand{\nameisEntry}{\mathit{isBinding}}
\newcommand{\isEntry}[1]{\nameisEntry\,#1}

\newcommand{\protad}{\mathit{OP}}
\newcommand{\protaddef}{
  \protad \eqdef
    &\SEND
    \opVar\;\locl\;\locr\;\El\;\Er \,
    &(\opVar, \locl, \locr)\,
    &\{\represents\locl\El \isep \represents\locr\Er\}
    .\\
    &\RECV
    \locu\,
    &(\locu)\,
    &\{\represents\locu{(\Node\opVar\El\Er)}\}
 }

\newtheorem{goal}[thm]{Goal}       
\newtheorem{hypo}[thm]{Hypothesis} 
\theoremstyle{plain}\newtheorem{statement}[thm]{Statement}

\let\hypersetup\undefined


\begin{document}

\title[Verifying Effect-Handler-Based Reverse-Mode AD]{Verifying an Effect-Handler-Based\texorpdfstring{\\}{} Define-By-Run Reverse-Mode AD Library}
%
%

\author[P.~E.~de Vilhena]{Paulo Emílio de Vilhena\lmcsorcid{0000-0001-7379-310X}}[a]
\address{Imperial College London, United Kingdom}
\email{p.de-vilhena@imperial.ac.uk}

\author[F.~Pottier]{François Pottier\lmcsorcid{0000-0002-4069-1235}}[b]
\address{Inria, France}
\email{francois.pottier@inria.fr}

\begin{abstract}
We apply program verification technology to the problem of specifying and
verifying automatic differentiation (AD) algorithms. We focus on
define-by-run, a style of AD where the program that must be differentiated
is executed and monitored by the automatic differentiation algorithm. We begin
by asking, ``what is an implementation of AD?'' and ``what does it mean for an
implementation of AD to be correct?'' We answer these questions both at an
informal level, in precise English prose, and at a formal level, using types
and logical assertions. After answering these broad questions, we focus on a
specific implementation of AD, which involves a number of subtle
programming-language features, including dynamically allocated mutable state,
first-class functions, and effect handlers. We present a machine-checked proof,
expressed in a modern variant of Separation Logic, of its correctness. We view
this result as an advanced exercise in program verification, with potential
future applications to the verification of more realistic automatic
differentiation systems and of other software components that exploit
delimited-control effects.

\end{abstract}

\maketitle

\section*{Introduction}

\Ad (AD) is an important family of algorithms and techniques whose aim is to
allow the efficient and exact evaluation of the derivative of a function that
is defined programmatically. As very well put by the authors of the Wikipedia
entry on the topic, \emph{``\ad exploits the fact that every computer program, no
matter how complicated, executes a sequence of elementary arithmetic
operations (addition, multiplication,~etc.). By applying the chain rule,
derivatives can be computed automatically, accurately, and using at most a
small constant factor more arithmetic operations than the original program.''}
%
%
Griewank and Walther's textbook~\cite{griewank-walther} offers a comprehensive
introduction to AD; Baydin \etal.~\cite{baydin-pearlmutter-radual-siskind-17}
survey its applications in machine learning.



There are two main families of \AD \emph{algorithms}, namely the
\emph{forward-mode} and \emph{reverse-mode} algorithms.
As noted above, a program that one wishes to differentiate performs a sequence
of elementary arithmetic operations. A~forward-mode algorithm processes this
sequence in order: the earliest arithmetic operation is examined first.
A~reverse-mode algorithm processes it in reverse order: the earliest
arithmetic operation is examined last.
%
Reverse mode is attractive when differentiating a function whose number of
outputs is several orders of magnitude smaller than the number of its inputs,
such as a~function of type $\mathbb{R}^n\rightarrow\mathbb{R}^m$ where $m \ll
n$~\cite[Chapter~1]{griewank-walther}.



As an independent distinction, one can also identify several families of
\emph{programming-language techniques} that allow deploying \AD in practice.
One broad class of techniques relies on \emph{program transformation}. The
source code of the program~$P$ that one wishes to differentiate is supplied to
an automatic differentiation tool, a special-purpose compiler,
which produces source code for a differentiated program~$P'$.
Another broad class relies on \emph{program monitoring}. In this approach,
no source code for~$P$ is needed, and no source code for~$P'$ is produced.
When the execution of~$P'$ is requested,
the program~$P$ is executed instead.
Its execution is monitored
in such a way that
the sequence of elementary arithmetic operations
performed by~$P$ can be observed
and the value of~$P'$ can eventually be computed.
%
In some communities, implementations of \AD based on program transformation
are referred to as \emph{define-then-run}, because the source code of $P'$ is
constructed in a separate phase, before $P$ and $P'$ are compiled and run,
whereas implementations based on program monitoring are known as
\emph{define-by-run}, because they do not involve such a phase separation.
These approaches have different pragmatic strengths and weaknesses.
Because they have access to and rely on the syntax of the source program,
the program-transformation approaches
offer more scope for optimization,
therefore potentially greater efficiency,
but require greater implementation effort 
and are limited to a fixed programming-language subset.
The program-monitoring approaches
are usually less efficient
but can be easier to implement
and are not restricted to a fixed set of syntactic constructs.

In this paper, we wish to use program verification technology to
\emph{specify} and \emph{verify} an implementation of \ad.
That is, we wish to answer two main questions.
Our first question is in fact two-pronged:
\hypertarget{qOneA}{(1A)}%
~\emph{what is an implementation of \AD?}
and
\hypertarget{qOneB}{(1B)}%
~\emph{what does it mean for an implementation of \AD to be correct?}
In other words, we ask what is the \emph{type}
and what is the \emph{specification}
of an \AD implementation.
Our second question~is,
\hypertarget{qTwo}{(2)}%
~\emph{how does one prove that a specific implementation of \AD is correct?}

\newcommand{\qOneA}{Question~\hyperlink{qOneA}{1A}\xspace}
\newcommand{\qOneB}{Question~\hyperlink{qOneB}{1B}\xspace}
\newcommand{\qTwo}{Question~\hyperlink{qTwo}{2}\xspace}

These are broad questions. Depending on which family of \AD algorithms and
which kind of implementation technique are considered, the difficulty of
answering these questions may vary. In the following, we narrow down the scope
of these questions and focus on specific situations where we are able to
propose original answers to these questions.




\subsection*{Specifying Define-By-Run \AD}


We believe that,
in  the program-transformation approach,
also known as define-then-run,
the above questions are by now quite well understood.
%
%
In this approach, an implementation of \AD is a compiler.
In other words, it is a function of type \oc|exp -> exp|, where \oc|exp| is
an algebraic~data type of abstract syntax trees.
Such a compiler is correct if, when it is applied to an abstract syntax tree
whose denotation is a mathematical expression~$E$, it~produces an abstract
syntax tree whose denotation is~$E'$, the mathematical derivative of $E$.
Admittedly, this high-level statement is somewhat sketchy and glosses over
some details: for instance, we have implicitly assumed that the expression~$E$
has one free variable~$\X$, and we have written~$E'$ for the derivative of~$E$
with respect to~$\X$.
Also, we have not defined the type \oc|exp| of programmatic expressions
or the type of mathematical expressions~$E$.
Nevertheless, this suggests how the correctness of an \AD
program transformation scheme
can be stated.
Many such schemes are proposed and proved correct in the literature;
we discuss some of them at the end of the paper
(\sref{sec:related}).



For this reason, throughout this paper, we focus on the
program-monitoring approach, also known as define-by-run.
In this setting, our first contribution
is an original answer to \qOneA. We remark that, \emph{in the
  program-monitoring approach, an implementation of \AD can
  still be viewed as a function of type \oc|exp -> exp|},
as in the previous paragraph,
\emph{albeit under a different definition of the type \oc|exp|}.
Indeed, we remark that the
\emph{Church-B{\"o}hm-Berarducci encoding},
also known as the \emph{tagless-final} representation,
is a suitable definition of \oc|exp| for this purpose.
This representation does not allow
inspecting the source code of a program,
but does allow executing a program
and monitoring its execution,
to a certain extent.
This is the topic of Section \sref{sec:intf}.

Under this definition of the type \oc|exp|, a~function~\oc|diff : exp -> exp|
should not be viewed as a compiler: it does not consume or produce source
code. It is an ordinary function, which can be compiled once and for all,
placed in a library, and
invoked by user programs that wish to exploit its functionality.



Building upon this approach,
our next contribution
is an original answer to \qOneB.
We propose a formal specification
for an implementation of define-by-run \AD,
that is,
for a function \oc|diff : exp -> exp|.
We give this specification
exactly the same general form
as the correctness statement
that was sketched earlier
when discussing define-then-run \AD:
\emph{when \oc|diff| is applied to a value of type \oc|exp| that denotes a
  mathematical expression~$E$, it must produce a value of type \oc|exp| that
  denotes the mathematical expression~$E'$}.
%
The crux is to define,
under our new definition of the type \oc|exp|,
what it means for a runtime value of type \oc|exp| to
denote a mathematical expression~$E$.
This is where our second contribution lies,
and this is the topic of Section~\sref{sec:spec}.


Our specification is expressed in higher-order \SL
\cite{reynolds-02,ohearn-tutorial-08,chargueraud-20,iris-lecture-notes}.
\SL is compositional: once the specification of a library has been fixed,
an~implementation of this library and a program that uses this library can be
independently verified.
Thus, an implementation of \oc|diff| is verified with respect to the
specification stated in the previous paragraph, without any knowledge of how
\oc|diff| might be used in a client program.
Conversely, a client program is verified based only on this specification,
without any knowledge of how \oc|diff| is implemented.
For example,
under the assumption that \oc|diff| satisfies this specification,
it is easy to verify that
if \oc|e| denotes~$E$
then \oc|diff (diff e)| denotes~$E''$,
the second derivative of~$E$.
In other words, it is immediately obvious that \emph{every correct
  implementation of \oc|diff| supports iterated differentiation}.
This is a strong guarantee, which would be difficult to obtain in a naive way,
by studying a specific implementation of \oc|diff| and by attempting to
understand what happens at runtime when two invocations of this function are
nested.

At this point, a~reader may ask whether our answers to \qOneA and \qOneB
-- that is, our proposed type and specification -- are reasonable.
To begin convincing this reader, we propose three minimalist
implementations of \AD, each of which fits in one page or less. They include a
forward-mode implementation, based on dual numbers (\fref{fig:fmad}); a
reverse-mode implementation that exploits a stack data structure
(\fref{fig:stad}); and a reverse-mode implementation that exploits
\emph{effect handlers}~\cite{pretnar-intro-15} and does not explicitly involve
a stack (\fref{fig:ad}). This is the topic of Section~\sref{sec:instances}.
Each of these implementations has type \oc|exp -> exp|, suggesting that this
type is indeed a plausible answer to \qOneA.

Is~our proposed specification also a plausible answer to \qOneB? To
substantiate this claim, it would be desirable to verify that all three toy
implementations are correct with respect to this specification.
However, presenting three proofs would be space-consuming. Furthermore, the
first two implementations use well-known techniques, so a reader who is
acquainted with \AD should easily be convinced (at an informal level) that
they are correct.
For this reason, we narrow down the scope of our investigation and focus on
the third implementation, which may seem more exotic.
%
%
How does one verify that it is correct?




\subsection*{Verifying Effect-Based Define-By-Run Reverse-Mode \AD}

This question is a specific instance of \qTwo. Answering this question is the
third contribution of this paper. \emph{We present a machine-checked proof
  that our effect-based implementation of~\AD is correct with respect
  to our specification of \AD.} This proof is carried out in a variant of
higher-order \SL.


Our effect-based implementation of \AD
is inspired by Wang and Rompf~\cite{wang-rompf-18} and by Wang
\etal.~\cite{wang-19}, who implement reverse-mode \AD using
dynamically allocated mutable state and the delimited-control operators
\texttt{shift} and \texttt{reset}~\cite{danvy-filinski-90}.
%
%
It is inspired also by Sivaramakrishnan~\cite{kc-18}
and by Sigal~\cite{sigal-21},
who replace \texttt{shift/reset} with
\emph{effect handlers}~\cite{pretnar-intro-15},
a more structured form of delimited control.

%
Wang \etal.~\cite{wang-19} publish detailed descriptions of several versions
of their code, but no proof of its correctness.
In fact, Wang and Rompf~\cite{wang-rompf-18} make a particularly striking and
provoking claim about their code: \emph{``Our implementation is so concise
  that it can serve as a specification of reverse mode AD''}. Although their
code is indeed concise and clearly structured (and, we believe, so is our code
in \fref{fig:ad}), we object that it involves a combination of several
nontrivial programming-language features, including first-class functions,
dynamically allocated mutable state, and delimited control.
Therefore, it is not at all obvious how and why this kind of code works. In
particular, we argue that our code (\fref{fig:ad}) is a concise and elegant
implementation that deserves to be verified with respect to the simple
specification that we have proposed.

To the best of our knowledge, this program verification challenge has not been
addressed to this day, and seems highly nontrivial. To begin with, the
literature offers very few proofs of programs that involve effect handlers,
and virtually no proofs of programs that involve both effect handlers and
primitive dynamically allocated mutable state.
%
Furthermore, because, in our setting, \oc|exp| is a second-order function type,
\oc|diff : exp -> exp| is a function of order three, which means that the
dialogue between \oc|diff| and the outside world is particularly complex. Yet,
our proposed specification is particularly simple: when applied to a value
that denotes~$E$, \oc|diff| must return a value that denotes~$E'$. This
specification allows no visible side effect: it states that \oc|diff| must
behave like a pure function. Thus, we face the challenge of proving that
the use of mutable state and delimited control is encapsulated and cannot be
observed by a user of \texttt{diff}.



Addressing this challenge is the third and main contribution of this paper.
For greater readability, the code that is presented in the paper is expressed
in a real-life programming language, namely \mocaml
4.12.0~\cite{sivaramakrishnan-21},
an experimental extension of OCaml with effect handlers.%
\footnote{As of version 5, effects and effect handlers have been integrated in
mainstream OCaml, albeit in the form of primitive functions: no syntactic
constructs have been introduced in order to perform or handle effects.}
Unfortunately, OCaml is a large programming language, whose semantics is only
loosely defined.
For this reason, we cannot reason directly about the code in \fref{fig:ad}.
Instead, we introduce \lang,
a core \lc equipped with effect handlers,
whose syntax and semantics are
defined inside the Coq proof assistant.
We manually transcribe the code of \fref{fig:ad} into \lang. (This manual
transcription step is unverified. The resulting \lang code is not shown
in the paper.)
Then, we use \hazel~\cite{de-vilhena-pottier-21},
a \SL for \lang,
to express our specification of define-by-run \AD
and to verify that our \lang code is correct
with respect to this specification.
\hazel, which we have introduced in previous
work~\cite{de-vilhena-pottier-21}, is built on top of the Iris
framework~\cite{iris,mosel-18},
and includes support for reasoning about effect handlers.
\hazel is defined inside Coq.
The soundness of its reasoning rules is machine-checked.
Our use of these reasoning rules is also machine-checked,
so, in the end, we obtain a fully machine-checked proof
of the correctness of our \lang code.

An electronic supplement to this paper \cite{repo} offers the definition of
the language~\lang, the definition and (machine-checked) soundness proof of
\hazel, and the (machine-checked) correctness proof of our \lang code. It
also includes a short description of the correspondence between the Coq
definitions and the paper~\cite{guide}.
The gap between \mocaml and \lang is discussed at the end of the paper
(\sref{sec:conclusion}).



For the sake of simplicity, throughout the paper, we restrict our attention to
expressions of one variable, and perform differentiation with respect to this
variable. Generalizing our ideas to handle expressions of several variables is
possible and should be straightforward, but would add a certain amount of
clutter to the code, definitions, and proofs shown in this paper; we prefer to
avoid it.


\subsection*{Summary}

We view this paper primarily as an advanced exercise in program specification
and verification in the presence of first-class functions,
dynamically allocated mutable state, and delimited-control effects. We hope
that it may also offer insights to readers who are interested in \AD and in
the connection between \AD and delimited-control effects.

\subsection*{Outline}

The paper is organized as follows. We propose a minimalist API (that is, a
type) for a define-by-run \AD library (\sref{sec:intf}). Then, we propose a
specification for such a library~(\sref{sec:spec}). We present three
implementations of this API (\sref{sec:instances}). We briefly introduce
effect handlers~(\sref{sec:effects}) and explain how they are exploited in our
third implementation of \AD (\sref{sec:impl}). In preparation for the
verification of this code, we recall a few basic properties of mathematical
expressions and differentiation (\sref{sec:math}), and briefly present a~\SL
equipped with support for reasoning about effects and
handlers~(\sref{sec:hazel}). Thus equipped, we offer a step-by-step
presentation of the proof of our main result~(\sref{sec:verif}). We review the
related work (\sref{sec:related}) and conclude (\sref{sec:conclusion}).

\section{A Type for Define-By-Run AD}
\label{sec:intf}
%

In this section, we propose a minimalist API for an \AD library and describe
a few examples of programs that use this API to construct and differentiate
expressions.



\subsection{The API}
\label{sec:intf:intf}

Our proposed API appears in \fref{fig:sig}. It is expressed as an OCaml module
signature. It begins with the definitions of two types, \oc|'v dict| and
\oc|exp|. Then, it requires the differentiation algorithm to be presented
as a function \oc|diff| of type \oc|exp -> exp|.
This API prescribes what functionality must be offered by an \AD library, but
does not mandate how the library must be implemented: as we will
see~(\sref{sec:instances}), a variety of algorithmic techniques and
programming-language features can be used in an implementation of \diff.

The type of \oc|diff| seems easy to read: differentiation transforms an
expression into an expression. For the reader to fully understand this type,
there remains for us to explain the definition of the type \oc|exp|, thereby
answering the questions: what is a mathematical expression? What~is the
machine representation of a mathematical expression?

\begin{figure}[p]
\begin{lstlisting}[keywordstyle=\bfseries\color{RubineRed}]
(* A record of the ring operations over a numeric type 'v. *)
type 'v dict =(*@ \label{sig:num} *)
  { zero : 'v; one : 'v; add : 'v -> 'v -> 'v; mul : 'v -> 'v -> 'v }

(* An expression of one variable in tagless-final style. *)
type exp =(*@ \label{sig:exp} *)
  { eval : (* forall *) 'v. 'v dict -> 'v -> 'v }(*@ \label{sig:eval} *)

(* The automatic differentiation algorithm. *)
val diff : exp -> exp(*@ \label{sig:diff} *)
\end{lstlisting}
\caption{An OCaml API for define-by-run \AD}
\label{fig:sig}
\end{figure}

\begin{figure}[p]
\begin{lstlisting}[keywordstyle=\bfseries\color{RubineRed}]
(* The expression (x+1)^3. *)
let e : exp =
  { eval =
      fun (type v) ({ zero; one; add; mul } : v dict) x ->(*@ \label{use:intro} *)
        let ( + ), ( * ) = add, mul in(*@ \label{use:overload} *)
        let cube x = x * x * x in(*@ \label{use:cube} *)
        cube (x + one)(*@ \label{use:meat} *)
  }

(* Its derivative. *)
let e' : exp = diff e(*@ \label{use:diff} *)

(* The ring of the floating-point numbers. *)
let float = { zero = 0.0; one = 1.0; add = ( +. ); mul = ( *. ) }

(* Evaluating e' in floating point at 4.0. *)
let () = assert (e'.eval float 4.0 = 75.0)(*@ \label{use:eval} *)
\end{lstlisting}
\caption{An example use of the API}
\label{fig:use}
\end{figure}

\begin{figure}[p]
\begin{lstlisting}[keywordstyle=\bfseries\color{RubineRed},linerange=BEGINMONOMIAL-ENDMONOMIAL,includerangemarker=false,firstnumber=0]
(* Reverse-Mode Automatic Differentiation with Effect Handlers. *)

(* This code works in Multicore OCaml 4.12.0. *)

(* The use of effect handlers in this code is based on the code by KC
   Sivaramakrishnan:

   https://github.com/ocaml-multicore/effects-examples/blob/master/algorithmic_differentiation.ml

   The packaging is different, though; we use expressions in tagless-final
   style.

   This code can be executed in the OCaml toplevel loop via the following
   commands:

   opam switch create 4.12.0+domains+effects --no-switch # (only once) opam
   exec --switch 4.12.0+domains+effects ocaml # runs the toplevel loop

 *)

(* -------------------------------------------------------------------------- *)
(** Type definitions. *)

type 'v dict =
  { zero : 'v; one : 'v; add : 'v -> 'v -> 'v; mul : 'v -> 'v -> 'v }

type exp =
  { eval : (* forall *) 'v. 'v dict -> 'v -> 'v }

module type AD = sig
  val diff : exp -> exp
end

(* -------------------------------------------------------------------------- *)
(** Forward-mode AD. *)

module ForwardMode : AD = struct

(* BEGINFMAD *)
let diff (e : exp) : exp = { eval =(*@ \label{line:fmad:e} *)
  fun (type v) ({ zero; one; add; mul } : v dict) (n : v) ->(*@ \label{line:fmad:v}\label{line:fmad:n} *)
    let ( + ), ( * ) = add, mul in
    let open struct(*@ \label{line:fmad:openstruct} *)
      type t = { v : v ; d : v }(*@ \label{line:fmad:type_t} *)
      let mk v d = { v; d }(*@ \label{line:fmad:mk} *)
      let dict =(*@ \label{line:fmad:num} *)
        let zero = mk zero zero(*@ \label{line:fmad:zero} *)
        and one = mk one zero(*@ \label{line:fmad:one} *)
        and add a b = mk (a.v + b.v) (a.d + b.d)(*@ \label{line:fmad:add} *)
        and mul a b = mk (a.v * b.v) (a.d * b.v + a.v * b.d)(*@ \label{line:fmad:mul} *)
        in { zero; one; add; mul }(*@ \label{line:fmad:dict} *)
      let x = mk n one(*@ \label{line:fmad:x} *)
      let y = e.eval dict x(*@ \label{line:fmad:y} *)
    end in
    y.d(*@ \label{line:fmad:read:result} *)
}
(* ENDFMAD *)

end

(* -------------------------------------------------------------------------- *)
(** Stack-based implementation of reverse-mode AD. *)

module StackBasedReverseMode : AD = struct

(* BEGINSTAD *)
let diff (e : exp) : exp = { eval =(*@ \label{line:stad:e} *)
  fun (type v) ({ zero; one; add; mul } : v dict) (n : v) ->(*@ \label{line:stad:v}\label{line:stad:n} *)
    let ( + ), ( * ) = add, mul in
    let open struct(*@ \label{line:stad:openstruct} *)
      (* The graph. *)
      type t = O | I | Var of { v : v ; mutable d : v }(*@ \label{line:stad:type_t} *)
      let mk n       = Var { v = n; d = zero }(*@ \label{line:stad:mk} *)
      let get_v u    = match u with O -> zero | I -> one  | Var u     -> u.v(*@ \label{line:stad:get_v} *)
      let get_d u    = match u with O | I -> assert false | Var u     -> u.d(*@ \label{line:stad:get_d} *)
      let update u i = match u with O | I -> ()    | Var u -> u.d <- u.d + i(*@ \label{line:stad:update} *)
      (* The stack. *)
      type op = Add | Mul(*@ \label{line:stad:type_op} *)
      type binding = Let of t * (t * op * t)(*@ \label{line:stad:type_letbinding} *)
      open Stack(*@ \label{line:stad:openstack} *)
      let stack : binding Stack.t = create()(*@ \label{line:stad:stack} *)
      (* The dictionary used in the forward phase. *)
      let dict =(*@ \label{line:stad:num} *)
        let zero = O
        and one = I
        and add a b =(*@ \label{line:stad:add} *)
          let u = mk (get_v a + get_v b) in
          push (Let (u, (a, Add, b))) stack; u
        and mul a b =(*@ \label{line:stad:mul} *)
          let u = mk (get_v a * get_v b) in
          push (Let (u, (a, Mul, b))) stack; u
        in { zero; one; add; mul }
      (* The forward phase. *)
      let x = mk n(*@ \label{line:stad:x} *)
      let y = e.eval dict x(*@ \label{line:stad:y} *)
      (* The backward phase. *)
      let () =(*@ \label{line:stad:backwardphase} *)
        update y one;(*@ \label{line:stad:update:y} *)
        while not (is_empty stack) do(*@ \label{line:stad:while} *)
          match pop stack with(*@ \label{line:stad:pop} *)
          | Let (u, (a, Add, b)) ->
              update a (get_d u);(*@ \label{line:stad:update:add:a} *)
              update b (get_d u)(*@ \label{line:stad:update:add:b} *)
          | Let (u, (a, Mul, b)) ->
              update a (get_d u * get_v b);(*@ \label{line:stad:update:mul:a} *)
              update b (get_d u * get_v a)(*@ \label{line:stad:update:mul:b} *)
        done(*@ \label{line:stad:while:done} *)
    end in
    get_d x(*@ \label{line:stad:read:result} *)
}
(* ENDSTAD *)

end

(* -------------------------------------------------------------------------- *)
(** Reverse-mode AD using effect handlers. *)

module EffectHandlerBasedReverseMode : AD = struct

(* BEGINEHAD *)
let diff (e : exp) : exp = (*@\codelabel{head}*) { eval =(*@ \label{line:e} *)
  fun (type v) ({ zero; one; add; mul } : v dict) (n : v) -> (*@\codelabel{body}*) (*@ \label{line:v}\label{line:n} *)
    let ( + ), ( * ) = add, mul in
    let open struct(*@ \label{line:openstruct} *)
      (* The graph. *)
      type t = O | I | Var of { v : v ; mutable d : v }(*@ \label{line:type_t} *)
      let mk n       = Var { v = n; d = zero }(*@ \label{line:mk} *)
      let get_v u    = match u with O -> zero | I -> one  | Var u     -> u.v(*@ \label{line:get_v} *)
      let get_d u    = match u with O | I -> assert false | Var u     -> u.d(*@ \label{line:get_d} *)
      let update u i = match u with O | I -> ()    | Var u -> u.d <- u.d + i(*@ \label{line:update} *)
      (* The dictionary. *)
      effect Add : t * t -> t(*@ \label{line:Add} *)
      effect Mul : t * t -> t(*@ \label{line:Mul} *)
      let zero' = O(*@ \label{line:zeroP} *)
      and one' = I(*@ \label{line:oneP} *)
      and add' a b = perform (Add (a, b))(*@ \label{line:add} *)
      and mul' a b = perform (Mul (a, b))(*@ \label{line:mul} *)
      let dict = { zero = zero'; one = one'; add = add'; mul = mul'}(*@ \label{line:dict} *)
    end in (*@ \label{line:endstruct} *)
    (* The forward and backward phases. *)
    (*@\codelabel{init}*)
    let x = mk n in(*@ \label{line:x} *)
    (*@\codelabel{heart}*)
    let _ =(*@ \label{line:handle} *)
      (*@\codelabel{handle}*) match (*@\codelabel{eval}*) e.eval dict x with(*@ \label{line:match}\label{line:eval} *)
      | (*@\codelabel{eff-add}*) effect (Add (a, b)) k ->(*@ \label{line:handle:Add} *)
          (*@\codelabel{add-body}*)
          let u = (*@\codelabel{add-fwd}*) mk (get_v a + get_v b) in(*@ \label{line:handle:Add:mk} *)
          let _ = (*@\codelabel{cont}*) continue k u in(*@ \label{line:continue1} *)
          (*@\codelabel{add-bwd}*)
          update a (get_d u);(*@ \label{line:handle:Add:update:a} *)
          update b (get_d u)(*@ \label{line:handle:Add:update:b} *)
      | (*@\codelabel{eff-mul}*) effect (Mul (a, b)) k ->(*@ \label{line:handle:Mul} *)
          let u = mk (get_v a * get_v b) in(*@ \label{line:handle:Mul:mk} *)
          let _ = continue k u in (*@ \label{line:continue2} *)
          update a (get_d u * get_v b);(*@ \label{line:handle:Mul:update:a} *)
          update b (get_d u * get_v a)(*@ \label{line:handle:Mul:update:b} *)
      | (*@\codelabel{ret}*) y ->(*@ \label{line:handle:Return} *)
          (*@\codelabel{seed}*) update y one(*@ \label{line:handle:one} *)
    in
    (*@\codelabel{done}*) get_d x(*@ \label{line:read:result} *)
}
(* ENDEHAD *)

end

(* -------------------------------------------------------------------------- *)
(** Some dictionaries. *)

let int   = { zero = 0; one = 1; add = ( + ); mul = ( * ) }
let float = { zero = 0.0; one = 1.0; add = ( +. ); mul = ( *. ) }

(* -------------------------------------------------------------------------- *)
(** Some expressions. *)

(* [f] represents the expression [x ↦ (x+1)^3]. *)

let f =
  { eval = fun {one; add; mul; _} x ->
      let ( + ), ( * ) = add, mul in
      let cube x = x * x * x in
      cube (one + x)
  }

(* Test. *)

let () =
  assert (f.eval int 2 = 27);
  assert (f.eval float 2.0 = 27.0);
  ()

(* [var] represents the expression x. *)

let var : exp =
  { eval =
      fun (type v) _dict (x : v) ->
        x
  }

(* [subst e1 e2] substitutes the expression [e1] for the (unique) variable
   in the expression [e2], yielding a new expression. *)

let subst (e1 : exp) (e2 : exp) : exp =
  { eval =
      fun (type v) dict (x : v) ->
        e2.eval dict (e1.eval dict x)
  }

(* Monomial and power. *)

(* BEGINMONOMIAL *)
(* The expression x^k. *)
let monomial (k : int) : exp = { eval =
  fun (type v) { one; mul; _ } (x : v) ->
    let ( * ) = mul in
    let result, x, k = ref one, ref x, ref k in
    while !k > 0 do(*@ \label{line:monomial:while} *)
      result := !result * (if !k mod 2 = 0 then one else !x);(*@ \label{line:monomial:update:result} *)
      x := !x * !x;(*@ \label{line:monomial:update:x} *)
      k := !k / 2
    done;(*@ \label{line:monomial:done} *)
    !result }
(* ENDMONOMIAL *)

let pow (k : int) (e : exp) : exp =
  subst e (monomial k)

(* Test. *)
let () =
  assert ((monomial 8).eval float 2.0 = 256.0);
  assert ((pow 2 (pow 4 var)).eval float 2.0 = 256.0);
  ()

(* -------------------------------------------------------------------------- *)
(** Testing an implementation of AD. *)

(* [Test(A)] tests that [A.diff] behaves correctly on some simple inputs. *)
module Test (A : AD) = struct

  open A

  (* [df] should be [x ↦ 3(x+1)^2]. *)
  let df = diff f

  (* [ddf] should be [x ↦ 6(x+1)]. *)
  let ddf = diff (diff f)

  let () =
    assert  (df.eval int   2  = 27);
    assert  (df.eval float 2. = 27.);
    assert (ddf.eval int   2  = 18);
    assert (ddf.eval float 2. = 18.);
    ()

  (* [monomial 4] is [x^4], so its derivative its [4.x^3], whose evaluation
     at 2 should yield 32. *)
  let () =
    assert ((diff (monomial 4)).eval float 2.0 = 32.0);
    ()

end

(* -------------------------------------------------------------------------- *)
(** Testing each of our three implementations. *)

(* A list of our implementations of AD. *)
let algorithms : (module AD) list = [
  (module ForwardMode);
  (module StackBasedReverseMode);
  (module EffectHandlerBasedReverseMode);
]

(* Test every instance in [algorithms]. *)
let () =
  List.iter (fun (module A : AD) ->
    let module T = Test(A) in ()
  ) algorithms

(* If we reach this point, then the tests have succeeded. *)
let () =
  print_endline "Success."
\end{lstlisting}
\caption{Another example use of the API: the function \monomial}
\label{fig:monomial}
\end{figure}

In this paper, we consider mathematical expressions that are constructed out
of the constants~0 and~1, addition, multiplication, and a single variable~$x$.
How might these mathematical expressions be represented in memory? As noted
earlier, a natural idea might be to represent them as abstract syntax trees.
One would do this by declaring \oc|exp| as an algebraic data type. However,
that would lead to an API for \emph{define-then-run} \AD, that is, for an
implementation of \AD as a program transformation, accepting source code and
producing source code. In this paper, instead, we wish to focus on
\emph{define-by-run} \AD, a~variant of~\AD that relies on a form of program
monitoring. To do so, we adopt an alternative representation of
expressions, namely the Church-B{\"o}hm-Berarducci
encoding~\cite{kiselyov-beyond-church-12},
also known as the tagless-final
representation~\cite{carette-kiselyov-shan-09,kiselyov-tagless-final-10}.
%
Remarkably, provided we define the type \oc|exp| in this way,
we can keep the convention that \oc|diff| has type
\oc|exp -> exp|.
That is, the type \oc|exp -> exp| can describe both define-then-run AD and
define-by-run AD!

In the tagless-final representation,
\emph{an expression is represented as a computation}.
Such a computation is given access to the four operations
(zero, one, addition, multiplication)
and to the value of the single variable
and must produce a value.
This is visible in \fref{fig:sig}, where an expression is represented as a
function of type \oc|'v dict -> 'v -> 'v| (line~\ref{sig:eval}). The first
argument, of type \oc|'v dict|, is a dictionary, that is, a record of four
fields, containing the implementations of the four operations. The second
argument, of type \oc|'v|, is the value of the variable. The result, also of
type \oc|'v|, is the value of the expression.

In fact, \emph{an expression is represented as a polymorphic computation}.
Indeed, the definition of the type \oc|exp| includes a universal
quantification over the type~\oc|'v|, thereby requiring every expression
to be polymorphic in the type \oc|'v|.%
\footnote{Technically, the type \oc|exp| is defined as a record
with one field, named \oc|eval|, which contains a polymorphic function:
for every type \oc|'v|, this function must have type \oc|'v dict -> 'v -> 'v|.}
This means that an expression must not care what kind of numbers it is applied
to.
Because polymorphism in OCaml is parametric~\cite{reynolds-83}, this
requirement limits the ways in which an expression can interact with
\emph{numbers} -- that is, with values of type \oc|'v|.
The only numbers to
which an expression initially has access are its second argument and the
dictionary fields \oc|zero| and \oc|one|. To create new numbers, an
expression must call the dictionary operations \oc|add| or \oc|mul|.
There is
no way for an expression to inspect a number.

This representation of expressions allows an expression to be evaluated in
many different ways, involving many different kinds of numbers. For
instance, by applying it to a suitable type of numbers and to a suitable
dictionary, one can evaluate it using integer arithmetic, evaluate it using
floating-point arithmetic, or convert it to a symbolic representation.
%
%
As will be demonstrated later on (\sref{sec:instances}, \sref{sec:impl}), this
representation of expressions allows several different implementations of \AD.

Our implementation of \rmad using effect handlers~(\sref{sec:impl})
evaluates the expression that one wishes to differentiate
under a nonstandard implementation of the four operations,
where addition and multiplication perform control effects.
Although this is not visible in the definition of the type \oc|exp|,%
\footnote{In \mocaml, at present, the type system does not keep track of
the effects that a function might perform. In other words, function types
are not annotated with effect information. Any function can in principle
perform any effect. An unhandled effect causes a runtime error.}
this implementation exploits the fact that \emph{an expression must be an
  effect-polymorphic computation}. That is, an expression must not care what
control effects (if any) are performed by the operations \oc|add| and
\oc|mul|.
This idea is explicitly spelled out in the next section (\sref{sec:spec}),
where we propose a specification for \diff.



\subsection{Example Uses of the API}
\label{sec:intf:examples}

%
A first example use of this API appears in \fref{fig:use}. It can be linked
with an arbitrary implementation of this API, that is, with an arbitrary
implementation of \diff.
This code
first builds a representation~\oc|e| of the mathematical expression
$(x+1)^3$. This construction is unfortunately rather verbose. Line~\ref{use:intro}
introduces the type parameter~\oc|v| and the five value parameters \oc|zero|,
\oc|one|, \oc|add|, \oc|mul|, and~\oc|x|. Line~\ref{use:overload} redefines
the infix operators~\,\oc|+| and \oc|*| as aliases for \oc|add| and \oc|mul|.
(One might wish to also redefine \oc|0| and \oc|1| as
aliases for \oc|zero| and \oc|one|, but OCaml does not allow this.)
The meat of the definition of~\oc|e|
is at lines~\ref{use:cube}--\ref{use:meat}.
Then, on line~\ref{use:diff}, \diff is applied to~\oc|e|,
yielding a representation~\oc|e'| of the
derivative of \oc|e| with respect to the variable~$x$.
Exactly what this function call does depends on how \diff is implemented. With
all three implementations shown in this paper, this function call terminates
immediately: it simply allocates and returns a~closure.
\Ad really takes places only at line~\ref{use:eval}, which
requests the evaluation of the expression~\oc|e'| using floating-point
arithmetic.
There, because we expect \oc|e'| to be equivalent to the expression
$3\cdot(x+1)^2$, and because we instantiate $x$ with the value~4, we expect the
result to be~75. Executing the code validates this expectation: with each of
the three implementations of \AD shown in this paper, the runtime assertion at
line~\ref{use:eval} succeeds.



A second example use of our API appears in \fref{fig:monomial}. The OCaml
expression \oc|monomial k| is intended to represent the mathematical
expression $x^k$.
%
%
%
This code is a textbook implementation of fast exponentiation.
It uses a number of nontrivial programming-language features, including
primitive operations on integers (division, modulus, comparisons), several
control constructs (a conditional construct, a loop),
and mutable references.
One might naively fear that this might prevent the application of~\diff to
\oc|monomial k|: indeed, in general, a program that uses these features is
not necessarily differentiable.
However, such a fear is unwarranted.
The use of these features in the definition of \monomial is an implementation
detail: it is not visible to an observer who monitors the behavior of
\oc|monomial k|. In particular, it cannot be observed by the function \diff
when \diff is applied to \oc|monomial k|.

We propose to reason about \oc|monomial k|,
and about an application of \diff to \oc|monomial k|,
in the following way.
To a user, how \oc|monomial| is implemented does not matter;
what matters is that
\oc|monomial k| represents the mathematical expression~$x^k$.
(The meaning of this claim is clarified in~\sref{sec:spec}.)
This is our proposed specification of the function \monomial.
This specification, together with our proposed specification of \diff,
guarantees that the function application \oc|diff (monomial k)| is safe
and that its result represents the mathematical expression~$k \cdot x^{k-1}$.

This reasoning illustrates the appeal of a compositional approach to program
verification.
Thinking in terms of abstract specifications makes it easy to
see that, if each of \diff and \monomial independently satisfies its
specification, then \oc|diff (monomial k)| must behave as intended. In
contrast, unfolding the concrete definitions of \diff and \monomial and
attempting to imagine what machine behavior results from the combination of
these definitions would be much more difficult.

This example also illustrates a~strength of define-by-run \AD over
define-then-run \AD. Even though an implementation of \diff has no knowledge
of primitive integers, conditionals, loops, or mutable references, it can
nevertheless be applied to a piece of OCaml code that uses these features,
provided this code obeys its contract, which is to represent a certain
mathematical expression~$E$. The expression~$E$ must inhabit a fixed subset of
the mathematical language, which (in this paper) includes the constants 0 and
1, addition, multiplication, and a single variable~$x$. Thus, whereas
define-then-run \AD can be applied only to a fixed subset of programs,
define-by-run \AD can be applied to an arbitrary program, as long as its
denotation inhabits a fixed subset of mathematical expressions.


\section{A Specification for Define-By-Run AD}
\label{sec:spec}
In the previous section (\sref{sec:intf}), we have presented an API for
define-by-run \AD. This API consists of a definition of the type $\tyExp$ and
a type for the function \diff. It is expressed in the
polymorphic-$\lambda$-calculus fragment of OCaml: it involves product types,
function types, and universal types. It describes the runtime representation
of the values that are exchanged between the function \diff and its
environment, but does not describe the intended meaning of these runtime
values.

We now go one step further and propose a specification for \diff. This
specification clarifies the connection between the runtime values that exist
in the machine's memory and the mathematical objects that these values
represent. Furthermore, it prescribes which values may or must be exchanged at
each point of the dialogue between \diff and its environment. It does not
mandate a specific implementation of \diff: as we shall see
(\sref{sec:instances}, \sref{sec:impl}), several implementations are possible.

For the moment, we express this specification in English, in an informal yet
precise style. Later on in the paper
(\sref{sec:verif:spec}), we show that
this specification can be formally expressed in Hazel~%
\cite{de-vilhena-pottier-21}, a variant of higher-order \SL equipped with
support for effects and effect handlers.

In the following, we write $\E$ for a mathematical expression in the fixed
universe defined by $\E ::= \X \mid 0 \mid 1 \mid \E+\E \mid \E\times\E$. We
write $\deriv\E$ for the derivative of~$\E$ with respect to the variable~$\X$.
(There is just one variable, named $\X$.) We use these mathematical expressions
in specifications only; they do not exist at runtime.

We write $\ev$ for a runtime value of type~$\tyExp$. We do not describe the
syntax or memory layout of runtime values, as it is irrelevant: what matters
is how runtime functions \emph{behave} when invoked.
%

The specification of \diff is very short:
\begin{statement}[Informal Specification of Differentiation]
\label{informal:spec:diff}
  If the value $\ev$ represents $\E$,
  then the function call $\App\diff\ev$
  diverges or returns a value $\ev'$ that represents $\deriv\E$.
\end{statement}

A reader who is curious to know how this statement is expressed in \SL may
wish to peek ahead at Statement~\ref{formal:spec:diff}.

%
Throughout the paper, we use a logic of partial correctness, which is why
divergence (nontermination) is not forbidden by our specification. By
convention, let us write \emph{yields} for \emph{diverges or returns}.
A reformulation of the above statement, which uses this short-hand and
avoids introducing the names $\ev$ and $\ev'$, is as follows:
%
%
\emph{when applied to a representation of~$\E$, the function~$\diff$ yields a
representation of~$\deriv\E$}.

This specification is extremely simple, but relies on a key auxiliary
definition, which we have not yet given. There remains to define the assertion
\emph{$\ev$~represents~$\E$}, that is, to specify what it means for a runtime
value $\ev$ to represent a mathematical expression~$\E$.
This is a bit more involved. Very roughly speaking, we wish to express the
idea that applying the function $\dotEval\ev$ to a dictionary of arithmetic
operations $(0, 1, \mathord+, \mathord\times)$ and to a number~$\num$ must
yield the value of the expression~$\E$ at the point $\X=\num$. However, such a
sentence glosses over several important points.

First, as noted earlier (\sref{sec:intf}), $\dotEval\ev$ must be polymorphic:
it must be able to compute with numbers of an arbitrary nature, provided
these numbers are equipped with sufficient structure. For our purposes, the
structure of a semiring suffices. Thus, our definition must involve a
universal quantification over a semiring~$\Num$. In the following, by
convention, when we write \emph{a mathematical number} or \emph{a number},
we mean \emph{an element of~$\Num$}.
We write $\mathord+$ and $\mathord\times$ for the
addition and multiplication operations of the semiring~$\Num$ and $0$ and $1$
for their neutral elements.

Second, Hoare Logic and \SL impose maintaining a distinction between a~runtime
value and the mathematical object that this value is meant to represent. Thus,
we must distinguish between runtime values such as $\zero$ and $\one$ and
numbers such as~$0$ and~$1$. We must also distinguish between the runtime
values $\add$ and $\mul$ and the semiring operations~$\mathord+$
and~$\mathord\times$, and explain how they must be related.

From these remarks, it follows that we must quantify not only over the
semiring~$\Num$, but also over the runtime values $\zero, \one, \add, \mul$
and over the relation that describes what it means for a runtime value to
represent a number. Quantifying over a relation may at first seem surprising,
but is in fact perfectly natural: such a higher-order quantification is
typically encountered in binary-logical-relation interpretations of universal
types.

A last key aspect is the treatment of control effects. The definition of
\emph{$\ev$~represents~$\E$} must specify which function invocations may perform
effects and what effects they may perform. As noted earlier~(\sref{sec:intf}),
we would like $\dotEval\ev$ to be effect-polymorphic. More precisely, we wish
to allow the functions \add and \mul to perform arbitrary effects.
Furthermore, we~wish to forbid the function $\dotEval\ev$ from handling these
effects or performing any effects of its own. Thus, the effects performed by
\add and \mul, and only these effects, can be observed outside $\dotEval\ev$.
Later on~(\sref{sec:hazel}, \sref{subsubsec:protocol}),
we introduce the notion of a \emph{protocol}~$\prot$,
which describes the effects that a function may perform.
For now, let us posit that, to express that~$\ev$ is effect-polymorphic,
the definition of \emph{$\ev$~represents~$\E$} must quantify over a protocol:
the universal quantification over~$\prot$ and the three occurrences of~$\prot$ in
the following definition reflect the idea that \emph{whatever effects \add and
\mul may perform, $\dotEval\ev$ may perform these effects (and no more)}.

\begin{statement}[Informal Runtime Representation of Expressions]
\label{informal:isExp}
A runtime value~$\ev$ represents an expression~$\E$ if
\begin{itemize}
\item for every semiring~$\Num$,
\item for every protocol~$\prot$,
\item 
      for every possible meaning of the assertion
      ``the runtime value~$\valnum$ represents the number~$\num$'',
\item 
      for all runtime values $\zero$, $\one$, $\add$, $\mul$ such that
      \begin{itemize}
      \item \zero represents $0$,
      \item \one represents $1$,
      \item when applied to representations of $\num$ and $\numb$, \\
            $\add$ yields a representation of $\num+\numb$ \\
            and may perform effects permitted by~$\prot$,
      \item when applied to representations of $\num$ and $\numb$, \\
            $\mul$ yields a representation of $\num\times\numb$ \\
            and may perform effects permitted by~$\prot$,
      \end{itemize}
\item for every number $\num$,
\item when applied to the record $\{\zero;\one;\add;\mul\}$
      and to a representation of the number~$\num$,
      the function $\dotEval\ev$
      yields a representation of the value
      of the expression $\E$ at $\X = \num$
      and may perform effects permitted by~$\prot$.
\end{itemize}
\end{statement}

Although this statement remains informal, it should give the reader a~fairly
exact preview of the formal definitions and statements that we present later
on (\sref{sec:verif:spec}). Only one technical aspect has been omitted above,
namely the use of Iris's \emph{persistence} modality in several places to
indicate that certain runtime values can be used as many times as one wishes.

\section{Forward-Mode and Stack-Based Reverse-Mode Implementations of AD}
\label{sec:instances}


In this section,
we briefly present two OCaml implementations of define-by-run \AD.
Each of them respects the API that we have proposed (\sref{sec:intf}),
and we believe (but we have not formally verified) that each of them
meets the specification that we have proposed (\sref{sec:spec}).
The first implementation performs forward-mode \AD.
The second implementation performs reverse-mode \AD
and uses a dynamically allocated mutable stack.

There are several reasons why we present these two implementations.
First, these examples help demonstrate that our API (\sref{sec:intf}) and
specification (\sref{sec:spec}) do admit several quite different
implementations.
%
%
Second, they allow us to recall some of the basic principles of forward-mode
and reverse-mode \AD.
Third, they prepare the reader to the presentation of a~third
implementation (\sref{sec:impl}), which exploits effect handlers
and performs reverse-mode \AD.
%

\begin{figure}[t]
\begin{lstlisting}[keywordstyle=\bfseries\color{RubineRed},linerange=BEGINFMAD-ENDFMAD,includerangemarker=false,firstnumber=0]
(* Reverse-Mode Automatic Differentiation with Effect Handlers. *)

(* This code works in Multicore OCaml 4.12.0. *)

(* The use of effect handlers in this code is based on the code by KC
   Sivaramakrishnan:

   https://github.com/ocaml-multicore/effects-examples/blob/master/algorithmic_differentiation.ml

   The packaging is different, though; we use expressions in tagless-final
   style.

   This code can be executed in the OCaml toplevel loop via the following
   commands:

   opam switch create 4.12.0+domains+effects --no-switch # (only once) opam
   exec --switch 4.12.0+domains+effects ocaml # runs the toplevel loop

 *)

(* -------------------------------------------------------------------------- *)
(** Type definitions. *)

type 'v dict =
  { zero : 'v; one : 'v; add : 'v -> 'v -> 'v; mul : 'v -> 'v -> 'v }

type exp =
  { eval : (* forall *) 'v. 'v dict -> 'v -> 'v }

module type AD = sig
  val diff : exp -> exp
end

(* -------------------------------------------------------------------------- *)
(** Forward-mode AD. *)

module ForwardMode : AD = struct

(* BEGINFMAD *)
let diff (e : exp) : exp = { eval =(*@ \label{line:fmad:e} *)
  fun (type v) ({ zero; one; add; mul } : v dict) (n : v) ->(*@ \label{line:fmad:v}\label{line:fmad:n} *)
    let ( + ), ( * ) = add, mul in
    let open struct(*@ \label{line:fmad:openstruct} *)
      type t = { v : v ; d : v }(*@ \label{line:fmad:type_t} *)
      let mk v d = { v; d }(*@ \label{line:fmad:mk} *)
      let dict =(*@ \label{line:fmad:num} *)
        let zero = mk zero zero(*@ \label{line:fmad:zero} *)
        and one = mk one zero(*@ \label{line:fmad:one} *)
        and add a b = mk (a.v + b.v) (a.d + b.d)(*@ \label{line:fmad:add} *)
        and mul a b = mk (a.v * b.v) (a.d * b.v + a.v * b.d)(*@ \label{line:fmad:mul} *)
        in { zero; one; add; mul }(*@ \label{line:fmad:dict} *)
      let x = mk n one(*@ \label{line:fmad:x} *)
      let y = e.eval dict x(*@ \label{line:fmad:y} *)
    end in
    y.d(*@ \label{line:fmad:read:result} *)
}
(* ENDFMAD *)

end

(* -------------------------------------------------------------------------- *)
(** Stack-based implementation of reverse-mode AD. *)

module StackBasedReverseMode : AD = struct

(* BEGINSTAD *)
let diff (e : exp) : exp = { eval =(*@ \label{line:stad:e} *)
  fun (type v) ({ zero; one; add; mul } : v dict) (n : v) ->(*@ \label{line:stad:v}\label{line:stad:n} *)
    let ( + ), ( * ) = add, mul in
    let open struct(*@ \label{line:stad:openstruct} *)
      (* The graph. *)
      type t = O | I | Var of { v : v ; mutable d : v }(*@ \label{line:stad:type_t} *)
      let mk n       = Var { v = n; d = zero }(*@ \label{line:stad:mk} *)
      let get_v u    = match u with O -> zero | I -> one  | Var u     -> u.v(*@ \label{line:stad:get_v} *)
      let get_d u    = match u with O | I -> assert false | Var u     -> u.d(*@ \label{line:stad:get_d} *)
      let update u i = match u with O | I -> ()    | Var u -> u.d <- u.d + i(*@ \label{line:stad:update} *)
      (* The stack. *)
      type op = Add | Mul(*@ \label{line:stad:type_op} *)
      type binding = Let of t * (t * op * t)(*@ \label{line:stad:type_letbinding} *)
      open Stack(*@ \label{line:stad:openstack} *)
      let stack : binding Stack.t = create()(*@ \label{line:stad:stack} *)
      (* The dictionary used in the forward phase. *)
      let dict =(*@ \label{line:stad:num} *)
        let zero = O
        and one = I
        and add a b =(*@ \label{line:stad:add} *)
          let u = mk (get_v a + get_v b) in
          push (Let (u, (a, Add, b))) stack; u
        and mul a b =(*@ \label{line:stad:mul} *)
          let u = mk (get_v a * get_v b) in
          push (Let (u, (a, Mul, b))) stack; u
        in { zero; one; add; mul }
      (* The forward phase. *)
      let x = mk n(*@ \label{line:stad:x} *)
      let y = e.eval dict x(*@ \label{line:stad:y} *)
      (* The backward phase. *)
      let () =(*@ \label{line:stad:backwardphase} *)
        update y one;(*@ \label{line:stad:update:y} *)
        while not (is_empty stack) do(*@ \label{line:stad:while} *)
          match pop stack with(*@ \label{line:stad:pop} *)
          | Let (u, (a, Add, b)) ->
              update a (get_d u);(*@ \label{line:stad:update:add:a} *)
              update b (get_d u)(*@ \label{line:stad:update:add:b} *)
          | Let (u, (a, Mul, b)) ->
              update a (get_d u * get_v b);(*@ \label{line:stad:update:mul:a} *)
              update b (get_d u * get_v a)(*@ \label{line:stad:update:mul:b} *)
        done(*@ \label{line:stad:while:done} *)
    end in
    get_d x(*@ \label{line:stad:read:result} *)
}
(* ENDSTAD *)

end

(* -------------------------------------------------------------------------- *)
(** Reverse-mode AD using effect handlers. *)

module EffectHandlerBasedReverseMode : AD = struct

(* BEGINEHAD *)
let diff (e : exp) : exp = (*@\codelabel{head}*) { eval =(*@ \label{line:e} *)
  fun (type v) ({ zero; one; add; mul } : v dict) (n : v) -> (*@\codelabel{body}*) (*@ \label{line:v}\label{line:n} *)
    let ( + ), ( * ) = add, mul in
    let open struct(*@ \label{line:openstruct} *)
      (* The graph. *)
      type t = O | I | Var of { v : v ; mutable d : v }(*@ \label{line:type_t} *)
      let mk n       = Var { v = n; d = zero }(*@ \label{line:mk} *)
      let get_v u    = match u with O -> zero | I -> one  | Var u     -> u.v(*@ \label{line:get_v} *)
      let get_d u    = match u with O | I -> assert false | Var u     -> u.d(*@ \label{line:get_d} *)
      let update u i = match u with O | I -> ()    | Var u -> u.d <- u.d + i(*@ \label{line:update} *)
      (* The dictionary. *)
      effect Add : t * t -> t(*@ \label{line:Add} *)
      effect Mul : t * t -> t(*@ \label{line:Mul} *)
      let zero' = O(*@ \label{line:zeroP} *)
      and one' = I(*@ \label{line:oneP} *)
      and add' a b = perform (Add (a, b))(*@ \label{line:add} *)
      and mul' a b = perform (Mul (a, b))(*@ \label{line:mul} *)
      let dict = { zero = zero'; one = one'; add = add'; mul = mul'}(*@ \label{line:dict} *)
    end in (*@ \label{line:endstruct} *)
    (* The forward and backward phases. *)
    (*@\codelabel{init}*)
    let x = mk n in(*@ \label{line:x} *)
    (*@\codelabel{heart}*)
    let _ =(*@ \label{line:handle} *)
      (*@\codelabel{handle}*) match (*@\codelabel{eval}*) e.eval dict x with(*@ \label{line:match}\label{line:eval} *)
      | (*@\codelabel{eff-add}*) effect (Add (a, b)) k ->(*@ \label{line:handle:Add} *)
          (*@\codelabel{add-body}*)
          let u = (*@\codelabel{add-fwd}*) mk (get_v a + get_v b) in(*@ \label{line:handle:Add:mk} *)
          let _ = (*@\codelabel{cont}*) continue k u in(*@ \label{line:continue1} *)
          (*@\codelabel{add-bwd}*)
          update a (get_d u);(*@ \label{line:handle:Add:update:a} *)
          update b (get_d u)(*@ \label{line:handle:Add:update:b} *)
      | (*@\codelabel{eff-mul}*) effect (Mul (a, b)) k ->(*@ \label{line:handle:Mul} *)
          let u = mk (get_v a * get_v b) in(*@ \label{line:handle:Mul:mk} *)
          let _ = continue k u in (*@ \label{line:continue2} *)
          update a (get_d u * get_v b);(*@ \label{line:handle:Mul:update:a} *)
          update b (get_d u * get_v a)(*@ \label{line:handle:Mul:update:b} *)
      | (*@\codelabel{ret}*) y ->(*@ \label{line:handle:Return} *)
          (*@\codelabel{seed}*) update y one(*@ \label{line:handle:one} *)
    in
    (*@\codelabel{done}*) get_d x(*@ \label{line:read:result} *)
}
(* ENDEHAD *)

end

(* -------------------------------------------------------------------------- *)
(** Some dictionaries. *)

let int   = { zero = 0; one = 1; add = ( + ); mul = ( * ) }
let float = { zero = 0.0; one = 1.0; add = ( +. ); mul = ( *. ) }

(* -------------------------------------------------------------------------- *)
(** Some expressions. *)

(* [f] represents the expression [x ↦ (x+1)^3]. *)

let f =
  { eval = fun {one; add; mul; _} x ->
      let ( + ), ( * ) = add, mul in
      let cube x = x * x * x in
      cube (one + x)
  }

(* Test. *)

let () =
  assert (f.eval int 2 = 27);
  assert (f.eval float 2.0 = 27.0);
  ()

(* [var] represents the expression x. *)

let var : exp =
  { eval =
      fun (type v) _dict (x : v) ->
        x
  }

(* [subst e1 e2] substitutes the expression [e1] for the (unique) variable
   in the expression [e2], yielding a new expression. *)

let subst (e1 : exp) (e2 : exp) : exp =
  { eval =
      fun (type v) dict (x : v) ->
        e2.eval dict (e1.eval dict x)
  }

(* Monomial and power. *)

(* BEGINMONOMIAL *)
(* The expression x^k. *)
let monomial (k : int) : exp = { eval =
  fun (type v) { one; mul; _ } (x : v) ->
    let ( * ) = mul in
    let result, x, k = ref one, ref x, ref k in
    while !k > 0 do(*@ \label{line:monomial:while} *)
      result := !result * (if !k mod 2 = 0 then one else !x);(*@ \label{line:monomial:update:result} *)
      x := !x * !x;(*@ \label{line:monomial:update:x} *)
      k := !k / 2
    done;(*@ \label{line:monomial:done} *)
    !result }
(* ENDMONOMIAL *)

let pow (k : int) (e : exp) : exp =
  subst e (monomial k)

(* Test. *)
let () =
  assert ((monomial 8).eval float 2.0 = 256.0);
  assert ((pow 2 (pow 4 var)).eval float 2.0 = 256.0);
  ()

(* -------------------------------------------------------------------------- *)
(** Testing an implementation of AD. *)

(* [Test(A)] tests that [A.diff] behaves correctly on some simple inputs. *)
module Test (A : AD) = struct

  open A

  (* [df] should be [x ↦ 3(x+1)^2]. *)
  let df = diff f

  (* [ddf] should be [x ↦ 6(x+1)]. *)
  let ddf = diff (diff f)

  let () =
    assert  (df.eval int   2  = 27);
    assert  (df.eval float 2. = 27.);
    assert (ddf.eval int   2  = 18);
    assert (ddf.eval float 2. = 18.);
    ()

  (* [monomial 4] is [x^4], so its derivative its [4.x^3], whose evaluation
     at 2 should yield 32. *)
  let () =
    assert ((diff (monomial 4)).eval float 2.0 = 32.0);
    ()

end

(* -------------------------------------------------------------------------- *)
(** Testing each of our three implementations. *)

(* A list of our implementations of AD. *)
let algorithms : (module AD) list = [
  (module ForwardMode);
  (module StackBasedReverseMode);
  (module EffectHandlerBasedReverseMode);
]

(* Test every instance in [algorithms]. *)
let () =
  List.iter (fun (module A : AD) ->
    let module T = Test(A) in ()
  ) algorithms

(* If we reach this point, then the tests have succeeded. *)
let () =
  print_endline "Success."
\end{lstlisting}
\caption{Forward-mode \AD in OCaml}
\label{fig:fmad}
\end{figure}


\subsection{Forward-Mode \AD}
\label{sec:instances:forward}

A forward-mode implementation of \diff appears in~\fref{fig:fmad}.
This code receives~%
(1) a representation~\oc|e| of a mathematical
expression~$\E$ (\lineref{fmad:e});~%
(2) a representation of a semiring~$\Num$,
in the form of a type~\oc|v| and a~dictionary~$\{ \zero; \one; \add; \mul \}$
(\lineref{fmad:v}); and~%
(3) a representation~\oc|n| of a number~$\num\in\Num$
(\lineref{fmad:n}).
Its aim, according to Statements~\ref{informal:spec:diff}
and~\ref{informal:isExp},
is to evaluate the mathematical expression $\deriv\E$
at $\num$.

Because the algorithm has access to the function \oc|e.eval|, it can
evaluate the expression~$\E$. Furthermore, because this function is
polymorphic, it can evaluate~$\E$
in an arbitrary semiring of its choosing.
The key idea of forward-mode \AD is to evaluate $\E$ in the
semiring~$\dual\Num$ of~\emph{dual numbers} over $\Num$.
A dual number is a pair of numbers.
Addition and multiplication of dual numbers are defined by
$(a, \dot{a}) + (b, \dot{b}) = (a + b, \dot{a} + \dot{b})$
and
$(a, \dot{a}) \times (b, \dot{b}) =
 (a \times b, \dot{a} \times b + a \times \dot{b})$.
The neutral elements of addition and multiplication are $(0, 0)$ and $(1, 0)$.
It is not difficult to see that,
by virtue of these definitions,
the value of~$\E$ at~$(\num, 1)$ in the semiring~$\dual\Num$
is a pair whose first component is
the value of~$\E$ at~$\num$ in the semiring~$\Num$
and whose second component is
the value of~$\deriv\E$ at~$\num$ in the semiring~$\Num$.
The second component of this pair is the desired result.

The code in~\fref{fig:fmad} implements this idea in a straightforward way.%
\footnote{``\oc|let open struct ... end|'' at \lineref{fmad:openstruct}
is an OCaml idiosyncrasy that allows toplevel definitions (of types, effects,
and values) to appear in the midst of an expression.}
A type~\tnumdual of dual numbers is defined
on \lineref{fmad:type_t}.
A dual number is represented as a record with two fields
named~\vfieldname{} and~\dfieldname{}, for ``value'' and ``derivative'',
each of which is a number of type~\oc|v|.
The operations on dual numbers and their neutral elements are defined
and grouped in a dictionary~\numStruct
on \linerange{fmad:num}{fmad:dict}.
%
%
This dictionary is used on \lineref{fmad:y} to evaluate the expression~$\E$.
This expression is evaluated at the dual number $(\num, 1)$,
represented by~\oc|x| (\lineref{fmad:x}).
As previously explained,
the result of this evaluation is a dual number
whose derivative component
is the desired result:
the value of $\deriv\E$ at $\num$.
This component is read and returned on \lineref{fmad:read:result}.


\subsection{Reverse-Mode \AD}
\label{sec:instances:reverse}

We opened this paper with a quote from the Wikipedia entry on \AD: \emph{``every
computer program executes a sequence of elementary arithmetic operations
(addition, multiplication,~etc.)''}. This sentence contains a key idea:
instead of thinking of a mathematical expression as a tree-structured object,
as suggested by the inductive syntax $\E ::= \X \mid 0 \mid 1 \mid \E+\E \mid
\E\times\E$, one can equivalently represent it as a~sequence of elementary
operations whose arguments and results are identified by names $\locl, \locr,
\locu,\locx,\locy,\ldots$
This view is described by the
following alternate syntax:
%
\newcommand{\SE}{S}
\[
  \SE ::= \bind\locu+\locl\locr \kwin \SE
     \mid \bind\locu\times\locl\locr \kwin \SE
     \mid \locy
\]
Although the leaves $\X$, $0$ and $1$ seem to have disappeared in this syntax,
they can be represented by three reserved names.
In this form, an expression~$\SE$ is a sequence of operations. Each operation
refers to its two operands by their names~$\locl$ and~$\locr$, which must have
been previously defined, and introduces the name~$\locu$ to stand for its
result. At the end of the sequence, a result, identified by its name~$\locy$,
is produced.

\begin{figure}[t]
\centering
%
\begin{subfigure}{.4\textwidth}
  \centering
  \begin{tikzpicture}[
    thick,
    align         = center,
    main/.style   = {draw, circle, minimum size=8mm},
    node distance = {1.7cm},
  ]
  \node[main] (U4)               {$u_4$}; 
  \node[main] (U2) [above of=U4] {$u_2$}; 
  \node[main]  (X) [above of=U2]   {$x$}; 
  %
  \node[main] (U3) [right of=U4] {$u_3$}; 
  \node[main] (U1) [above of=U3] {$u_1$}; 
  \node[main]  (I) [above of=U1]  {\tI};
  %
  \draw[->] (U4) to [out=120, in=240, looseness=1.] (U2);
  \draw[->] (U4) to [out= 60, in=300, looseness=1.] (U2);
  %
  \draw[->] (U2) to [out=120, in=240, looseness=1.]  (X);
  \draw[->] (U2) to [out= 60, in=300, looseness=1.]  (X);
  %
  \draw[->] (U3) to [out=110, in=340, looseness=1.] (U2);
  \draw[->] (U3) -- (U1);
  %
  \draw[->] (U1) to [out=110, in=340, looseness=1.]  (X);
  \draw[->] (U1) --  (I);
  \end{tikzpicture}
  \caption{DAG view}
  \label{subfig:graph}
\end{subfigure}
%
\begin{subfigure}{.4\textwidth}
  \centering
  \newcommand{\kwlet}{\ensuremath{\mathsf{let}\;}}
  \renewcommand{\kwin}{\ensuremath{\;\mathsf{in}}}
  \renewcommand{\tMul}{\ensuremath{\times}} 
    \[\begin{tabular}{c@{\;}c@{\;}c@{\;}c@{\;}c@{\;}c@{\kwin}}
        \kwlet & $u_1$ &=&   $x$ &\tMul& \tI \\
        \kwlet & $u_2$ &=&   $x$ &\tMul&  $x$ \\
        \kwlet & $u_3$ &=& $u_2$ &\tMul& $u_1$ \\
        \kwlet & $u_4$ &=& $u_2$ &\tMul& $u_2$ \\
        $u_3$
    \end{tabular}\]
  \caption{Sequential view}
  \label{subfig:stack}
\end{subfigure}
\caption{DAG and stack built by the forward evaluation of \oc|monomial 3|}
\label{fig:operations}
\end{figure}

\begin{figure}[t]
  \setlength{\tabcolsep}{7pt}
  \renewcommand{\arraystretch}{1.2}
  \begin{tabular}{|c||c||c|c|c|c|c|c|}
    \cline{3-7}
    \multicolumn{2}{c|}{} &
    \multicolumn{5}{c|}{$\longrightarrow$ Time $\longrightarrow$} \\
    \hhline{--|=====|}
          & $\locx$   & $0$ &  $0$ &   $0$ & $2n^2$ & $3n^2$ \\
          & $\locu_1$ & $0$ &  $0$ & $n^2$ & $n^2$  & $\#$   \\
    Space & $\locu_2$ & $0$ &  $0$ &   $n$ &  $\#$  & $\#$   \\
          & $\locu_3$ & $0$ &  $1$ &  $\#$ &  $\#$  & $\#$   \\
          & $\locu_4$ & $1$ & $\#$ &  $\#$ &  $\#$  & $\#$   \\
    \hline
  \end{tabular}
\caption{Derivatives computed during the backward phase of \oc|monomial 3|}
\label{fig:variables}
\end{figure}

A~central quality of this alternate presentation of expressions is that it
offers at the same time a view of an expression as a \emph{sequence} of
operations and a view of an expression as a \emph{directed acyclic graph}
(DAG). In the latter view, each vertex is identified by a name, and there are
edges from~$\locu$ to~$\locl$ and from~$\locu$ to~$\locr$ if~$\locu$ is the
result of an operation whose operands are~$\locl$ and~$\locr$.
The vertex~$\locy$ is the root vertex of the graph.
Figures~\ref{subfig:graph} and~\ref{subfig:stack} offer an example of these
views.
The sequential view is valuable because it allows traversing the graph in the
forward direction (in dependency order) or in the reverse direction (in
reverse dependency order), as desired. The graphical view is valuable because
it keeps track of and allows taking advantage of sharing.

Reverse-mode \AD exploits this presentation of expressions.
It is organized in two phases.
In the first phase, known as the \emph{forward phase}, the
expression that one wishes to differentiate is converted into this form: a
sequential view and a DAG view are explicitly constructed in memory.
%
%
%
%
The second phase, known as the \emph{backward phase}, is where differentiation
really takes place. During this phase, the graph is traversed in reverse
dependency order:
each vertex is processed in turn.
During this traversal,
a sequential expression~$\SE$
is gradually constructed:
it evolves (grows) with time.
The following \emph{backward invariant} is maintained:
at each point in time, for every vertex~$\locu$ that
has not yet been processed, the \pd $\partial\SE/\partial\locu$
is computed and stored at vertex~$\locu$.
Initially,~$\SE$ is just~$\locy$,
the root vertex.
Then, every time a~vertex~$\locu$ is processed,
the expression~$\SE$ is extended with a binding for~$\locu$:
for example, if~$\locu$ is a name for~$\locl\mathop{+}\locr$,
then~$\SE$ is updated
to~$\bind\locu+\locl\locr\kwin\SE$.
The backward phase ends when every vertex has been
processed except for~$\locx$,
the vertex that stands for the variable~$\X$.
At this point, the expression~$\SE$ is equivalent
to the expression~$\E$ that one wishes to differentiate.
The backward invariant then implies that
the vertex~$\locx$ stores the derivative of~$\E$
with respect to~$\locx$. This is the desired result.

To illustrate the backward phase,
\fref{fig:variables} shows the evolution of derivatives computed
during this phase of the differentiation of \oc|monomial 3|.
We assume that, during the forward phase,
this expression has been converted to the sequential view of
Figure~\ref{subfig:stack}.
Furthermore, we assume that
the derivative of \oc|monomial 3| is
queried at a number~$n$.
%
%
At each vertex and at each point in time, one number is stored.
Thus, one column in~\fref{fig:variables} shows a snapshot of the
numbers stored at all vertices at a given point in time;
whereas one line in~\fref{fig:variables} shows the evolution through time of
the number stored at a given vertex.
Transitioning from a column to its right neighbor
corresponds to processing one vertex.
A processed vertex is marked with a hash symbol \#.
(Once a vertex has been processed,
the value that is stored at this vertex loses its meaning
and is no longer read or updated.)
In the first column, we see that each vertex~$\locu$
stores the \pd of~$\locu_3$ (the root vertex)
with respect to~$\locu$ at~$n$.
In the second column, the vertex~$\locu_4$
has been processed, so every unprocessed vertex~$\locu$
stores the \pd
of~$\bind{\locu_4}\times{\locu_2}{\locu_2}\kwin{\locu_3}$
with respect to~$\locu$ at~$n$.
By extending this reasoning to the remaining columns,
one can see that, in the last column,
the vertex~$\locx$ stores the \pd of the
sequential expression of Figure~\ref{subfig:stack}
with respect to~$\locx$ at~$n$.


\subsection{Stack-Based Reverse-Mode \AD}
\label{sec:instances:reverse:code}

\begin{figure}[p]
\begin{lstlisting}[keywordstyle=\bfseries\color{RubineRed},linerange=BEGINSTAD-ENDSTAD,includerangemarker=false,firstnumber=0]
(* Reverse-Mode Automatic Differentiation with Effect Handlers. *)

(* This code works in Multicore OCaml 4.12.0. *)

(* The use of effect handlers in this code is based on the code by KC
   Sivaramakrishnan:

   https://github.com/ocaml-multicore/effects-examples/blob/master/algorithmic_differentiation.ml

   The packaging is different, though; we use expressions in tagless-final
   style.

   This code can be executed in the OCaml toplevel loop via the following
   commands:

   opam switch create 4.12.0+domains+effects --no-switch # (only once) opam
   exec --switch 4.12.0+domains+effects ocaml # runs the toplevel loop

 *)

(* -------------------------------------------------------------------------- *)
(** Type definitions. *)

type 'v dict =
  { zero : 'v; one : 'v; add : 'v -> 'v -> 'v; mul : 'v -> 'v -> 'v }

type exp =
  { eval : (* forall *) 'v. 'v dict -> 'v -> 'v }

module type AD = sig
  val diff : exp -> exp
end

(* -------------------------------------------------------------------------- *)
(** Forward-mode AD. *)

module ForwardMode : AD = struct

(* BEGINFMAD *)
let diff (e : exp) : exp = { eval =(*@ \label{line:fmad:e} *)
  fun (type v) ({ zero; one; add; mul } : v dict) (n : v) ->(*@ \label{line:fmad:v}\label{line:fmad:n} *)
    let ( + ), ( * ) = add, mul in
    let open struct(*@ \label{line:fmad:openstruct} *)
      type t = { v : v ; d : v }(*@ \label{line:fmad:type_t} *)
      let mk v d = { v; d }(*@ \label{line:fmad:mk} *)
      let dict =(*@ \label{line:fmad:num} *)
        let zero = mk zero zero(*@ \label{line:fmad:zero} *)
        and one = mk one zero(*@ \label{line:fmad:one} *)
        and add a b = mk (a.v + b.v) (a.d + b.d)(*@ \label{line:fmad:add} *)
        and mul a b = mk (a.v * b.v) (a.d * b.v + a.v * b.d)(*@ \label{line:fmad:mul} *)
        in { zero; one; add; mul }(*@ \label{line:fmad:dict} *)
      let x = mk n one(*@ \label{line:fmad:x} *)
      let y = e.eval dict x(*@ \label{line:fmad:y} *)
    end in
    y.d(*@ \label{line:fmad:read:result} *)
}
(* ENDFMAD *)

end

(* -------------------------------------------------------------------------- *)
(** Stack-based implementation of reverse-mode AD. *)

module StackBasedReverseMode : AD = struct

(* BEGINSTAD *)
let diff (e : exp) : exp = { eval =(*@ \label{line:stad:e} *)
  fun (type v) ({ zero; one; add; mul } : v dict) (n : v) ->(*@ \label{line:stad:v}\label{line:stad:n} *)
    let ( + ), ( * ) = add, mul in
    let open struct(*@ \label{line:stad:openstruct} *)
      (* The graph. *)
      type t = O | I | Var of { v : v ; mutable d : v }(*@ \label{line:stad:type_t} *)
      let mk n       = Var { v = n; d = zero }(*@ \label{line:stad:mk} *)
      let get_v u    = match u with O -> zero | I -> one  | Var u     -> u.v(*@ \label{line:stad:get_v} *)
      let get_d u    = match u with O | I -> assert false | Var u     -> u.d(*@ \label{line:stad:get_d} *)
      let update u i = match u with O | I -> ()    | Var u -> u.d <- u.d + i(*@ \label{line:stad:update} *)
      (* The stack. *)
      type op = Add | Mul(*@ \label{line:stad:type_op} *)
      type binding = Let of t * (t * op * t)(*@ \label{line:stad:type_letbinding} *)
      open Stack(*@ \label{line:stad:openstack} *)
      let stack : binding Stack.t = create()(*@ \label{line:stad:stack} *)
      (* The dictionary used in the forward phase. *)
      let dict =(*@ \label{line:stad:num} *)
        let zero = O
        and one = I
        and add a b =(*@ \label{line:stad:add} *)
          let u = mk (get_v a + get_v b) in
          push (Let (u, (a, Add, b))) stack; u
        and mul a b =(*@ \label{line:stad:mul} *)
          let u = mk (get_v a * get_v b) in
          push (Let (u, (a, Mul, b))) stack; u
        in { zero; one; add; mul }
      (* The forward phase. *)
      let x = mk n(*@ \label{line:stad:x} *)
      let y = e.eval dict x(*@ \label{line:stad:y} *)
      (* The backward phase. *)
      let () =(*@ \label{line:stad:backwardphase} *)
        update y one;(*@ \label{line:stad:update:y} *)
        while not (is_empty stack) do(*@ \label{line:stad:while} *)
          match pop stack with(*@ \label{line:stad:pop} *)
          | Let (u, (a, Add, b)) ->
              update a (get_d u);(*@ \label{line:stad:update:add:a} *)
              update b (get_d u)(*@ \label{line:stad:update:add:b} *)
          | Let (u, (a, Mul, b)) ->
              update a (get_d u * get_v b);(*@ \label{line:stad:update:mul:a} *)
              update b (get_d u * get_v a)(*@ \label{line:stad:update:mul:b} *)
        done(*@ \label{line:stad:while:done} *)
    end in
    get_d x(*@ \label{line:stad:read:result} *)
}
(* ENDSTAD *)

end

(* -------------------------------------------------------------------------- *)
(** Reverse-mode AD using effect handlers. *)

module EffectHandlerBasedReverseMode : AD = struct

(* BEGINEHAD *)
let diff (e : exp) : exp = (*@\codelabel{head}*) { eval =(*@ \label{line:e} *)
  fun (type v) ({ zero; one; add; mul } : v dict) (n : v) -> (*@\codelabel{body}*) (*@ \label{line:v}\label{line:n} *)
    let ( + ), ( * ) = add, mul in
    let open struct(*@ \label{line:openstruct} *)
      (* The graph. *)
      type t = O | I | Var of { v : v ; mutable d : v }(*@ \label{line:type_t} *)
      let mk n       = Var { v = n; d = zero }(*@ \label{line:mk} *)
      let get_v u    = match u with O -> zero | I -> one  | Var u     -> u.v(*@ \label{line:get_v} *)
      let get_d u    = match u with O | I -> assert false | Var u     -> u.d(*@ \label{line:get_d} *)
      let update u i = match u with O | I -> ()    | Var u -> u.d <- u.d + i(*@ \label{line:update} *)
      (* The dictionary. *)
      effect Add : t * t -> t(*@ \label{line:Add} *)
      effect Mul : t * t -> t(*@ \label{line:Mul} *)
      let zero' = O(*@ \label{line:zeroP} *)
      and one' = I(*@ \label{line:oneP} *)
      and add' a b = perform (Add (a, b))(*@ \label{line:add} *)
      and mul' a b = perform (Mul (a, b))(*@ \label{line:mul} *)
      let dict = { zero = zero'; one = one'; add = add'; mul = mul'}(*@ \label{line:dict} *)
    end in (*@ \label{line:endstruct} *)
    (* The forward and backward phases. *)
    (*@\codelabel{init}*)
    let x = mk n in(*@ \label{line:x} *)
    (*@\codelabel{heart}*)
    let _ =(*@ \label{line:handle} *)
      (*@\codelabel{handle}*) match (*@\codelabel{eval}*) e.eval dict x with(*@ \label{line:match}\label{line:eval} *)
      | (*@\codelabel{eff-add}*) effect (Add (a, b)) k ->(*@ \label{line:handle:Add} *)
          (*@\codelabel{add-body}*)
          let u = (*@\codelabel{add-fwd}*) mk (get_v a + get_v b) in(*@ \label{line:handle:Add:mk} *)
          let _ = (*@\codelabel{cont}*) continue k u in(*@ \label{line:continue1} *)
          (*@\codelabel{add-bwd}*)
          update a (get_d u);(*@ \label{line:handle:Add:update:a} *)
          update b (get_d u)(*@ \label{line:handle:Add:update:b} *)
      | (*@\codelabel{eff-mul}*) effect (Mul (a, b)) k ->(*@ \label{line:handle:Mul} *)
          let u = mk (get_v a * get_v b) in(*@ \label{line:handle:Mul:mk} *)
          let _ = continue k u in (*@ \label{line:continue2} *)
          update a (get_d u * get_v b);(*@ \label{line:handle:Mul:update:a} *)
          update b (get_d u * get_v a)(*@ \label{line:handle:Mul:update:b} *)
      | (*@\codelabel{ret}*) y ->(*@ \label{line:handle:Return} *)
          (*@\codelabel{seed}*) update y one(*@ \label{line:handle:one} *)
    in
    (*@\codelabel{done}*) get_d x(*@ \label{line:read:result} *)
}
(* ENDEHAD *)

end

(* -------------------------------------------------------------------------- *)
(** Some dictionaries. *)

let int   = { zero = 0; one = 1; add = ( + ); mul = ( * ) }
let float = { zero = 0.0; one = 1.0; add = ( +. ); mul = ( *. ) }

(* -------------------------------------------------------------------------- *)
(** Some expressions. *)

(* [f] represents the expression [x ↦ (x+1)^3]. *)

let f =
  { eval = fun {one; add; mul; _} x ->
      let ( + ), ( * ) = add, mul in
      let cube x = x * x * x in
      cube (one + x)
  }

(* Test. *)

let () =
  assert (f.eval int 2 = 27);
  assert (f.eval float 2.0 = 27.0);
  ()

(* [var] represents the expression x. *)

let var : exp =
  { eval =
      fun (type v) _dict (x : v) ->
        x
  }

(* [subst e1 e2] substitutes the expression [e1] for the (unique) variable
   in the expression [e2], yielding a new expression. *)

let subst (e1 : exp) (e2 : exp) : exp =
  { eval =
      fun (type v) dict (x : v) ->
        e2.eval dict (e1.eval dict x)
  }

(* Monomial and power. *)

(* BEGINMONOMIAL *)
(* The expression x^k. *)
let monomial (k : int) : exp = { eval =
  fun (type v) { one; mul; _ } (x : v) ->
    let ( * ) = mul in
    let result, x, k = ref one, ref x, ref k in
    while !k > 0 do(*@ \label{line:monomial:while} *)
      result := !result * (if !k mod 2 = 0 then one else !x);(*@ \label{line:monomial:update:result} *)
      x := !x * !x;(*@ \label{line:monomial:update:x} *)
      k := !k / 2
    done;(*@ \label{line:monomial:done} *)
    !result }
(* ENDMONOMIAL *)

let pow (k : int) (e : exp) : exp =
  subst e (monomial k)

(* Test. *)
let () =
  assert ((monomial 8).eval float 2.0 = 256.0);
  assert ((pow 2 (pow 4 var)).eval float 2.0 = 256.0);
  ()

(* -------------------------------------------------------------------------- *)
(** Testing an implementation of AD. *)

(* [Test(A)] tests that [A.diff] behaves correctly on some simple inputs. *)
module Test (A : AD) = struct

  open A

  (* [df] should be [x ↦ 3(x+1)^2]. *)
  let df = diff f

  (* [ddf] should be [x ↦ 6(x+1)]. *)
  let ddf = diff (diff f)

  let () =
    assert  (df.eval int   2  = 27);
    assert  (df.eval float 2. = 27.);
    assert (ddf.eval int   2  = 18);
    assert (ddf.eval float 2. = 18.);
    ()

  (* [monomial 4] is [x^4], so its derivative its [4.x^3], whose evaluation
     at 2 should yield 32. *)
  let () =
    assert ((diff (monomial 4)).eval float 2.0 = 32.0);
    ()

end

(* -------------------------------------------------------------------------- *)
(** Testing each of our three implementations. *)

(* A list of our implementations of AD. *)
let algorithms : (module AD) list = [
  (module ForwardMode);
  (module StackBasedReverseMode);
  (module EffectHandlerBasedReverseMode);
]

(* Test every instance in [algorithms]. *)
let () =
  List.iter (fun (module A : AD) ->
    let module T = Test(A) in ()
  ) algorithms

(* If we reach this point, then the tests have succeeded. *)
let () =
  print_endline "Success."
\end{lstlisting}
\caption{Stack-based reverse-mode \AD in OCaml}
\label{fig:stad}
\end{figure}



Having presented the key ideas of the reverse-mode approach to \ad,
let us now discuss an implementation of this approach.
This implementation of reverse-mode \ad appears in~\fref{fig:stad}.
It is stack-based: it uses a stack to construct a DAG view and a sequential
view of the expression that one wishes to differentiate.
It begins like our forward-mode implementation
(\sref{sec:instances:forward}):
it receives
a representation~\oc|e| of a mathematical expression~$\E$ (\lineref{stad:e});
a representation of a semiring~$\Num$,
in the form of a type~\oc|v| of numbers
and a~dictionary~$\{ \zero; \one; \add; \mul \}$
(\lineref{stad:v});
and a representation~\oc|n| of a number~$\num\in\Num$
(\lineref{stad:n}).
%
%



A type~\tnumdual of vertices is declared on \lineref{stad:type_t}.
A vertex is
either~\oc|O|, which denotes the constant~0;
or~\oc|I|, which denotes the constant~1;
or a heap-allocated record,
carrying the tag~\oc|Var|.
The address of this record effectively serves as the name of the vertex.
This record holds two fields, named \vfieldname{} and \dfieldname{},
for ``value'' and ``derivative''.
%
%
The \vfield is immutable and is initialized during the forward phase.
The \dfield is mutable and is updated (possibly several times) during
the backward phase.
These two fields allow us to associate two numbers, \oc|u.v| and \oc|u.d|,
with each vertex~\oc|u|.
%
%
Four auxiliary functions help construct and access vertices.
The function \oc|mk| (\lineref{stad:mk})
allocates and initializes a new vertex.
The functions \oc|get_v| and \oc|get_d|
(\linerange{stad:get_v}{stad:get_d})
read the \vfieldname{}~and \dfields of a vertex,
handling the constants~\oc|O| and \oc|I| in a suitable way.%
\footnote{
  The function \oc|get_d| is applied during the backward phase to the source
  vertex of an edge. The vertices~\oc|O| and~\oc|I| are never the source of
  an edge, so \oc|get_d| is never applied to them. The OCaml expression
  \oc|assert false| is used to denote these dead branches.}
The function \oc|update| (\lineref{stad:update})
updates the \oc|d| field of a vertex
by adding the value~\oc|i| to this field.
Applying \oc|update| to the vertices~\oc|O| or~\oc|I|
does nothing. 



A stack is allocated at \lineref{stad:stack}. This stack is mutable and is
initially empty. It stores a sequence of \emph{bindings} of the form
$\bind\locu\opVar\locl\locr$, where $\locu,\locl,\locr$ are vertices
and where $\opVar$ is \oc|Add| or \oc|Mul|
(\linerange{stad:type_op}{stad:type_letbinding}).
Each such binding records an addition or multiplication operation, whose
result is the vertex~$\locu$ and whose operands are the vertices~$\locl$
and~$\locr$. It indicates the existence of two graph edges, from~$\locu$
to~$\locl$ and from~$\locu$ to~$\locr$.
This data structure is known in the literature as a \emph{tape} or
a \emph{Wengert list}.

The functions \oc|add| (\lineref{stad:add})
and \oc|mul| (\lineref{stad:mul})
create a new vertex~\oc|u|,
extend the stack with a new binding for~\oc|u|,
and return this vertex.
The field \oc|u.v| receives the sum or the product
of the fields \oc|a.v| and \oc|b.v|.
The constants~\oc|O| and~\oc|I| and
the functions~\oc|add| and~\oc|mul| are grouped
in a dictionary~\numStruct,
for use during the forward phase.




After these preparations, the forward phase can take place.
%
First, a~vertex~$x$
is created to stand for the variable~$\X$
(\lineref{stad:x}).
%
%
Then, the expression~$\E$ is evaluated
under the dictionary~\numStruct,
yielding a vertex~$y$
(\lineref{stad:y}).
This evaluation serves two main purposes.
First, the expression~$\E$ is converted into the
sequential representation~$\SE$ that was discussed
earlier. Indeed, at the end of the forward phase,
the bindings stored in the stack,
together with the root vertex~$y$,
form such a sequential representation.
%
Second, the expression~$\E$ is evaluated
in the semiring~$\Num$
at $\X = \num$.
Indeed,
the~\vfield of every vertex stores the
numeric value
that corresponds to this vertex.




Then comes the backward phase
(\linerange{stad:backwardphase}{stad:while:done}).
Writing 1 into the \dfield of the vertex~$y$
(\lineref{stad:update:y}) establishes the
backward invariant that we have sketched earlier
(\sref{sec:instances:reverse}).
Indeed, for every vertex~$\locu$, the \pd
``$\partial\locy/\partial\locu$'' is~1 if the vertices~$\locu$ and~$\locy$ are
the same vertex, and it is~0 if they are distinct vertices.
The backward invariant serves as a loop invariant
for the loop that follows.
In this loop, each vertex~$\locu$ is processed in turn, in reverse order:
the most recently created vertices are processed first.
%
%
As long as the stack is nonempty,
an entry is popped off the stack (\lineref{stad:pop}).
This entry records whether the vertex~$u$ was the result of an addition or
multiplication operation, and what were the operands~$a$ and~$b$ of this
operation.
In each case, information is propagated along the edges
from~$\locu$ to~$\locl$ and from~$\locu$ to~$\locr$:
the \dfields associated with the vertices~$\locl$ and~$\locr$
are updated.
%
%
These updates, which we do not explain in detail for now,
serve to maintain the backward invariant.
At the end of the loop, the backward invariant
implies that the \pd ``$\partial\locy/\partial\locx$''
is stored in the \dfield of~the vertex~$x$.
The code retrieves and returns this final result
(\lineref{stad:read:result}).

\section{Effects and Handlers}
\label{sec:effects}

\newcommand{\delimitedfootnote}{\footnote{%
The literature offers a wide variety of delimited-control operators, that is,
operators that allow capturing delimited
continuations~\cite{felleisen-88,danvy-filinski-90,sitaram-93}.
Effect handling is equivalent in expressive power to many of these
operators~\cite{forster-kammar-lindley-pretnar-19}.}}




\subsection{Effect handling}

Effect handling can be understood as a generalization of exception handling, a
familiar feature of many high-level programming languages, including Lisp,
CLU~\cite{liskov-snyder-79}, Ada, Modula-3, C\raisebox{0.2mm}{\hbox{\tt++}},
Standard ML, OCaml, Java, and many more.
Exception handling allows the execution of a computation to be monitored by a
\emph{handler}. The computation may at any time decide to \emph{throw an
  exception}.
In such an event, the computation is interrupted and the handler takes
control.
%
%
In the early days of exception handling, it was debated whether a computation
that throws an exception should be terminated or possibly resumed after the
handler has run. A~consensus emerged in favor of the first option, the
\emph{termination model}, because it was perceived to be easier to reason about
and easier to implement efficiently than the second option, the
\emph{resumption model}.
Ryder and Soffa~\cite{ryder-soffa-03} offer a historical account.

Like exception handling, effect handling involves the interplay of a
computation and a handler. At any time, the computation can interrupt itself
by \textit{performing an effect}. Control is then transferred to the handler.
As a crucial new feature, the handler receives a first-class function, also
known as a \emph{delimited continuation},\delimitedfootnote{} which represents
the suspended computation: invoking this function resumes the computation.
If the computation is resumed, then another instance of the handler is
installed, so the dialogue continues: the computation can perform another
effect, causing control to be again transferred to the handler, and so on.

A continuation is an ordinary function, which an effect handler can use in a
variety of ways. If the continuation is not invoked at all, then the
computation is stopped. If the continuation is invoked
at the end of the handler, then the computation is resumed after the handler
has run. If the continuation is invoked somewhere in the middle of the
handler, then part of the handler runs \emph{before the computation is
  resumed} and part of it runs \emph{after the computation has finished}. The
example that we shall present shortly (\sref{sec:effects:example}),
as well as our effect-based \rmad algorithm (\sref{sec:impl}),
exploits this pattern.

Yet other uses of the continuation can be imagined. In some applications, the
continuation is not invoked by the handler, but is returned by the handler or
stored by the handler in memory for use at a later time.
%
In other applications, such as backtracking search, a~continuation is invoked
several times. This means that a computation that suspends itself \emph{once}
can be resumed \emph{more than once}. This use of continuations is powerful,
but requires care: it breaks the property that \emph{a block of code, once
entered, is exited at most once}, and thereby compromises the \emph{frame
rule}~\cite{de-vilhena-pottier-21}, one of the most fundamental reasoning rules
of \SL. Both \mocaml and our reasoning rules~\cite{de-vilhena-pottier-21}
require that a continuation be invoked at most once. In our \rmad algorithm,
every continuation is invoked exactly once.


Effect handlers are found in several research programming languages, such as
Eff \cite{bauer-pretnar-15,eff},
Effekt~\cite{effekt-language-20},
Frank~\cite{lindley-mcbride-mclaughlin-17},
Koka~\cite{leijen-koka-14,koka},
Links~\cite{hillerstrom-lindley-atkey-20},
and \mocaml~\cite{dolan-17,sivaramakrishnan-21}.
%
%
They have also been implemented as a library in mainstream programming
languages such as Scala~\cite{effekt-20}.

\subsection{Example}
\label{sec:effects:example}

We now illustrate effect handlers via a simple example. Although this example
may seem somewhat artificial, it deserves to be understood, as it exploits
effect handling exactly in the same way as the \rmad algorithm that we wish to
study.

\mocaml offers three basic constructs for effect handling.
The statement \oc|perform v| interrupts execution and transfers control
to the nearest enclosing handler, which receives the value~\oc|v|
and a continuation~\oc|k|. The statement \oc|continue k w| invokes the
continuation~\oc|k|: the suspended computation is resumed, just as if
\oc|perform v| had returned the value~\oc|w|. Finally, the \oc|match|
construct wraps a computation in a handler.

The example in \fref{fig:ask} exploits all of these constructs in combination.
It is a complete \mocaml program, whose output appears in \fref{fig:ask:out}.

In line~\ref{ask:Ask}, the effect \oc|Ask| is declared,
with signature \oc|int -> int|. This means that
the expression \oc|perform (Ask x)|
requires \oc|x| to have type \oc|int|
and has type \oc|int|.
In line~\ref{ask:ask},
\oc|ask x| is defined as a shorthand for
\oc|perform (Ask x)|.
Thus, the function \oc|ask| has type \oc|int -> int|.

\begin{figure}
\begin{lstlisting}[keywordstyle=\bfseries\color{RubineRed}]
open Printf
effect Ask : int -> int                                                         (*@ \label{ask:Ask} *)
let ask x = perform (Ask x)                                                     (*@ \label{ask:ask} *)
let handle (client : unit -> int) =                                             (*@ \label{ask:handle} *)
  match client() with
  | effect (Ask x) k ->                                                         (*@ \label{ask:effect} *)
      let y = x + 1 in
      printf "I am queried at %d and I am replying %d.\n" x y;                  (*@ \label{ask:print1} *)
      continue k y;                                                             (*@ \label{ask:continue} *)
      printf "Earlier, I have been queried at %d and I have replied %d.\n" x y  (*@ \label{ask:print2} *)
  | result ->                                                                   (*@ \label{ask:result} *)
      printf "The computation is finished and returns %d.\n" result             (*@ \label{ask:finished} *)
let () =
  handle (fun () -> ask 2 + ask 7)                                              (*@ \label{ask:client} *)
\end{lstlisting}
\caption{A simple demonstration of effect handlers}
\label{fig:ask}
\end{figure}

\begin{figure}
\begin{lstlisting}[language=none,numbers=none]
I am queried at 7 and I am replying 8.
I am queried at 2 and I am replying 3.
The computation is finished and returns 11.
Earlier, I have been queried at 2 and I have replied 3.
Earlier, I have been queried at 7 and I have replied 8.
\end{lstlisting}
\caption{Output of the program in \fref{fig:ask}}
\label{fig:ask:out}
\end{figure}

The function \oc|handle| at line~\ref{ask:handle} executes the computation
\oc|client()| under a handler for the effect \oc|Ask|. The handler takes the
form of a \oc|match| construct whose first branch (line~\ref{ask:effect})
takes control when the computation performs the effect \oc|Ask| and whose
second branch (line~\ref{ask:result}) takes control when the computation
finishes.

The \oc|effect| branch at line~\ref{ask:effect} specifies how the handler
behaves when the client performs an effect. The handler receives the integer
value \oc|x| that was passed as an argument to \oc|ask| as well as a
continuation \oc|k|. It defines \oc|y| as \oc|x+1| and applies the
continuation~\oc|k| to~\oc|y| (line~\ref{ask:continue}) between two
\oc|printf| statements. Thus, the first \oc|printf| statement is executed
before the suspended computation is resumed, whereas the second \oc|printf|
statement takes effect only after the suspended computation terminates. It is
crucial to remark that the execution of \oc|continue k y| may involve
further effects and their handling.

The second branch, at line~\ref{ask:result}, specifies what to do after the
client terminates normally and returns a value, \oc|result|. The \oc|printf|
statement at line~\ref{ask:finished} displays this value.

We are now in a position to understand why the function application
\oc|handle (fun () ->| \oc|ask 2 + ask 7)|
at line~\ref{ask:client} produces the output shown in \fref{fig:ask:out}.
\mocaml happens to follow a right-to-left evaluation order, so the function
call \oc|ask 7| takes place first, causing control to be transferred to the
handler, with \oc|x| bound to \oc|7|. The handler immediately prints
``\texttt{I am queried at 7...}'', then resumes the client by calling
\oc|continue k 8| at line~\ref{ask:continue}. (The \oc|printf| statement at
line~\ref{ask:print2}, whose effect is to print ``\texttt{Earlier, I have been
  queried at 7...}'', is delayed until this call returns.) The client then
resumes its work and reaches the function call \oc|ask 2|, again causing
control to be transferred to the handler. This is in fact a new instance of
the handler, where this time \oc|x| is bound to \oc|2|. This second effect is
handled like the previous one. The handler immediately prints ``\texttt{I am
  queried at 2...}'', then resumes the client by calling \oc|continue k 3|.
(The \oc|printf| statement at line~\ref{ask:print2}, whose effect is to print
``\texttt{Earlier, I have been queried at 2...}'', is delayed until this call
returns.) The client now computes \oc|3 + 8| and terminates with the
value~\oc|11|. The termination of the client is handled by a third and last
instance of the handler: there, control reaches the \oc|printf| statement at
line~\ref{ask:finished}, producing the third line of output. This third
instance of the handler is then finished and disappears. The previous two
instances of the handler, whose activation records still exist on the control
stack, then complete their execution. Thus, the two \oc|printf|
statements that were delayed earlier are allowed to take effect.
The most recent handler instance
completes first: this explains the order in which the last two
lines of output appear.

In summary, the execution of this code is divided in two phases. During the
first phase, which lasts as long as the client runs, the client performs a
sequence of effects, and the \oc|printf| statements at line~\ref{ask:print1}
are executed in order, while the \oc|printf| statements at
line~\ref{ask:print2} are accumulated on the control stack. During the second
phase, which begins when the client terminates, the \oc|printf| statements
that have been delayed are popped off the control stack and are executed in
reverse order.
Using Danvy and Goldberg's terminology~\cite{danvy-goldberg-05},
the first phase occurs at \textit{call time}
whereas the second phase occurs at \textit{return time}.

\section{Effect-Based Reverse-Mode AD}
\label{sec:impl}
\begin{figure}[thp]
\begin{lstlisting}[keywordstyle=\bfseries\color{RubineRed},linerange=BEGINEHAD-ENDEHAD,includerangemarker=false,firstnumber=0]
(* Reverse-Mode Automatic Differentiation with Effect Handlers. *)

(* This code works in Multicore OCaml 4.12.0. *)

(* The use of effect handlers in this code is based on the code by KC
   Sivaramakrishnan:

   https://github.com/ocaml-multicore/effects-examples/blob/master/algorithmic_differentiation.ml

   The packaging is different, though; we use expressions in tagless-final
   style.

   This code can be executed in the OCaml toplevel loop via the following
   commands:

   opam switch create 4.12.0+domains+effects --no-switch # (only once) opam
   exec --switch 4.12.0+domains+effects ocaml # runs the toplevel loop

 *)

(* -------------------------------------------------------------------------- *)
(** Type definitions. *)

type 'v dict =
  { zero : 'v; one : 'v; add : 'v -> 'v -> 'v; mul : 'v -> 'v -> 'v }

type exp =
  { eval : (* forall *) 'v. 'v dict -> 'v -> 'v }

module type AD = sig
  val diff : exp -> exp
end

(* -------------------------------------------------------------------------- *)
(** Forward-mode AD. *)

module ForwardMode : AD = struct

(* BEGINFMAD *)
let diff (e : exp) : exp = { eval =(*@ \label{line:fmad:e} *)
  fun (type v) ({ zero; one; add; mul } : v dict) (n : v) ->(*@ \label{line:fmad:v}\label{line:fmad:n} *)
    let ( + ), ( * ) = add, mul in
    let open struct(*@ \label{line:fmad:openstruct} *)
      type t = { v : v ; d : v }(*@ \label{line:fmad:type_t} *)
      let mk v d = { v; d }(*@ \label{line:fmad:mk} *)
      let dict =(*@ \label{line:fmad:num} *)
        let zero = mk zero zero(*@ \label{line:fmad:zero} *)
        and one = mk one zero(*@ \label{line:fmad:one} *)
        and add a b = mk (a.v + b.v) (a.d + b.d)(*@ \label{line:fmad:add} *)
        and mul a b = mk (a.v * b.v) (a.d * b.v + a.v * b.d)(*@ \label{line:fmad:mul} *)
        in { zero; one; add; mul }(*@ \label{line:fmad:dict} *)
      let x = mk n one(*@ \label{line:fmad:x} *)
      let y = e.eval dict x(*@ \label{line:fmad:y} *)
    end in
    y.d(*@ \label{line:fmad:read:result} *)
}
(* ENDFMAD *)

end

(* -------------------------------------------------------------------------- *)
(** Stack-based implementation of reverse-mode AD. *)

module StackBasedReverseMode : AD = struct

(* BEGINSTAD *)
let diff (e : exp) : exp = { eval =(*@ \label{line:stad:e} *)
  fun (type v) ({ zero; one; add; mul } : v dict) (n : v) ->(*@ \label{line:stad:v}\label{line:stad:n} *)
    let ( + ), ( * ) = add, mul in
    let open struct(*@ \label{line:stad:openstruct} *)
      (* The graph. *)
      type t = O | I | Var of { v : v ; mutable d : v }(*@ \label{line:stad:type_t} *)
      let mk n       = Var { v = n; d = zero }(*@ \label{line:stad:mk} *)
      let get_v u    = match u with O -> zero | I -> one  | Var u     -> u.v(*@ \label{line:stad:get_v} *)
      let get_d u    = match u with O | I -> assert false | Var u     -> u.d(*@ \label{line:stad:get_d} *)
      let update u i = match u with O | I -> ()    | Var u -> u.d <- u.d + i(*@ \label{line:stad:update} *)
      (* The stack. *)
      type op = Add | Mul(*@ \label{line:stad:type_op} *)
      type binding = Let of t * (t * op * t)(*@ \label{line:stad:type_letbinding} *)
      open Stack(*@ \label{line:stad:openstack} *)
      let stack : binding Stack.t = create()(*@ \label{line:stad:stack} *)
      (* The dictionary used in the forward phase. *)
      let dict =(*@ \label{line:stad:num} *)
        let zero = O
        and one = I
        and add a b =(*@ \label{line:stad:add} *)
          let u = mk (get_v a + get_v b) in
          push (Let (u, (a, Add, b))) stack; u
        and mul a b =(*@ \label{line:stad:mul} *)
          let u = mk (get_v a * get_v b) in
          push (Let (u, (a, Mul, b))) stack; u
        in { zero; one; add; mul }
      (* The forward phase. *)
      let x = mk n(*@ \label{line:stad:x} *)
      let y = e.eval dict x(*@ \label{line:stad:y} *)
      (* The backward phase. *)
      let () =(*@ \label{line:stad:backwardphase} *)
        update y one;(*@ \label{line:stad:update:y} *)
        while not (is_empty stack) do(*@ \label{line:stad:while} *)
          match pop stack with(*@ \label{line:stad:pop} *)
          | Let (u, (a, Add, b)) ->
              update a (get_d u);(*@ \label{line:stad:update:add:a} *)
              update b (get_d u)(*@ \label{line:stad:update:add:b} *)
          | Let (u, (a, Mul, b)) ->
              update a (get_d u * get_v b);(*@ \label{line:stad:update:mul:a} *)
              update b (get_d u * get_v a)(*@ \label{line:stad:update:mul:b} *)
        done(*@ \label{line:stad:while:done} *)
    end in
    get_d x(*@ \label{line:stad:read:result} *)
}
(* ENDSTAD *)

end

(* -------------------------------------------------------------------------- *)
(** Reverse-mode AD using effect handlers. *)

module EffectHandlerBasedReverseMode : AD = struct

(* BEGINEHAD *)
let diff (e : exp) : exp = (*@\codelabel{head}*) { eval =(*@ \label{line:e} *)
  fun (type v) ({ zero; one; add; mul } : v dict) (n : v) -> (*@\codelabel{body}*) (*@ \label{line:v}\label{line:n} *)
    let ( + ), ( * ) = add, mul in
    let open struct(*@ \label{line:openstruct} *)
      (* The graph. *)
      type t = O | I | Var of { v : v ; mutable d : v }(*@ \label{line:type_t} *)
      let mk n       = Var { v = n; d = zero }(*@ \label{line:mk} *)
      let get_v u    = match u with O -> zero | I -> one  | Var u     -> u.v(*@ \label{line:get_v} *)
      let get_d u    = match u with O | I -> assert false | Var u     -> u.d(*@ \label{line:get_d} *)
      let update u i = match u with O | I -> ()    | Var u -> u.d <- u.d + i(*@ \label{line:update} *)
      (* The dictionary. *)
      effect Add : t * t -> t(*@ \label{line:Add} *)
      effect Mul : t * t -> t(*@ \label{line:Mul} *)
      let zero' = O(*@ \label{line:zeroP} *)
      and one' = I(*@ \label{line:oneP} *)
      and add' a b = perform (Add (a, b))(*@ \label{line:add} *)
      and mul' a b = perform (Mul (a, b))(*@ \label{line:mul} *)
      let dict = { zero = zero'; one = one'; add = add'; mul = mul'}(*@ \label{line:dict} *)
    end in (*@ \label{line:endstruct} *)
    (* The forward and backward phases. *)
    (*@\codelabel{init}*)
    let x = mk n in(*@ \label{line:x} *)
    (*@\codelabel{heart}*)
    let _ =(*@ \label{line:handle} *)
      (*@\codelabel{handle}*) match (*@\codelabel{eval}*) e.eval dict x with(*@ \label{line:match}\label{line:eval} *)
      | (*@\codelabel{eff-add}*) effect (Add (a, b)) k ->(*@ \label{line:handle:Add} *)
          (*@\codelabel{add-body}*)
          let u = (*@\codelabel{add-fwd}*) mk (get_v a + get_v b) in(*@ \label{line:handle:Add:mk} *)
          let _ = (*@\codelabel{cont}*) continue k u in(*@ \label{line:continue1} *)
          (*@\codelabel{add-bwd}*)
          update a (get_d u);(*@ \label{line:handle:Add:update:a} *)
          update b (get_d u)(*@ \label{line:handle:Add:update:b} *)
      | (*@\codelabel{eff-mul}*) effect (Mul (a, b)) k ->(*@ \label{line:handle:Mul} *)
          let u = mk (get_v a * get_v b) in(*@ \label{line:handle:Mul:mk} *)
          let _ = continue k u in (*@ \label{line:continue2} *)
          update a (get_d u * get_v b);(*@ \label{line:handle:Mul:update:a} *)
          update b (get_d u * get_v a)(*@ \label{line:handle:Mul:update:b} *)
      | (*@\codelabel{ret}*) y ->(*@ \label{line:handle:Return} *)
          (*@\codelabel{seed}*) update y one(*@ \label{line:handle:one} *)
    in
    (*@\codelabel{done}*) get_d x(*@ \label{line:read:result} *)
}
(* ENDEHAD *)

end

(* -------------------------------------------------------------------------- *)
(** Some dictionaries. *)

let int   = { zero = 0; one = 1; add = ( + ); mul = ( * ) }
let float = { zero = 0.0; one = 1.0; add = ( +. ); mul = ( *. ) }

(* -------------------------------------------------------------------------- *)
(** Some expressions. *)

(* [f] represents the expression [x ↦ (x+1)^3]. *)

let f =
  { eval = fun {one; add; mul; _} x ->
      let ( + ), ( * ) = add, mul in
      let cube x = x * x * x in
      cube (one + x)
  }

(* Test. *)

let () =
  assert (f.eval int 2 = 27);
  assert (f.eval float 2.0 = 27.0);
  ()

(* [var] represents the expression x. *)

let var : exp =
  { eval =
      fun (type v) _dict (x : v) ->
        x
  }

(* [subst e1 e2] substitutes the expression [e1] for the (unique) variable
   in the expression [e2], yielding a new expression. *)

let subst (e1 : exp) (e2 : exp) : exp =
  { eval =
      fun (type v) dict (x : v) ->
        e2.eval dict (e1.eval dict x)
  }

(* Monomial and power. *)

(* BEGINMONOMIAL *)
(* The expression x^k. *)
let monomial (k : int) : exp = { eval =
  fun (type v) { one; mul; _ } (x : v) ->
    let ( * ) = mul in
    let result, x, k = ref one, ref x, ref k in
    while !k > 0 do(*@ \label{line:monomial:while} *)
      result := !result * (if !k mod 2 = 0 then one else !x);(*@ \label{line:monomial:update:result} *)
      x := !x * !x;(*@ \label{line:monomial:update:x} *)
      k := !k / 2
    done;(*@ \label{line:monomial:done} *)
    !result }
(* ENDMONOMIAL *)

let pow (k : int) (e : exp) : exp =
  subst e (monomial k)

(* Test. *)
let () =
  assert ((monomial 8).eval float 2.0 = 256.0);
  assert ((pow 2 (pow 4 var)).eval float 2.0 = 256.0);
  ()

(* -------------------------------------------------------------------------- *)
(** Testing an implementation of AD. *)

(* [Test(A)] tests that [A.diff] behaves correctly on some simple inputs. *)
module Test (A : AD) = struct

  open A

  (* [df] should be [x ↦ 3(x+1)^2]. *)
  let df = diff f

  (* [ddf] should be [x ↦ 6(x+1)]. *)
  let ddf = diff (diff f)

  let () =
    assert  (df.eval int   2  = 27);
    assert  (df.eval float 2. = 27.);
    assert (ddf.eval int   2  = 18);
    assert (ddf.eval float 2. = 18.);
    ()

  (* [monomial 4] is [x^4], so its derivative its [4.x^3], whose evaluation
     at 2 should yield 32. *)
  let () =
    assert ((diff (monomial 4)).eval float 2.0 = 32.0);
    ()

end

(* -------------------------------------------------------------------------- *)
(** Testing each of our three implementations. *)

(* A list of our implementations of AD. *)
let algorithms : (module AD) list = [
  (module ForwardMode);
  (module StackBasedReverseMode);
  (module EffectHandlerBasedReverseMode);
]

(* Test every instance in [algorithms]. *)
let () =
  List.iter (fun (module A : AD) ->
    let module T = Test(A) in ()
  ) algorithms

(* If we reach this point, then the tests have succeeded. *)
let () =
  print_endline "Success."
\end{lstlisting}
\caption{Effect-based reverse-mode \AD in Multicore OCaml}
\label{fig:ad}
\end{figure}

We now propose a third implementation \AD,
which, like the previous two,
obeys the API (\sref{sec:intf})
and the specification (\sref{sec:spec})
proposed earlier.
It is based on Sivaramakrishnan's
implementation~\cite{kc-18}, which itself was inspired by Wang \etal.'s
work~\cite{wang-rompf-18,wang-19}.
Its code appears in \fref{fig:ad}.

The overall structure of the function \diff
is the same as in our earlier two examples
(Figures~\ref{fig:fmad} and~\ref{fig:stad}).
%
%
The representation of vertices and the auxiliary functions on vertices
(\linerange{type_t}{update}) are the same as in our earlier reverse-mode
implementation (\fref{fig:stad}).
What is new here is that the stack disappears.
In our earlier implementation, the call \oc|e.eval dict x|
(\lineref{stad:y} of \fref{fig:stad})
represents just the forward phase.
It fills up the stack,
which is then emptied via an explicit loop.
Here, in contrast, the dictionary~\numStruct is defined in such a way
that the call \oc|e.eval dict x|
(\lineref{eval})
represents both the forward phase and the backward phase.
No loop is necessary.

The definition of the dictionary~\numStruct is almost trivial.
Two effects, \oc|Add| and \oc|Mul|, are declared
at \linerange{Add}{Mul}.
The operations~\oc|add| and \oc|mul| perform these effects
(\linerange{add}{mul}).
The meaning of these operations is defined by the way in which
these effects are handled.

The bulk of the computation,
which takes place between \linesand{handle}{handle:one},
exploits effects in the following way.
Because the operations~\oc|dict.add| and \oc|dict.mul|
perform \oc|Add| and \oc|Mul| effects,
and because the function call \oc|e.eval dict x|
can invoke these operations,
it can itself
perform \oc|Add| and \oc|Mul| effects.
We handle these effects by wrapping this function call in
a handler (\lineref{match}).
The structure of this handler is analogous to that found in our
previous example (\fref{fig:ask}). Because the continuation is invoked in the
middle of the handler's code (\linesand{continue1}{continue2}),
the execution of the algorithm is divided in two
phases: a \emph{forward phase} and a \emph{backward phase}.

The forward phase lasts as long as the function call \oc|e.eval dict x| runs. During
this phase,
the operations~\oc|dict.add| and \oc|dict.mul|
can be invoked,
causing a
sequence of effects to take place. When an effect occurs, it is serviced by
the handler code that precedes the \oc|continue| statement
(\linesand{handle:Add:mk}{handle:Mul:mk}). Control is
then immediately handed back to the computation~\oc|e.eval dict x|,
while the execution
of the handler code that follows the \oc|continue| statement is postponed.

The backward phase begins when the function call \oc|e.eval dict x| terminates and
returns a vertex~\oc|y|,
which represents its final result.
At this point, the handler receives control
(\lineref{handle:Return}) and writes the number~$1$ into
\oc|y.d|.
%
Then, the control stack is unwound and all of the handler code whose execution
was postponed earlier is executed. Thus, for each \oc|Add| effect that
took place earlier, the code at
\linerange{handle:Add:update:a}{handle:Add:update:b} is
executed, and for each \oc|Mul| effect that took place earlier, the code at
\linerange{handle:Mul:update:a}{handle:Mul:update:b} is
executed.
%
%
Once this is over, the code block at \linerange{handle}{handle:one} is exited.
Control moves on to the last line, where the desired result is
found in the \oc|d| field of the vertex~\oc|x|.

\section{Mathematical Expressions}
\label{sec:math}
In preparation for the formal part of this paper, we define mathematical
expressions and prove some of their basic properties.
First, we introduce the syntax and evaluation of mathematical expressions
(\sref{sec:math:expr}). Then, we present an alternative syntax, where a
mathematical expression is viewed as a sequence of operations
(\sref{sec:math:altexpr}). Finally, we define partial derivatives and present
several forms of the Chain Rule (\sref{sec:math:chain}).

In our informal specification (\sref{sec:spec}),
we have focused on expressions of one variable.
We have written~$\E$ for a mathematical expression defined by $\E ::= \X \mid
0 \mid 1 \mid \E+\E \mid \E\times\E$, and we have written $\deriv\E$ for the
derivative of~$\E$ with respect to the variable~$\X$.
However, our proof requires reasoning about expressions of several variables
$\imath, \jmath, \ldots$ and about partial derivatives $\pp\E\jmath$.
Therefore, from here on, we work with
expressions of several variables.
We take a symbolic view of expressions and derivation:
an expression is regarded as an abstract syntax tree;
partial derivation $\pp\cdot\jmath$
is regarded as a transformation of an expression
into an expression.

\subsection{Expressions; Evaluation; Free Semiring}
\label{sec:math:expr}
\label{sec:math:semiring}



An \emph{expression}~$\E$ is an abstract syntax tree built out of
(1)~the constants~$\Zero$ and~$\One$;
(2)~applications of the binary arithmetic operators~$\Add$ and~$\Mul$,
also known as \emph{nodes};
and (3)~other numeric constants or variables, also known as \emph{leaves},
drawn from some set~$\iSet$.

\begin{defi}[Expressions]
Let $\iSet$ be a finite or infinite set.
Let $\imath$ and $\jmath$ range over $\iSet$.
The set $\DExpr{\iSet}$ of expressions~$\E$
whose variables are drawn from $\iSet$ is defined as follows:
\[\begin{array}{r@{}r@{\;}c@{\;}l}
                & \opVar & ::= & \Add \mid \Mul \\
\DExpr\iSet \ni & \E     & ::= &
  \Leaf\imath       \mid
  \Zero             \mid
  \One              \mid
  \Node\opVar\E\E
\end{array}\]
\end{defi}

When we wish to reason about expressions of one variable,
we fix a name~$\X$ and we instantiate $\iSet$ with the
singleton set $\SX$.
Thus, $\DExprSX$ denotes the set of expressions of one variable.
In Coq, the name~$\X$ is encoded as the unit value \oc|tt|,
and the singleton set $\SX$ is encoded as the \oc|unit| type,
whose single inhabitant is \oc|tt|.

An expression can be \textit{evaluated} in an arbitrary semiring
$(\Num, 0, \mathord+, 1, \mathord\times, \mathord\equiv)$,
where~$\equiv$ is an equivalence relation on the carrier set~$\Num$,
and where the axioms of a semiring hold with respect to this equivalence relation.
%
%
%
\emph{Evaluating} an expression~$\E\in\DExpr\iSet$
under an \emph{environment}~$\env$,
a~mapping of variables in~$\iSet$ to numbers in~$\Num$,
yields a number $\evalmap\E\env\in\Num$,
the value of this expression.

\begin{defi}[Expression Evaluation]
\label{def:evalmap}
  Evaluation~$\evalmapname$ is inductively defined as follows:
  \[\begin{array}{rcl}
        \evalmap{\Leaf\imath}\env &\eqdef& \env(\imath) \\
                \evalmap\Zero\env &\eqdef& 0 \\
                \evalmap\One\env  &\eqdef& 1 \\
    \evalmap{\Node\Add\el\er}\env &\eqdef& \evalmap\el\env + \evalmap\er\env \\
    \evalmap{\Node\Mul\el\er}\env &\eqdef& \evalmap\el\env \times \evalmap\er\env \\
  \end{array}\]
\end{defi}



When equipped with the syntactic constructors~$\Add$ and~$\Mul$ as addition
and multiplication, with the constants~$\Zero$ and~$\One$ as neutral elements,
and with a suitable definition of equivalence, the set $\DExpr\iSet$ forms a
semiring, known as the \emph{free semiring}.
%
\begin{lem}[Free Semiring]
\label{def:free:semiring}
Let~$\iSet$ be a set.
Define~$\Equiv{\DExpr\iSet}{}{}$
as the smallest equivalence relation on $\DExpr\iSet$
for which the semiring axioms hold.
Then,
$(\DExpr\iSet, \Zero, \Add, \One, \Mul, \Equiv{\DExpr\iSet}{}{})$
forms a semiring.
\end{lem}

The free semiring plays a role in the verification of the first phase of
\rmAD, whose purpose is to construct a sequential view and a DAG view of an
expression~$\E$ (\sref{sec:instances:reverse}). During this phase, the
expression~$\E$, which can be evaluated in an arbitrary semiring
(\sref{sec:intf:intf}), is evaluated in the free semiring
(\sref{subsubsec:aux:semiring}).


\subsection{Sequential View of Expressions}
\label{sec:math:altexpr}



As pointed out earlier (\sref{sec:instances:reverse}),
when reasoning about reverse-mode \AD,
it can be useful to have an alternate view of expressions
as sequences of elementary operations.
There, we proposed the abstract syntax
\(
  \SE ::= \bind\locu{\mathop\opVar}\locl\locr \kwin \SE
     \mid \locy
\).
It is in fact more convenient to use
the equivalent presentation
$\SE ::= \efill\K\locy$,
where a \emph{context}~$\K$ is defined
as a list of \emph{bindings}~$\binding$:

\begin{defi}[Bindings; contexts]\label{def:let-expr}
  Bindings and contexts are defined as follows:
  \[
  \begin{array}{r@{\;}c@{\;}l}
    \binding & ::= &
      \typicalbinding \\
    \K & ::= &
      \listnil \; \mid \;
      \listcons{\binding}\K
  \end{array}
  \]
\end{defi}

The identifiers~$\locu$, $\locl$, $\locr$, $\locy$ can be thought of as names
for auxiliary variables. In our Coq proof, they are drawn from the set~$\Val$
of the runtime values; in the paper, this detail does not matter much.
We write $\binding$ for a single binding and $\K$ for a list of
bindings, where, by convention, the left end of the list represents the
earliest binding and the right end represents the newest binding.
We use a semicolon to denote all three forms of concatenation,
that is, $\listcons\binding\K$ for ``cons'',
$\listsnoc\K\binding$ for ``snoc'',
and $\K\listconcat\K'$ for general concatenation.
Moreover, we use $\defs{K}$ to denote the list of identifiers
defined by the context $K$, that is,
$\defs{\listcons\typicalbinding\K} = \listcons\locu{\defs{\K}}$ and
$\defs{\listnil} = \listnil$.

The hole $\listnil$ at the right end of a list of bindings $\K$ can be viewed
as a placeholder, waiting to be filled with an identifier~$\locy$.
This is why we refer to~$\K$ as a \emph{context}.
A pair of a context~$\K$ and an identifier~$\locy$ forms an alternative
representation of an expression: it is a linear representation, where every
subexpression is designated by an identifier, and where the order of
construction is explicit. The identifier~$\locy$ designates the root of the
expression.
The operation of filling a context~$\K$ with an identifier~$\locy$, defined
next, converts an expression
represented as a pair $(\K, \locy)$ to an
ordinary tree-structured expression.

\begin{defi}[Filling]
\label{def:interp}
The function $\interpname$
is inductively defined as follows:
%
\[\begin{array}{rcl@{\qquad}l}
\interp\listnil\locy
  &\eqdef& \Leaf\locy &                            \\
\interp{(\listsnoc\K\typicalbinding)}\locy
  &\eqdef&
    \Node\opVar{\interp\K\locl}{\interp\K\locr}
                      &\;\text{if}\; \locu=\locy   \\
\interp{(\listsnoc\K{\typicalbinding})}\locy
  &\eqdef&
    \interp\K\locy    &\;\text{otherwise}
\end{array}\]
\end{defi}

If~$\K$ is a context and~$\locy\in\Val$ is an identifier,
then~$\interp\K\locy$ is an expression in~$\DExpr\Val$,
that is, an expression whose leaves are identifiers.



Our last definition is the extension of an environment $\env\in\Val\ar\Num$
with a context~$\K$, resulting in an updated environment
$\envExtension\env\K\in\Val\ar\Num$.
An intuition is that $\envExtension\env\K$ can be obtained by starting from
$\env$ and by executing the bindings in $\K$, in succession, from left
to right.
We could reflect this intuition by giving an inductive definition of
$\envExtension\env\K$.
Instead, we give the following direct definition, which is equivalent:

\begin{defi}[Extension of an Environment]
\label{def:extension}
  $\envExtension\env\K$ is the environment that maps
  every identifier~$\locy$
  to $\evalmap{\interp\K\locy}\env$.
\end{defi}

\subsection{Partial Derivatives; Chain Rule}
\label{sec:math:chain}



The partial derivative of an expression $\E\in\DExpr\iSet$
with respect to a variable $\jmath\in\iSet$
is an expression $\pp\E\jmath\in\DExpr\iSet$.

\begin{defi}[Partial Derivative]
\label{def:pd}
Partial derivation $\pp\cdot\cdot$ is inductively defined as follows:
\[\begin{array}{rcl}
  \pp{(\Leaf\imath)}\jmath &\eqdef& \One \qquad \text{if} \; \imath = \jmath \\
  \pp{(\Leaf\imath)}\jmath &\eqdef& \Zero \qquad \text{otherwise}             \\
  \pp{(\Zero)}\jmath &\eqdef& \Zero \\
  \pp{(\One)}\jmath  &\eqdef& \Zero \\
  \pp{(\Node\Add\el\er)}\jmath &\eqdef&
      \Node{\,\Add\,}
        {\pp\el\jmath}
        {\pp\er\jmath}                                                   \\
  \pp{(\Node\Mul\el\er)}\jmath &\eqdef&
      \Node{\;\Add\;}
        {\Node{\,\Mul}{\pp\el\jmath}{\er}}
        {\Node\Mul{\el}{\pp\er\jmath}}
\end{array}\]
\end{defi}

One recognizes in the above definition the well-known laws that indicate how
to compute a \pd of a variable, a constant, a sum, and a product. Most
mathematicians would view the above equations as a set of laws that can be
\emph{proved}, based on a more semantic definition of derivation. We take
these laws as the \emph{definition} of derivation, because this is sufficient
for our purposes, and removes the need for us to engage in deeper mathematics.

%
%

\begin{defi}[Derivative]
\label{def:sd}
  Let $E \in \DExprSX$ be a univariate expression.
  The derivative of~$\E$, written $\deriv\E$, is
  also a univariate expression,
  defined by $\deriv\E = \pp\E\X$.
\end{defi}


In traditional accounts of Calculus,
the Chain Rule states how to compute the
derivative of the composition of two differentiable functions.
In our formalism,
the Chain Rule states how to compute the partial
derivative of the expression~$\evalmap\E\F$.
%
%
This expression is the result of evaluating~$\E$
in the free semiring~$\DExpr\jSet$ under an environment~$\F$
that maps every variable~$\imath\in\iSet$ to an expression~$\F(\imath)$.
It can also be understood as the result of substituting
$\F(\imath)$ for $\imath$ in $\E$, for every $\imath\in\iSet$,
or as the sequential composition
``$\kw{let}\;[\imath=\F(\imath)]_{\imath\in\iSet}\;\kw{in}\;\E$''.



\begin{lem}[Chain Rule]
\label{lemma:chain:rule}
Let~$\Num$ be a semiring.
Let~$\iSet$ and~$\jSet$ be sets.
Let~$\E\in\DExpr\iSet$ be an expression
whose variables inhabit~$\iSet$.
Let~$\wtt\F{\typeArrow\iSet{\DExpr\jSet}}$
be a map of~$\iSet$ into the free
semiring~$\DExpr\jSet$.
Let the environment~$\wtt\envB{\typeArrow\jSet\Num}$
be a map of~$\jSet$ into~$\Num$.
Then, the following equation holds:
\[\begin{array}{r@{\;\quad}c@{\quad}l}
\Diff\envB{\evalmap\E\F}\jmath
& \equiv_\Num &
  \sum\limits_{\imath \in \vars\E}
    \Diff{\lambda\imath.\evalmap{\F(\imath)}\envB}\E\imath
      \times
    \Diff\envB{\F(\imath)}\jmath
\end{array}\]
\end{lem}

The left-hand side of this equation is
the partial derivative of~$\evalmap\E{F}$ with
respect to a variable~$\jmath\in\jSet$,
evaluated at~$\envB$.
The sum in the right-hand side is indexed
by the set~$\vars\E$.
This set is a finite subset of~$\iSet$:
it is the set of the variables~$\imath\in\iSet$
that occur in the expression~$\E$.
%
%
%


%
When reasoning about the backward phase of \rmAD,
we use the following specialized version of the Chain Rule:

\begin{lem}[Left-End Chain Rule]
\label{lemma:diff:filling}
Let~$\Num$ be a semiring.
Let~$\locl$,~$\locr$,~$\locu$,~$\x$,~$\y \in \Val$ be
auxiliary variables.
Let~$\opVar$ be an operation.
Let~$\Kl$ and~$\Kr$ be contexts.
Let~$\wtt\env{\typeArrow\Val\Num}$ be an environment.
Let~$\binding$ stand for the binding~$\typicalbinding$.
Suppose~$\x\neq\locu$.
Then, the following equation holds:
\[\begin{array}{r@{\quad}c@{\quad}r@{\;}l}
\Diff{\envExtension\env\Kl}
     {\interp{(\listcons\binding\Kr)}\y}\x
& \Equiv\Num{}{} & &
  \Diff{\envExtension\env{\listsnoc\Kl\binding}}
       {\interp\Kr\y}\x
\\[2mm]
& & + &
  \Diff{\envExtension\env{\listsnoc\Kl\binding}}
       {\interp\Kr\y}\locu                        \,\times\,
  \Diff{\envExtension\env\Kl}
       {\Node\opVar{(\Leaf\locl)}{(\Leaf\locr)}}\x
\end{array}\]
\end{lem}

The left-hand side is
the partial derivative
of~$\interp{(\listcons\binding\Kr)}\y$
with respect to~$\x$,
and the right-hand side
involves the partial derivatives
of $\interp\Kr\y$
with respect to~$\x$ and~$\locu$.
Therefore, this lemma concerns the addition of
a binding~$\binding$ at the \emph{left} end
of a sequence of bindings~$\Kr$.
This is typical of reverse mode; if we wanted
to reason about forward mode, we would instead
wish to add a binding
at the \emph{right} end of a sequence.


\lemmaref{lemma:diff:filling}
is obtained from \lemmaref{lemma:chain:rule}
by instantiating both~$\iSet$ and~$\jSet$ with~$\Val$
and by instantiating~$\E$,~$\F$, and~$\envB$ as follows:
%
\begin{itemize}
\item
$\wtt\E{\DExpr\Val}$ is the expression $\interp\Kr\y$;
\item
$\wtt\F{\typeArrow\Val{\DExpr\Val}}$ maps $\locu$ to
$\Node\opVar{(\Leaf\locl)}{(\Leaf\locr)}$
and is the identity elsewhere;
\item
$\wtt\envB{\typeArrow\Val\Num}$ is
$\envExtension\env\Kl$.
\end{itemize}

This lemma is key to the verification
of the backward phase of \rmAD,
where it justifies how
the \dfields of auxiliary variables
are incremented.
Consider a~vertex~$\locu$, constructed during the forward phase as the result
of an arithmetic operation~$\opVar$ whose operands are two earlier
vertices~$\locl$ and~$\locr$.
%
According to the backward invariant (\defref{def:backward:inv}),
during the backward phase,
when the vertex~$\locu$ is about to be processed,
its \dfield holds the partial derivative
$\pp{\interp\Kr\y}\locu$,
for some context~$\Kr$ and value~$\y$.
Similarly,
the~\dfields of the vertices~$\locl$ and $\locr$ hold
the partial derivatives
$\pp{\interp\Kr\y}\locl$
and
$\pp{\interp\Kr\y}\locr$.
After processing the vertex~$\locu$,
for the backward invariant to be preserved,
the~\dfields of the vertices~$\locl$ and $\locr$ must hold
the partial derivatives
$\pp{\interp{(\listcons\binding\Kr)}\y}\locl$
and
$\pp{\interp{(\listcons\binding\Kr)}\y}\locr$,
where
$\binding$ stands for the binding~$\typicalbinding$.
\lemmaref{lemma:diff:filling},
instantiated once with $\x := \locl$
and once with $\x := \locr$,
tells us exactly
what update instructions must be performed.

\section{A Program Logic for Effect Handlers}
\label{sec:hazel}
Traditional \SL~\cite{reynolds-02,ohearn-19} allows
specifications to be written
at a pleasant level of abstraction,
combining rigor and generality.
A program specification expressed in \SL is rigorous,
as it has a well-defined mathematical meaning.
It is general, because it does not mention the areas of
memory that the program must not or need not access:
it describes only the data structures that
the program needs to access or modify.

The program logic \hazel, proposed by the authors in
previous work~\cite{de-vilhena-pottier-21},
extends this methodology to support programs
that involve effects and effect handlers.
\hazel consists of two main components,
namely: (1)~a core programming language with support for
effect handlers, equipped with a small-step operational semantics;
and (2)~a program logic, where the standard notion of
\textit{weakest precondition}~\cite[\S6]{iris}
is parameterized with a
\textit{protocol}.
A~protocol can be understood as
a contract between a program that performs effects
and a context that handles these effects.


\subsection{\lang and its Operational Semantics}

To reason about programs in a rigorous way, a necessary step
is to define what programs are and how they behave.
In other words, one must define the syntax
and the dynamic semantics of the programming language of interest.
In this paper, we are interested in reasoning about \mocaml programs.
However, proposing a formal dynamic semantics for all of
Multicore OCaml
would be a challenging endeavor.
For this reason,
we focus on a core calculus, a subset of \mocaml,
which can express
our effect-based implementation of \AD~(\sref{sec:impl}).
A~small set of features, including
first-class functions,
references,
effect handlers and
one-shot continuations,
suffices.
In previous work~\cite{de-vilhena-pottier-21},
we have defined such a calculus,
dubbed \lang (for ``heaps and handlers''),
and have endowed it with a small-step operational semantics.
We do not recall the syntax or operational semantics of \lang,
whose details are not relevant here.
For the purposes of the present paper,
the following basic facts should suffice.
There is an infinite
set~$\typeLoc$ of memory locations.
The set~$\typeVal$ of values
contains the unit value,
memory locations,
binary products,
binary sums,
first-class functions,
and first-class continuations.
The heap is modeled as a
finite map of memory locations to values.

There are differences between \lang and \mocaml,
most of which we believe are inessential.
Perhaps the most critical difference is that an effect
in \mocaml carries a name and a value,
and an \oc|effect| declaration dynamically
generates a fresh name,
whereas an effect in \lang is nameless:
it carries just a value.
For the time being,
we gloss over this aspect,
and discuss it in \sref{sec:conclusion}.


\subsection{The program logic \hazel}

An operational semantics defines the behavior of programs,
but does not provide a convenient means
\emph{to describe} or \emph{to reason about}
this behavior at a high level of abstraction.
A program logic addresses this shortcoming.
It offers a specification language
in which one can describe how a program behaves
in the eye of an outside
observer, without exposing the details of its inner workings.

\hazel is a program logic for \lang.
%
\hazel is both an instance and an extension of the program logic
Iris~\cite{iris}. Iris is programming-language-independent: we
instantiate it for~\lang. Iris traditionally has no support for
effect handlers: we extend it with such support.
The main novel ingredient of \hazel's specification language is
a richer \textit{weakest precondition} predicate,
which is parameterized with a \emph{protocol}.

A traditional weakest precondition predicate,
as found in propositional dynamic logic \cite{pdl}
or in Iris~\cite[\S6]{iris},
takes the form~$\wpnomask\expr\post$,
where~$\expr$ is a program
(or an~expression that is part of a larger program)
and~$\post$ is a postcondition.
A postcondition is a~predicate
that describes the result value and the final state:
in short,
the assertion $\wpnomask\expr\post$
guarantees that the program~$\expr$ can be safely executed
(that is, it will not crash)
and that, if execution terminates,
then,
in the final state,
the result value~$\val$
satisfies the assertion~$\post(\val)$.
A~condition~$P$ on the initial state,
also known as a~precondition,
can be expressed via an implication:
$\propA \sepimp \wpnomask\expr\post$
means that, if initially the assertion~$P$ holds,
then it is safe to execute~$\expr$ and,
once a value~$\val$ is returned,
$\post(\val)$ holds.
In Iris, this implication is an \emph{affine} assertion,
which means that it
represents a permission to execute~$e$ \emph{at most once}.
When this assertion is wrapped in
the \emph{persistence} modality~$\pers$,
it represents a permission to execute~$\expr$
as many times as one wishes.
Thus, the persistent assertion~%
$\pers\;(\propA \sepimp \wpnomask\expr\post)$
is equivalent to the traditional Hoare triple~%
$\{\propA\} \,\expr\, \{\post\}$~\cite[\S6]{iris}.
The distinction between affine and persistent assertions
matters especially in \hazel because first-class
continuations in \lang are one-shot:
an attempt to invoke a continuation twice
causes a runtime failure. \hazel statically rules
out this kind of failure:
when proving the correctness of an algorithm, \hazel
requires the user to prove that every continuation
is invoked at most once.

In \lang, a program can not only diverge,
or terminate and return a value,
but can also interrupt itself by performing an effect.
For this reason, in contrast with a traditional
weakest precondition, an
\emph{extended weakest precondition} in \hazel takes the form~%
$\ewpnomask\expr\prot\post$, where~$\expr$ is a program,~%
$\post$ is a postcondition, and~$\prot$ is a \emph{protocol}.
A protocol describes the effects that a program may perform: it can be
thought of as a contract between the program and the
effect handler that encloses it.
In short, the assertion~$\ewpnomask\expr\prot\post$ means that (1)
it is safe to execute~$e$,
that (2) if a value~$\val$ is returned,
then~$\post(\val)$ holds, and
that (3) if~$\expr$ performs an effect,
then this effect respects the protocol~$\prot$.

What is a protocol?
We answer this question only partially at this point,
because the specification of \diff (\sref{sec:verif:spec}) involves very few
protocols. One important concrete protocol is the \textit{empty
  protocol}~$\protbot$, which forbids all effects:
the assertion $\ewpnomask\expr\protbot\post$
guarantees that the program~$\expr$
does not perform any effect.
Another important idiom is the use of an \emph{abstract protocol},
that is, a universally quantified protocol variable~$\prot$.
Abstract protocols are typically used to describe
\emph{effect-polymorphic} higher-order functions.
An expression in tagless-final style (\sref{sec:intf})
is an example of such a function.

\section{Formal Verification of Effect-Based Reverse-Mode AD}
\label{sec:verif}


We are equipped with a basic theory of mathematical expressions and
derivation (\sref{sec:math}) and with a program logic (\sref{sec:hazel}).
Therefore, we can now translate our earlier informal specification of \AD
(Statement~\ref{informal:spec:diff}) into a formal specification and prove
that our effect-based implementation of \rmAD~(\sref{sec:impl}) meets this
specification.
We follow this path.
We propose a formal specification of \AD
(\sref{sec:verif:spec})
and we document our machine-checked proof
that this implementation
satisfies this specification
(\sref{sec:verif:proof}).


\subsection{Specification}
\label{sec:verif:spec}

Our formal specification of \ad is a
straightforward
rendition in \hazel of
our earlier informal specification
(Statement~\ref{informal:spec:diff}).
\begin{statement}[Specification of Differentiation]
\label{formal:spec:diff}
The specification of the function \diff is expressed as follows:
\[\begin{array}{l}
  \pers\;
  \forall\,\ev,\,\E. \;
    \isExp\ev\E \sepimp
      \ewpnomask
        {(\App\diff\ev)}{\protbot}
        {\ev'.\;\isExp{\ev'}{\deriv\E}}
\end{array}\]
\end{statement}

The use of the empty protocol~$\protbot$ in this statement indicates that the
function call~$\App\diff\ev$ does not perform any effect.
This is in accord with the informal Statement~\ref{informal:spec:diff}, where
$\App\diff\ev$ is allowed to diverge or return a value, but is not allowed to
perform an effect.

This statement relies on the binary predicate~$\isExpname$,
which relates a runtime value~$\ev$ to a
mathematical expression~$\E$.
The assertion~$\isExp\ev\E$ reflects the
idea that~\emph{$\ev$~represents~$\E$\,},
for which we gave an informal definition
in Statement~\ref{informal:isExp}.
We now propose a formal definition:

\begin{defi}[Runtime Representation of an Expression]
\label{formal:isExp}
The predicate
$\wtt{\isExpname}{\typeArrow\Val{\typeArrow\DExprSX\iProp}}$
is defined as follows:
\[\begin{array}{l}
\isExp\expr\E \eqdef \crqb
  {\pers\;\forall\,\Num,\,\prot,\,\implementsname,\,
                   \zero,\,\one,\,\add,\,\mul.\crb
    {\numSpec
       \zero\one\add\mul\Num\prot\implementsname \sepimp\crb
      {\forall\,\tn,\,\num.\;\;\implements\tn\num\sepimp
         \ewpnomask
           {(\AppTwo{\dotEval\ev}\explicitNumStruct\tn)}
           {\prot}
           {\y.\;\implements\y{\evalmap\E{\parenvX\num}}}
      }
    }
  }
\end{array}\]
\end{defi}

The persistence modality $\pers$
that appears at the beginning of this definition
strengthens the meaning of
$\isExp\expr\E$ and makes it a persistent assertion.
Thus, once $\isExp\expr\E$ has been established,
this fact can be used as many times as one wishes,
as opposed to at most once.
As we will see shortly,
$\isExp\expr\E$ implies that $\expr$ is a function:
the persistence modality indicates that this
function can be called as many times as one wishes.%
\footnote{The persistence modality in the definition of~$\isExpname$ is in
fact optional. With this modality, Statement~\ref{formal:spec:diff} means that
$\App\diff\ev$ may evaluate the expression~$\ev$ several times and returns an
expression~$\ev'$ that can safely be evaluated several times. Without this
modality, Statement~\ref{formal:spec:diff} means that $\App\diff\ev$
evaluates~$\ev$ at most once and returns an expression~$\ev'$ that must be
evaluated at most once. We have verified in Coq that both variants of the
statement are true. (A Boolean parameter is used to avoid any duplication.) As
far as we can see, neither variant of the statement implies the other. We
thank one of the reviewers for pointing out the existence of the
variant without the persistence modality.}

Following this modality, comes a series of
universally quantified variables,
of the same nature and
in the same order
as in Statement~\ref{informal:isExp}.
An expression~$\expr$ in tagless-final style
is polymorphic in
a type of numbers
(a~semiring~$\Num$),
in the runtime representation of these numbers
(a predicate $\implementsname$),
in the implementation of the semiring operations
(the~values $\zero, \one, \add, \mul$),
and in a protocol~$\prot$.
%
%

The predicate $\implementsname$
relates a runtime value~$\tx$ and a number~$\num\in\Num$
and means that
``\emph{the value~$\tx$ represents the number~$\num$}''.
The universal quantification on $\implementsname$
makes it an abstract predicate:
the expression~$\expr$ must not know how numbers
are represented.
It may assume that the runtime values $\zero,\, \one,\, \add,\, \mul$
are correct implementations of the semiring operations
$0, 1, \mathord+, \mathord\times$:
this is expressed by the hypothesis
$\numSpec\zero\one\add\mul\Num\prot\implementsname$,
whose definition follows shortly.

The last line in the definition of~$\isExpname$
states that
applying $\dotEval\ev$
to two arguments,
namely the dictionary $\explicitNumStruct$
and a representation
of the number~$\num$,
must either
produce a representation
of the number~$\evalmap\E{\parenvX\num}$
or perform an effect permitted by~$\prot$.
The number~$\evalmap\E{\parenvX\num}$
is the result of evaluating the expression~$\E$
in the semiring~$\Num$ at the point~$\num$.
We have restricted our attention to expressions~$\E$
of a single variable, which is why an expression is
evaluated at a point~$\num\in\Num$.

%
\begin{defi}[Runtime Representation of a Semiring]
\label{def:numSpec}
The predicate $\numSpecname$ 
is defined as follows:
\[\begin{array}{@{}l@{}}
\numSpec\zero\one\add\mul\Num\prot\implementsname \eqdef \crqb{
  \begin{array}{@{}l@{\;}r@{}}
    \pers\;\implements\zero{0}                      & \land \\

    \pers\;\implements\one{1}                       & \land \\

    \pers\;\forall\,\locl,\,\locr,\,\num,\,\numb.\;
      \implements\locl\num \sepimp
        \implements\locr\numb \sepimp
          \ewpnomask{(\AppTwo\add\locl\locr)}\prot{\locu.\;
            \implements\locu{(\num + \numb)}
          }                                       & \land \\

    \pers\;\forall\,\locl,\,\locr,\,\num,\,\numb.\;
      \implements\locl\num \sepimp
        \implements\locr\numb \sepimp
          \ewpnomask{(\AppTwo\mul\locl\locr)}\prot{\locu.\,
            \implements\locu{(\num\times\numb)}
          }                                       & \land \\
    \pers\;\forall\,\locl,\,\num,\,\numb. \;
      \implements\locl\num \sepimp
        \Equiv\Num\num\numb \sepimp
          \implements\locl\numb
                                                  & \land \\
    \pers\;\forall\,\locl,\,\num. \;
      \implements\locl\num \sepimp \pers\;\implements\locl\num
                                                  &
  \end{array}
}
\end{array}\]
%

\end{defi}

The right-hand side of this definition
is a conjunction of the following claims:
(1)~the value~$\zero$ represents~$0$;
(2)~the value~$\one$ represents~$1$;
(3)~for all numbers $\num$ and $\numb$,
    when applied to representations of $\num$ and~$\numb$,
    $\add$ computes a representation of $\num+\numb$
    and may perform effects permitted by~$\prot$;
(4)~for all numbers $\num$ and $\numb$,
    when applied to representations of $\num$ and~$\numb$,
    $\mul$ computes a representation of $\num\times\numb$
    and may perform effects permitted by~$\prot$;
(5)~$\implementsname$ is compatible with
    the equivalence relation~$\Equiv\Num{}{}$, that is,
    if the numbers~$\num$ and~$\numb$ are equivalent,
    then a representation of~$\num$ is also a
    representation of~$\numb$;
(6) a representation of a number is persistent.
A look back at Statement~\ref{informal:isExp} confirms that
claims 1--4 were present there already. Claims 5--6 are more
technical and were omitted there.


\subsection{Verification}
\label{sec:verif:proof}

In this subsection,
we present a formal proof of
Statement~\ref{formal:spec:diff},
where we take $\diff$ to be a manual transcription in \lang
of the reverse-mode algorithm of \fref{fig:ad}.
To make it manageable,
we organize this proof in many segments.

In the beginning (\sref{subsubsec:hypo:goal}),
we step over the first few lines of the code in \fref{fig:ad}.
%
Then, we define a \SL predicate $\isVarname$, which describes auxiliary
variables (\sref{subsubsec:aux}), and step over the allocation of the first
auxiliary variable at label \extracoderef{init} (\sref{subsubsec:x}). This
brings us to the label \extracoderef{heart}.

At this point,
we allocate a piece of \emph{ghost state},
a key ingredient of the proof (\sref{subsubsec:ghost}).
%
%
This ghost state keeps a record of the interaction between the differentiation
algorithm and the expression that is being differentiated. Its evolution is
monotonic.
It appears in the main two invariants (\sref{subsubsec:invariants}).
The \emph{forward invariant} holds during the forward phase;
the \emph{backward invariant} holds during the backward phase.
They describe the content of
the \vfield and \dfield of
every auxiliary variable.

Still at this point,
we introduce another \SL predicate,
$\representsmodname$,
which equips vertices with semiring structure
(\sref{subsubsec:aux:semiring}).
This predicate plays a role in the definition of
the \emph{protocol} (\sref{subsubsec:protocol})
that describes the interaction between
the handlee and the effect handler.

After these definitions,
we resume our step-by-step inspection of the code.
We successively examine the combination of the handlee and the handler
(\sref{subsubsec:heart}),
the handlee alone (\sref{subsubsec:handlee}),
and the handler alone (\sref{subsubsec:handler}),
focusing in turn on the return branch
(\sref{subsubsec:return})
and on the effect branch
(\sref{subsubsec:add}).
Along the way, we point out where \hazel's key reasoning rules are used. We do
not show or explain these rules: for more information on Iris and \hazel, we
refer the reader to the papers where these logics are
presented~\cite{iris,de-vilhena-pottier-21}.


\subsubsection{Initial Hypotheses and Goal}
\label{subsubsec:hypo:goal}

The first step in the proof of Statement~\ref{formal:spec:diff} is to
introduce the metavariables~$\expr$ and~$\E$ and the assumption~$\isExp\ev\E$.
This assumption is persistent, so it remains present until the end of the
proof.
\begin{hypo}
\label{hyp:isExp}
We assume that $\expr$ is a value
and that $\E$ is a mathematical expression.
We assume that the persistent assertion
\(\isExp\ev\E\)
holds.
\end{hypo}
The goal is then reduced to the following assertion:
\begin{equation*}
\ewpnomask
  {(\App\diff\ev)}
  {\protbot}
  {\ev'.\;\isExp{\ev'}{\deriv\E}}
\end{equation*}
By definition of \diff,
the function call~$\App\diff\ev$ reduces (in one step)
to a value,
namely,
the value \oc|\{ eval = ... \}|
that appears
at label~\uniquecoderef{head}%
\footnote{In the electronic version of this paper, a label such as
\extracoderef{head} is a hyperlink towards this label in \fref{fig:ad}.
In the reverse direction, a label in \fref{fig:ad} is a hyperlink to
the point in the proof where this label is reached.}
in \fref{fig:ad},
%
in which the variable~\oc|e| is replaced
with the value~$\expr$.

In the following, for the sake of readability,
we elide the substitution $[\expr/\oc|e|]$.
We write that from here on, the variable~\oc|e|
denotes the value $\expr$. We allow ourselves
to write a goal where the variable~\oc|e| occurs,
and we leave it to the reader to understand that,
in such a goal, \oc|e| means~$\expr$.
More generally, every time we reason about a reduction step that causes a
substitution $[\val/\tx]$ to arise, we elide this substitution.
As this paper is accompanied with a machine-checked proof,
we adopt a certain level of informality and
choose to avoid the noise caused by these substitutions.
%
%
Thus,
whereas
in our machine-checked proof
the current goal always involves a closed term,
in the paper
our hypotheses, goals, and definitions have free
variables.
The reader must understand that each such variable
actually denotes a certain closed value to which this variable has been
bound at an earlier point in the proof.
%


With this convention in mind, we may write that
$\App\diff\ev$ reduces to the value (\extracoderef{head}),
eliding the substitution $[\expr/\oc|e|]$.
Thus, the goal can be transformed to:
\[\isExp{(\extracoderef{head})}{\deriv\E}\]
where, by convention, we use the label \extracoderef{head} as a short-hand for
the subexpression that is identified by this label in \fref{fig:ad}.%

The next step is to unfold~$\isExpname$ in this goal.
The goal becomes:
%
\[\begin{array}{l}
\pers\;
\forall\,\Num,\,\protR,\,\implementsnameR,\,
         \zeroR,\,\oneR,\,\addR,\,\mulR.
\crb{
  \numSpec\zeroR\oneR\addR\mulR\Num\protR\implementsnameR
  \sepimp\crb{
    \forall\,\nR,\,\num \in \Num.\;\;
    \implementsR\nR\num
    \sepimp\crb{
      \ewpnomask{(\AppTwo{\dotEval{(\extracoderef{head})}}\explicitNumStructR\nR)}
        \protR
        {\y.\;\implementsR\y{\evalmap{\deriv\E}{\parenvX\num}}
      }
    }
  }
}
\end{array}\]

We then introduce all of the metavariables and assumptions
that appear in the first three lines of this goal. Again,
these assumptions are persistent.

\begin{hypo}
\label{hyp:diff}
We assume
a semiring $\Num$,
a protocol $\protR$,
a predicate $\implementsnameR$,
four values $\zeroR,\, \oneR,\, \addR,\, \mulR$,
a value $\nR$,
and
a number $\num\in\Num$.
We assume that the following two persistent assertions hold:
\[\begin{array}{l}
\numSpec\zeroR\oneR\addR\mulR\Num\protR\implementsnameR
\\
\implementsR\nR\num
\end{array}\]
\end{hypo}



The goal is now reduced to:
\begin{equation*}
\ewpnomask
  {(\AppTwo{\dotEval{(\extracoderef{head})}}\explicitNumStructR\nR)}
  {\protR}
  {\y.\;
     \implementsR\y{\evalmap{\deriv\E}{\parenvX\num}}
  }
\end{equation*}

This expression reduces (in several steps) to the expression identified
by the label \uniquecoderef{body}:
\begin{equation*}
\ewpnomask
  {(\extracoderef{body})}
  {\protR}
  {\y.\;
     \implementsR\y{\evalmap{\deriv\E}{\parenvX\num}}
  }
\end{equation*}

Again, in this goal, we have elided a substitution of actual arguments for
formal parameters: from here on, the variables $\zero, \one, \add, \mul, \tn$
are bound to the values $\zeroR, \oneR, \addR, \mulR, \nR$.

Between the label \extracoderef{body} and the label \extracoderef{init}, the
code in \fref{fig:ad} contains a series of definitions. Each of these
definitions binds a variable to a value: therefore, it has no side effect and
reduces in one step, giving rise to a substitution of a value for a variable.
In keeping with our convention, in the paper, we elide these substitutions.
Thus, the previous goal reduces to the following goal:

\begin{goal}
\label{goal:init}
The goal is now:
\begin{equation*}
\ewpnomask
  {(\uniquecoderef{init})}
  {\protR}
  {\y.\;
     \implementsR\y{\evalmap{\deriv\E}{\parenvX\num}}
  }
\end{equation*}
\end{goal}

In words, the goal is to prove that the expression identified by the label
\extracoderef{init} in \fref{fig:ad} obeys the protocol~$\protR$ and
eventually returns a representation of the number
$\evalmap{\deriv\E}{\parenvX\num}$, which is the value of the
expression~$\deriv\E$ at the point $\num$.


\subsubsection{Auxiliary Variables}
\label{subsubsec:aux}


%

During its forward and backward phases, the algorithm allocates
objects of type~\tnumdual~(\fref{fig:ad}, \lineref{type_t}).
We refer to a value of this type as a \emph{vertex}
(\sref{sec:instances:reverse}).
In our proof
(which is based not the OCaml code,
 but on the \lang code),
such a value is either
the value~\oc|O|
or
the value~\oc|I|
or
an \emph{auxiliary variable} (\sref{sec:math:altexpr}),
that is,
a value of the form~$\tVar\,\pair\numm\loc$,
where
$\numm$ is the content of the \dfield
and
$\loc$ is the address of the \vfield.

To describe the content of an auxiliary variable in a concise fashion,
we introduce a predicate~$\isVarname$. In short,
the assertion
$\isVar\locu{\tov\numb}{\tod\numb}$
means that $\locu$ is an auxiliary variable whose \vfieldname{}~and
\dfields contain (representations of) the numbers $\tov\numb$ and
$\tod\numb$.
As is usual in \SL, this assertion also represents the unique
ownership of the auxiliary variable~$\locu$. In other words,
it represents a unique permission to read and update this
auxiliary variable.
%
\begin{defi}
\label{def:isVar}
The predicate~%
$\isVarname$
is defined as follows:
\[\begin{array}{l}
  \isVar\locu{\tov\numb}{\tod\numb} \eqdef
  \crqb{
    \exists\,\numm,\,\loc,\,\numd.\;
    \crqb{
      \ISEP\left\{
      \begin{array}{lll}
      \locu = \tVar\,\pair\numm\loc  & \isep & \loc\mapsto \numd \\
        \implementsR\numm{\tov\numb} & \isep &
        \implementsR{\numd}{\tod\numb}
      \end{array}
      \right.
    }
  }
\end{array}\]
\end{defi}

In this definition, $\locu$ is a value
and
$\tov\numb$ and $\tod\numb$ are numbers,
that is,
inhabitants of the semiring~$\Num$,
which was introduced in \hypref{hyp:diff}.



\subsubsection{The First Auxiliary Variable}
\label{subsubsec:x}

Let us now come back to the current goal, that is,
\goalref{goal:init}.
The subexpression at label \extracoderef{init}
begins with the definition \oc|let x = mk n|.
This definition has a side effect: it allocates
a fresh auxiliary variable in the heap.
The variable~$\tx$ becomes bound to a certain value,
%
which we do not explicitly describe: instead, in
keeping with our convention, we use the variable~$\tx$ to stand for it.
Therefore, we can say that this fresh auxiliary variable is described
by the assertion $\isVar\tx\num{0}$.%
\footnote{As the reader may recall, the variable $\tn$ denotes the value
  $\nR$, which, according to \hypref{hyp:diff}, is a~runtime
  representation of the number~$\num$.}

Because this assertion is not persistent, we do not present it as a displayed
Hypothesis. Instead, we show it as part of the current goal.

\begin{goal}
\label{goal:heart}
The previous goal, namely \goalref{goal:init}, is replaced with:
\begin{equation*}
\isVar\tx\num{0} \sepimp
\ewpnomask
  {(\uniquecoderef{heart})}
  {\protR}
  {\y.\;
     \implementsR\y{\evalmap{\deriv\E}{\parenvX\num}}
  }
\end{equation*}
\end{goal}
In other words, we are now at label \extracoderef{heart}
and we have allocated one auxiliary variable~$\tx$.


\subsubsection{Ghost Theory}
\label{subsubsec:ghost}

During their forward phase,
both
the stack-based~(\sref{sec:instances:reverse:code})
and the effect-based~(\sref{sec:impl})
implementations of \rmAD
construct a sequential view of the expression~$\E$.
To describe this view at the logical level,
we have defined a \emph{context}~$\K$
to be a list of \emph{bindings}~$\binding$
(\sref{sec:math:altexpr}).

While reasoning about the execution of the forward phase, we would like to
speak of the \emph{current context}, which records the bindings created so
far. Furthermore, we would like to remark that the current context evolves
in a monotonic manner: a new binding can be added at its right end, but an
existing binding is never removed. Thus, once a binding~$\binding$ appears in
the current context, this fact remains true forever.

These informal considerations can be made precise in Iris
(therefore also in \hazel)
using ghost state.
%
%
%
In our setting, a single ghost cell, which holds the current context,
suffices for this purpose. For the sake of brevity and simplicity, we do not
describe the low-level implementation of our ghost state: we refer the reader
to Timany and Birkedal's work \cite{timany-birkedal-21} for general principles
and to our Coq proof \cite{repo} for more detail. Instead, we describe the
abstract API offered by our ghost state. This API is given by the
following lemma, whose proof forms a small library that is separate
from our main proof.

\begin{lem}[Custom Ghost State]
\label{lemma:ghost:theory}
The following closed assertion is valid:

\[
%
%
\upd\;
\exists\,\namecurrCtx, \nameisEntry.\;
\ISEP
\left\{
\begin{array}{@{}l@{}}
%
%
\currCtx\listnil
\\[2mm]

%
%
\TirNameStyle{NewBinding} \\[-2.2mm]
  \pers\;
  \forall\K,\binding.\;
  \currCtx\K \sepimp \upd\;
    \ISEP\left\{
    \begin{array}{@{}l@{}}
      \currCtx{(\listsnoc\K\binding)} \\
      \isEntry\binding
    \end{array}
    \right.
\\[2mm]
%
%
\TirNameStyle{ExploitBinding} \\
  \pers\;
  \forall\K,\binding.\;
  \currCtx\K \sepimp
    \isEntry{\binding} \sepimp
      \binding \in \K
\\[2mm]
%
%
\TirNameStyle{PersistBinding} \\
  \pers\;
  \forall\binding.\;
  \isEntry\binding \sepimp
    \pers\;\isEntry\binding
\end{array}
\right.
\]

\end{lem}

The symbol $\upd$ is the \emph{update} modality:
the assertion $\upd\,P$ means
that it is possible to update the ghost state in such a way that the assertion
$P$ holds. Thus, this lemma can be read as follows: provided a certain ghost
update is performed (namely, the allocation of a ghost cell which holds the
current context), it is possible to define two abstract predicates
$\namecurrCtx$ and $\nameisEntry$ such that:
\begin{enumerate}
\item initially $\currCtx\listnil$ holds,
      that is,
      the current context is empty;
\item as many times as one wishes,
      $\currCtx\K$ can be transformed via a ghost update into
      $\currCtx{(\listsnoc\K\binding)} \isep \isEntry\binding$,
      thus extending the current context with a new binding;
\item the conjunction $\currCtx\K \isep \isEntry\binding$
      implies $\binding\in\K$,
      which means that if $\binding$ has been once
      inserted into the current context, then it remains forever in the
      current context;
\item the assertion $\isEntry\binding$ is persistent.
\end{enumerate}

The assertion~$\currCtx\K$ states that
the current context is~$\K$.
It is not persistent: the current context changes over time.
The assertion~$\isEntry\binding$ states that the
binding~$\binding$ appears in the current context.
It is persistent: a binding, once created, exists forever
in the current context.

\lemmaref{lemma:ghost:theory} states that one \emph{can} allocate
a piece of ghost state that has the properties described above.
The next step in our main proof is to
\emph{actually} 
allocate this ghost state. We do so by
eliminating the ghost update
and the existential quantifiers
that appear in the statement of \lemmaref{lemma:ghost:theory}.
This enriches our main proof with new persistent hypotheses:


\begin{hypo}
\label{hyp:ghost:state}
We assume that the predicates $\namecurrCtx$ and $\nameisEntry$
satisfy the laws \RULE{NewBinding}, \RULE{ExploitBinding}, and
\RULE{PersistBinding}.
\end{hypo}

The assertion $\currCtx\listnil$ is not persistent, so we do not list it as
part of \hypref{hyp:ghost:state}. Instead, we show it as part of the
current goal.

\begin{goal}
\label{goal:heart:isContext}
The previous goal, namely \goalref{goal:heart}, is replaced with:
\begin{equation*}
\isVar\tx\num{0} \sepimp
\currCtx\listnil \sepimp
\ewpnomask
  {(\extracoderef{heart})}
  {\protR}
  {\y.\;
     \implementsR\y{\evalmap{\deriv\E}{\parenvX\num}}
  }
\end{equation*}
\end{goal}

In other words, we are still at the program point \extracoderef{heart},
we have allocated the auxiliary variable~$\tx$,
and we have created a ghost current context, which is currently empty.
It is now time to spell out the invariants of the forward phase
and of the backward phase.


\subsubsection{Forward and Backward Invariants}
\label{subsubsec:invariants}

We are now ready to indicate exactly what values are stored,
during the forward and backward phases,
in the \vfield and \dfield
of each auxiliary variable.
For this purpose, we introduce two invariants.
The \emph{forward invariant} describes
the content of these fields
during the forward phase;
the \emph{backward invariant} describes
their content
during the backward phase.


The environment $\basicenv$, defined next, appears in the definitions of the
forward and backward invariants. Its role is as follows. The variables
$\zeroP$, $\oneP$, $\tx$
(bound at lines~\ref{line:zeroP}, \ref{line:oneP} and~\ref{line:x})
stand for certain values.%
\footnote{%
  $\zeroP$ and $\oneP$ denote the values \oc|O| and \oc|I|, while
  $\tx$    denotes the memory location allocated at \lineref{x}.
}
We wish to express the idea that the values
$\zeroP$ and $\oneP$ represent the numbers $0$ and $1$, respectively;
and that the value $\tx$ represents the number~$\num\in\Num$,
introduced in \hypref{hyp:diff}.
The environment $\basicenv$ serves this purpose:

\begin{defi}
\label{def:basicenv}
  Let $\wtt\basicenv{\typeArrow\Val\Num}$
  be an environment that maps
  the value $\zeroP$ to 0,
  the value $\oneP$ to 1,
  and the value $\tx$ to $\num$.
\end{defi}



The environment $\basicenv$ is useful because our invariants involve
expressions in $\DExpr\Val$, that is, expressions whose leaves are values. We
typically evaluate these expressions in the environment~$\basicenv$ or in an
extension of this environment. For this purpose, we define two shorthands:

\begin{defi}[Expression Evaluation Shorthands]
  We write $\evalbasic\E$ as a shorthand for $\evalmap\E\basicenv$.
  We write $\evalext\E\K$ as a shorthand for
  $\evalmap\E{\envExtension\basicenv\K}$.
\end{defi}


We can now give the definitions of the forward and backward invariants.




The forward invariant states that, during the forward phase:
\begin{itemize}
\item every auxiliary variable~$\locu$ created so far
      represents the mathematical expression $\interp\K\locu$,
      where the current context~$\K$ records all of the operations
      performed so far; and
\item the \vfield of the auxiliary variable~$\locu$
      stores the value of this expression.
\end{itemize}
This is expressed as follows:

\begin{defi}[Forward Invariant]
\label{def:forward:inv}
The forward invariant
$\ForwardInv\K$,
an assertion parameterized
by a context~$\K$,
is defined as follows:
\[\begin{array}{rcl}
\ForwardInv\K &\eqdef& \\
   \currCtx\K &\isep&
   \left(\begin{array}{l}
     \forall\locu\in\{\tx\}\cup\defs\K.
     \crqb{
       \vars{\interp\K\locu}\subseteq\{\zeroP,\oneP,\tx\}\isep \\
       \isVar\locu{\evalbasic{\interp\K\locu}}{0}
     }
   \end{array}\right)
\end{array}\]
\end{defi}

%

The first conjunct, $\currCtx\K$,
asserts that~$\K$ is the current context
(\sref{subsubsec:ghost}).
The second conjunct
asserts that~$\K$ is \emph{well-formed}:
that is,
filling~$\K$ with an auxiliary variable~$\locu$,
where~$\locu\in\{\tx\}\cup\defs\K$,
yields an expression~$\interp\K\locu$
whose leaves are (a subset of) the values
$\zeroP$,~$\oneP$, and~$\tx$.
Therefore, it makes sense to evaluate the expression $\interp\K\locu$
under the environment $\basicenv$.
The forward invariant asserts that the result of this evaluation,
namely the number $\evalbasic{\interp\K\locu}$,
is stored in the \vfield of the auxiliary variable~$\locu$.


Let us now move on to the backward invariant.

During the forward phase,
a sequence
of arithmetic operations
takes place,
which we represent by the current context~$\K$,
a sequence of bindings.
At the end of the forward phase,
this current context becomes fixed.
During the backward phase, the bindings in~$\K$
are examined and processed
in reverse order:
that is,
the rightmost bindings in the list~$\K$
are processed first.
Therefore,
during the backward phase,
it is natural to split~$\K$
into two contexts,
that is,
to write~$\K$ as a concatenation $\Kl\listconcat\Kr$,
where~$\Kl$ contains the
\emph{pending bindings},
which have not yet been processed,
and~$\Kr$ contains the
\emph{processed bindings}.

The backward invariant is defined as follows:

\begin{defi}[Backward Invariant]
\label{def:backward:inv}
The backward invariant~$\BackwardInv\Kl$,
an assertion parameterized
by a context~$\Kl$,
is defined as follows:
\[\begin{array}{rcl}
\BackwardInv\Kl          &\eqdef&
  \exists\,\Kr,\,\y.\;\BackwardInvBody\Kl\Kr\y              \\[3mm]
\BackwardInvBody\Kl\Kr\y &\eqdef&
  \ISEP\left\{\begin{array}{ll}
    \glabel{A} &
      \Equiv\Num
        {\evalmap{\deriv\E}{\parenvX\num}}
        {\Diffbasic{\interp{(\Kl\listconcat\Kr)}\y}\tx}     \\[2mm]
    \glabel{B} &
    \begin{array}{@{}l}
      \forall\,\locu\in\{\tx\}\cup\defs\Kl.                 \\
        \quad\isVar\locu
          {\evalbasic{\interp\Kl\locu}}
          {\Diffext\Kl{\interp\Kr\y}\locu}
    \end{array}
  \end{array}\right.
\end{array}\]
\end{defi}

The backward invariant is parameterized by
the pending bindings~$\Kl$
and begins with an existential quantification
over the processed bindings~$\Kr$ and
over the root auxiliary variable~$\y$.
%
It states that:
\begin{enumerate}
\item[(\gref{A})]
      the number that the algorithm aims to compute,
      namely the value of the symbolic derivative~$\deriv\E$,
      is the value of the partial derivative $\pp{\interp\K\y}\x$.%
\item[(\gref{B})]
     the \dfield of every \emph{pending}
     auxiliary variable~$\locu$
     stores
     the partial derivative of the expression~$\interp\Kr\y$
     with respect to~$\locu$.
\end{enumerate}

Point~\gref{A} intuitively follows from the fact that the expression
$\interp\K\y$ is the expression~$\E$. Technically, however, we cannot write
$\interp\K\y=\E$ because $\interp\K\y$ inhabits $\DExpr\Val$ (it is an
expression whose leaves are values) whereas $\E$~inhabits $\DExprSX$ (it is an
expression whose sole leaf is the variable~$\X$).

Point~\gref{B} reveals a fundamental distinction between processed and
pending auxiliary variables during the backward phase.
A processed auxiliary variable $\locu\in\defs\Kr$ is never read or written
again: in fact, its ownership has been abandoned, as it is not mentioned in
the invariant.
At the logical level,
because the binding that defines~$\locu$ is part of $\Kr$,
this binding influences the meaning of the expression~$\interp\Kr\y$.
Thus, the auxiliary variable~$\locu$ is merely a
name for a subexpression of the expression~$\interp\Kr\y$.
In contrast, a pending auxiliary variable
$\locu\in\{\tx\}\cup\defs\Kl$
can be thought of as a variable in a mathematical sense:
indeed, it can be a leaf of the expression~$\interp\Kr\y$,
and the value of the partial derivative~$\pp{\interp\Kr\y}\locu$
is stored in memory
in the \dfield of the auxiliary variable~$\locu$.


\subsubsection{Abstract Semiring Structure for Vertices}
\label{subsubsec:aux:semiring}


We have just given much detail about the representation of vertices
in memory and about their role during the forward and backward phases.
However, in the eyes of
the expression \oc|e : exp| with which the differentiation algorithm
interacts,
vertices
must be presented as abstract objects
equipped with semiring structure.
Indeed, an expression \oc|e : exp| is polymorphic in a semiring.

To equip vertices with semiring structure,
we exploit the idea that a vertex represents an expression
in~$\DExprSX$.
Expressions have semiring structure:
indeed,
$\DExprSX$ is a semiring,
the free semiring (\sref{sec:math:semiring}).

To express the idea
that a vertex~$\locu$ represents an expression
$\E\in\DExprSX$,
we must define
an assertion~$\representsmod\locu\E$,
where $\locu\in\Val$ is a vertex
and $\E\in\DExprSX$ is an expression.
Anticipating on the fact that
we will need to instantiate the parameter
$\implementsname$ in the definition of~$\numSpecname$
(\defref{def:numSpec})
with $\representsmodname$,
we want $\representsmodname$
to be compatible with the relation $\ExprEquivname$
and
to be persistent.

We would like the definition of $\representsmodname$
to express the intuitive idea that
a vertex~$\locu$
represents the expression~$\interp\K\locu$,
where $\K$ is the current context.
However, in the definition of $\representsmodname$,
we cannot rely on the assertion $\currCtx\K$,
because it is not persistent:
it represents the unique ownership of the ghost current context.
Fortunately, we can instead rely on
the predicate~$\nameisEntry$,
which is persistent.
In so doing, we exploit the fact that the current
context evolves in a monotonic manner.

\begin{defi}[Abstract View of Vertices]
\label{def:represents}
The assertion~$\representsmod\locu\E$
and the assertion $\represents\locu\E$
are inductively defined as follows:
\[\begin{array}{rcl}
\representsmod\locu\E
&\eqdef&
  \exists\,\T.\quad
  \represents\locu\T
  \quad\isep\quad
  \Equiv\DExprSX\T\E
\\[2mm]
%
\represents\locu\Zero                 &\eqdef& \locu =
  \zeroP                                           \\
%
\represents\locu\One                  &\eqdef& \locu =
  \oneP                                            \\
%
\represents\locu{(\Leaf\X)}           &\eqdef& \locu =
  \tx                                              \\
%
\represents\locu{(\Node\opVar\El\Er)} &\eqdef&
  \exists\,\locl,\,\locr.\;
    \begin{array}[t]{l}
      \isEntry{(\typicalbinding)}  \isep \\
      \represents\locl\El          \isep
      \represents\locr\Er
    \end{array}
\end{array}\]
\end{defi}

The vertices~$\zeroP$,~$\oneP$ and~$\tx$
respectively represent
the expressions~$\Zero$,~$\One$, and~$\Leaf\X$.
If a vertex~$\locu$ is the result of an
arithmetic operation~$\opVar$
whose operands were the vertices~$\locl$ and~$\locr$,
where~$\locl$ represents~$\El$
and~$\locr$ represents~$\Er$,
then
$\locu$
represents the expression~$\Node\opVar\El\Er$.
In summary, the assertion $\represents\locu\E$
means that
decoding the DAG that exists in memory, 
beginning at the vertex~$\locu$,
yields the tree~$\E$;
and
$\representsmodname$
is just the compatible closure of
$\representsname$.

At this point, the reader might expect us to prove that
we have successfully equipped vertices with semiring structure,
by establishing an $\numSpecname$ assertion.
Indeed, we are almost ready to do this;
we will do so in \lemmaref{lemma:aux:semiring}.
Before we can state this lemma, however,
we must provide a description of
the effects that the functions~$\addP$ and $\mulP$
are allowed to perform.
In \hazel, this is done by defining a \emph{protocol}.


\subsubsection{An Effect Protocol}
\label{subsubsec:protocol}

The arithmetic operations~%
$\addP$ (\lineref{add}) and~%
$\mulP$ (\lineref{mul})
perform effects.
In \hazel, an effect is described by a protocol~\cite{de-vilhena-pottier-21},
which describes the precondition and the postcondition
of the \oc|perform| instruction.
A protocol serves as a contract
that mediates the interaction
between a handlee and a handler.
We must now define the protocol
that is in use in the \rmAD algorithm of \fref{fig:ad}.

The functions~$\addP$ and $\mulP$ are trivial:
their bodies consist of a single \oc|perform| instruction.
This remark helps us guess how to define the protocol:
the protocol should literally paraphrase
the specifications
that we wish to give for the functions~$\addP$ and~$\mulP$.

What should these specifications be?
In the eyes of the expression \oc|e : exp| with which the differentiation
algorithm interacts,
a vertex is an abstract object,
which represents an expression (\sref{subsubsec:aux:semiring}).
We wish to state that $\addP$,
applied to two arguments~$\locl$ and~$\locr$
that represent two expressions~$\El$ and~$\Er$,
produces a result~$\locu$
that represents the expression~$\Node\Add\El\Er$.
We wish to make a similar statement about $\mulP$.

This leads us to define the protocol~$\protad$ (for ``operation'')
in the following way.

\begin{defi}[Protocol]
\label{def:protad}
The protocol~$\protad$ is defined as follows:
\[\begin{array}{@{}r@{\;}l@{\;}l@{\;}l@{}}
  \protaddef
\end{array}\]
\end{defi}

This is a \emph{send-receive} protocol~\cite{de-vilhena-pottier-21}.
The first line describes the message that is sent by the handlee;
the second line describes the reply that is sent by the handler.
Each line begins with a series of binders,
followed with the value that must be sent
(on the first line, the triple $(\opVar, \locl, \locr)$;
 on the second line, the value~$\locu$),
followed with an assertion
that must be satisfied.

Intuitively,
this protocol states that,
to perform an effect~$\opVar$
with operands~$\locl$ and~$\locr$,
there must exist expressions~$\El$ and~$\Er$,
such that~$\locl$ represents~$\El$
and~$\locr$ represents~$\Er$; and that,
if these conditions are met,
then a vertex~$\locu$
representing~$\Node\opVar\El\Er$
can be expected
as a result of performing this effect.
%

Thanks to this definition,
we can now state and prove
that we have successfully equipped
our vertices
with semiring structure:

\begin{lem}[Semiring Structure for Vertices]
\label{lemma:aux:semiring}
This persistent assertion holds:
\[
  \numSpec\zeroP\oneP\addP\mulP\DExprSX\protad\representsmodname
\]
\end{lem}

The proof is easy. A look at \defref{def:numSpec} helps recall the six
proof obligations: the values $\zeroP, \oneP, \addP, \mulP$ must implement the
four semiring operations; the predicate $\representsmodname$ must be compatible
and persistent.


\subsubsection{At the Heart of the Algorithm}
\label{subsubsec:heart}

After a long digression, it may seem,
we come back to the main narrative
of the proof.
Our current goal is still \goalref{goal:heart:isContext}:
\begin{equation*}
\isVar\tx\num{0} \sepimp
\currCtx\listnil \sepimp
\ewpnomask
  {(\extracoderef{heart})}
  {\protR}
  {\y.\;
     \implementsR\y{\evalmap{\deriv\E}{\parenvX\num}}
  }
\end{equation*}

Our next step is to establish that, at this point, the forward invariant
holds. It is not difficult to check that the entailment
\(
\currCtx\listnil \isep
\isVar\tx\num{0}
\vdash
\ForwardInv\listnil
\)
holds,
so the current goal can be changed to:
\begin{equation*}
\ForwardInv\listnil \sepimp
\ewpnomask
  {(\extracoderef{heart})}
  {\protR}
  {\y.\;
     \implementsR\y{\evalmap{\deriv\E}{\parenvX\num}}
  }
\end{equation*}

Because the expression at label \extracoderef{heart} is the sequential
composition of two subexpressions, which are respectively identified
by the labels \uniquecoderef{handle} and \uniquecoderef{done},
this goal can be decomposed
(via the sequencing rule of \SL)
into the following two goals:

\begin{goal}
\label{goal:handle}
The correctness of the forward and backward phases
is expressed by this goal:
\[\begin{array}{l}
\ForwardInv\listnil\sepimp
  \ewpnomask
    {(\extracoderef{handle})}
    {\protR}
    {\_.\;\BackwardInv\listnil}
\end{array}\]
\end{goal}


\begin{goal}
\label{goal:done}
The correctness of the final read
is expressed by this goal:
\[\begin{array}{l}
\BackwardInv\listnil \sepimp
\ewpnomask
  {(\extracoderef{done})}
  {\protR}
  {\y.\;\implementsR\y{\evalmap{\deriv\E}{\parenvX\num}}}
\end{array}\]
\end{goal}

We have ingenuously guessed that the assertion $\BackwardInv\listnil$
holds at the program point \extracoderef{done}.

The proof of \goalref{goal:done} is easy.
Indeed,
when $\Kl$ is empty,
the backward invariant implies that
the desired result,
namely the number~$\evalmap{\deriv\E}{\parenvX\num}$,
is equal to
the number~$\Diffbasic{\interp\Kr\y}\tx$,
%
%
which
is stored in the \dfield
of the auxiliary variable~$\tx$.

Therefore, the sole outstanding goal
is \goalref{goal:handle}.
It is arguably the most interesting goal in this entire proof,
because both the forward and the backward phases occur
during the execution of the subexpression
identified by the label~\extracoderef{handle},
and because
this subexpression is a \oc|match ... with effect ...| construct,
that is, a handlee wrapped in a deep handler.
The ability to reason about this construct is
a unique feature of \hazel,
in contrast with plain Iris.
%
%
%
For this purpose,
\hazel offers the following reasoning rule~\cite{de-vilhena-pottier-21}:

\[\inferrule[\nameReasoningTryWithDeep]{
    \ewpnomask{\oc|e|}\prot\post
      \and
    \isDeepHandler{}
      {\oc|h| }{\oc|r|}
      {\prot}{\prot'}
      {\post}{\post'}
  }{
    \ewpnomask{(\oc|match e with h \| r|)}{\prot'}{\post'}
  }
\]

In the syntax of \lang, a handler consists of two branches:
an effect branch~\oc|h|
and
a return branch~\oc|r|.
This reasoning rule
allows the handlee~\oc|e|
and the handler~$(\oc|h| \mid \oc|r|)$
to be separately verified,
provided that they agree on a protocol~$\prot$,
  which describes the effects that the handlee performs
  and that the handler's effect branch intercepts;
and on a postcondition~$\post$,
  which describes the value that the handlee returns
  and that the handler's return branch receives.
The rule's first premise corresponds to
the verification of the handlee.
The rule's second premise,
a \emph{deep-handler judgment},
requires the verification of the handler.

We now apply the reasoning rule~\ReasoningTryWithDeep
to \goalref{goal:handle},
choosing to instantiate
the metavariable~$\prot$
with the protocol~$\protad$
and the metavariable~$\post$
with the postcondition $\lambda\y.\;\representsmod\y\E$.
The metavariables~$\prot'$ and ~$\post'$
must be instantiated
with the protocol~$\protR$
and
with the postcondition $\lambda\_.\;\BackwardInv\listnil$.
This results in two goals,
\goalref{goal:eval}
and
\goalref{goal:handler},
which we separately examine.


\subsubsection{Reasoning About the Handlee}
\label{subsubsec:handlee}

The first goal that results from the application
of~\ReasoningTryWithDeep
is the following:

\begin{goal}
\label{goal:eval}
The correctness of the handlee
is expressed as follows:
\[\begin{array}{l}
\ewpnomask
  {(\uniquecoderef{eval})}
  {\protad}
  {\y.\;\representsmod\y\E}
\end{array}\]
\end{goal}

This goal requires verifying that
the code at label \extracoderef{eval}
computes a value~$\y$ that represents the expression~$\E$.
During its execution, this code
may perform effects according to the protocol~$\protad$.

A look at \fref{fig:ad} shows that
the code at label \extracoderef{eval} is the function call
\oc|e.eval dict x|.

\goalref{goal:eval} is proved as follows.
In the assertion~$\isExp\ev\E$ (\hypref{hyp:isExp}),
we unfold $\isExpname$ (\defref{formal:isExp})
and we instantiate the universally quantified metavariables
$\Num$, $\prot$, $\implementsname$,
$\zero$, $\one$, $\add$, and $\mul$
respectively with the free semiring $\DExprSX$,
the protocol~$\protad$,
the predicate~$\representsmodname$,
and the values $\zeroP$, $\oneP$, $\addP$, and $\mulP$.
We obtain the following fact:
\newcommand{\tnS}{\tn}
\newcommand{\numS}{\num}
\begin{equation*}
\begin{array}{l}
\numSpec\zeroP\oneP\addP\mulP\DExprSX\protad\representsmodname
\sepimp \crb{
  \forall\,\tnS.\;\forall\numS\in\DExprSX.\; \crb{
    \representsmod\tnS\numS \sepimp
    \crb{
      \ewpnomask
        {(\extracoderef{eval})}
        {\protad}
        {\y.\;\representsmod\y{\evalmap\E{\parenvX\numS}}}
    }
  }
}
\end{array}
\end{equation*}

Thanks to \lemmaref{lemma:aux:semiring}, the requirement on the first line holds.
We instantiate the metavariables~$\tnS$ and~$\numS$ on the second line
with~$\tx$ and~$\Leaf\X$.
By Definition~\ref{def:represents},
the instantiated assertion on the third line,
$\representsmod\tx{(\Leaf\X)}$,
simplifies to~$\tx = \tx$.
Thanks to the identity~$\evalmap\E{\parenvX{\Leaf\X}} = \E$
(which is proved by induction on~$\E$),
the fourth line becomes \goalref{goal:eval}.



\subsubsection{Reasoning About the Handler}
\label{subsubsec:handler}

The second goal that results from the application of
\ReasoningTryWithDeep
is a \emph{deep-handler judgment}.
In general,
a deep-handler judgment has the following shape:
\begin{equation}
\label{handler:judgment:shape}
\isDeepHandler{}
  {\oc|h|}{\oc|r|}
  {\prot}{\prot'}
  {\pred}{\pred'}
\end{equation}
We omit the definition of the deep-handler judgment,
which the reader can find
in a previous paper by the authors~\cite{de-vilhena-pottier-21}.
For the purposes of this proof,
it is sufficient to know that
this judgment includes correctness statements
for both the effect branch~\oc|h|
and the return branch~\oc|r|.
%

To present the second goal that results from the application of
\ReasoningTryWithDeep,
we introduce an abbreviation,~$\isHandlername$.
In anticipation for an inductive proof,
we parameterize this abbreviation with a context~$\Kl$.

\begin{defi}[Correct Handler]
\[\begin{array}{l}
\isHandler\Kl \eqdef \crqb{
  \isDeepHandlerVertical
    {\jointeeff
      {\uniquecoderef{eff-add}}
      {\uniquecoderef{eff-mul}}
    }
    {\uniquecoderef{ret}}
    {\protad}
    {\protR}
    {\y.\;\representsmod\y\E}
    {\_.\;\BackwardInv\Kl}
  }
\end{array}\]
\end{defi}

The assertion $\isHandler\Kl$ states that our effect handler handles effects
according to the protocol~$\protad$ and is itself allowed to perform effects
according to the protocol~$\protR$. It also states that the handlee must
return a value~$\y$ that represents the expression~$\E$ and that the handler
(that is, the handler's effect branch and return branch)
must terminate in a state where the bindings that have not yet been processed
(during the backward phase) are $\Kl$.

This abbreviation allows us to state the goal as follows:

\begin{goal}
\label{goal:handler}
The correctness of the first instance of the handler
is expressed as follows:
\[\ForwardInv\listnil \sepimp \isHandler\listnil\]
\end{goal}

This is the sole outstanding goal.

We remark that the application of \ReasoningTryWithDeep
requires the handlee and the handler to be separately verified.
Therefore,
the assertion~$\ForwardInv\listnil$,
which was present before the rule was applied
(\goalref{goal:handle}),
must be transferred either to the handlee or to the handler.
This assertion is not needed
while reasoning about the handlee (\sref{subsubsec:handlee});
therefore, we have transferred it to the handler.
This explains why it appears in \goalref{goal:handler}.

In order to establish \goalref{goal:handler},
we transform it into a more general statement,
which is amenable to a proof by induction:

\begin{goal}
\label{goal:handler:general}
The correctness of an arbitrary instance of the handler
is expressed as follows:
\begin{equation*}
\forall\,\Kl.\;
  \ForwardInv\Kl \sepimp \isHandler\Kl
\end{equation*}
\end{goal}

Although we do not show the definition of the deep-handler judgment,
we can reveal that it is a
\emph{non-separating conjunction} of
a statement about the effect branch
and
a statement about the return branch.
Thus, the assertion~$\ForwardInv\Kl$ serves
as a precondition both
for the effect branch
and
for the return branch.

The universal quantification on $\Kl$ allows us to reason
about a point in time
where the effect branch or the return branch is entered
and the current context is~$\Kl$.
The occurrence of~$\Kl$
in~$\ForwardInv\Kl$
means that,
when the handler is invoked,
a sequence of arithmetic operations
described by~$\Kl$
has been performed already.
The occurrence of~$\Kl$
in~$\isHandler\Kl$
means that,
when the effect branch or the return branch terminates,
the bindings that have \emph{not yet been processed}
in the backward phase
are exactly the bindings in~$\Kl$.
This reflects the well-bracketed manner in which
each instance of the handler is executed:
the bindings that have already been constructed (in the forward phase)
when the execution of the handler begins
are also the bindings that have not yet been processed (in the backward phase)
when the execution of the handler ends.

To prove \goalref{goal:handler:general},
we use
\emph{Löb induction}~\cite{iris}.
The reason why induction is required
is the recursive nature of deep handlers:
a deep handler is syntactic sugar
for a recursive function
in which a shallow handler is installed~\cite{de-vilhena-pottier-21}.
%
%
The application of Löb's induction principle
gives rise to the following induction hypothesis:
\begin{hypo}[Induction Hypothesis]
\label{hypo:lob}
\[
\later\;(\forall\,\Kl.\;
  \ForwardInv\Kl \sepimp \isHandler\Kl)
\]
\end{hypo}

The symbol~$\later$ is the
\emph{later} modality~\cite{iris}.
%
%
It prevents circular proofs:
in its absence,
\hypref{hypo:lob}
would coincide with \goalref{goal:handler:general},
so the proof would be finished---but
that would of course be an unsound form of reasoning.
The later modality can be eliminated only after
the program has performed at least one execution step,
thereby forbidding this kind of circular reasoning.

The current goal is still \goalref{goal:handler:general}.
This goal begins with a universal quantification on~$\Kl$,
which we now introduce. This yields the following hypothesis
and goal:
\begin{hypo}
  We assume that a context~$\Kl$ is given.
\end{hypo}
\begin{goal}
The goal is now:
\begin{equation*}
  \ForwardInv\Kl \sepimp \isHandler\Kl
\end{equation*}
\end{goal}

Unfolding $\isHandlername$,
unfolding the definition of the deep-handler judgment,
and applying the introduction rule for a non-separating conjunction
reduces this goal
to two subgoals:
the \emph{verification of the return branch}
\extracoderef{ret}
and
the \emph{verification of the effect branch}
$\jointeeff{\extracoderef{eff-add}}{\extracoderef{eff-mul}}$.
The former is discussed below
(\goalref{goal:seed}).
The latter is subdivided into
one subgoal related to the addition branch
\extracoderef{eff-add}
(\goalref{goal:add:branch})
and one subgoal related to the multiplication branch
\extracoderef{eff-mul},
whose description we omit,
as it is analogous.


%
\subsubsection{The Return Branch}
\label{subsubsec:return}

The body of the return branch is
the expression labeled \extracoderef{seed}.
%
\begin{goal}
\label{goal:seed}
The correctness of the return branch
is expressed as follows:
\[\begin{array}{l}
\ForwardInv\Kl\sepimp\crb
  {\representsmod\y\E\sepimp\crb
     {\ewpnomask
        {(\uniquecoderef{seed})}
        {\protR}
        {\_.\;\BackwardInv\Kl}
     }
  }
\end{array}\]
\end{goal}

The return branch is invoked when the function call \oc|e.eval dict x|
terminates. At this moment, the forward phase ends and the backward phase
begins. The return branch consists of just one instruction,
namely \oc|update y one|.
Under the assumption that the forward invariant holds
and that $\Kl$ is the current context,
we must prove that this instruction
establishes the backward invariant,
where every binding in $\Kl$ is pending.
We may also assume that the value~$\y$ satisfies
the postcondition of the handlee,
which is $\lambda\y.\;\representsmod\y\E$.


Confronting $\ForwardInv\Kl$ and $\representsmod\y\E$
allows us to deduce $\y \in \defs\Kl$.
This step exploits the reasoning rule \RULE{ExploitBinding}.
What is more, confronting these assertions allows us to
establish a link between
the expression~$\E$
and the expression~$\interp\Kl\y$.
In the interest of space,
we omit the details.

To justify that \oc|update y one|
can be safely executed,
an $\isVarname$ assertion,
which represents
a permission to update the \dfield of~$\y$,
must be presented.
Because~$\y \in \defs\Kl$ holds,
this permission can be obtained from the
forward invariant~$\ForwardInv\Kl$.

After executing \oc|update y one|,
this $\isVarname$ assertion is updated
so as to reflect the fact that
the \dfield of~$\y$ now contains the number~$1$,
and the goal is to establish the postcondition:
%
\[\exists\,\Kr,\,\y.\;\BackwardInvBody\Kl\Kr\y\]
To do so,
the existentially quantified
variables~$\Kr$ and~$\y$
are instantiated with~$\listnil$ and~$\y$.
We omit the details of the remainder of the proof of \goalref{goal:seed}.
In short, verifying that the backward invariant holds
involves checking that the \dfield of the auxiliary variable~$\y$
contains the number~$1$ and that the \dfield of every other auxiliary
variable~$\locu$ contains the number~$0$.
Indeed, very roughly speaking,
$\pp\y\y$ is 1 and $\pp\y\locu$ is 0.

\subsubsection{The Addition Effect Branch}
\label{subsubsec:add}

The body of the effect branch,
in the case of addition,
is the expression labeled \extracoderef{add-body}.
%
\begin{goal}
\label{goal:add:branch}
The correctness of the addition effect branch
is expressed as follows:
\[\begin{array}{l}
\forall\,\locl,\,\locr,\,\El,\,\Er,\,\kont.\; \crb
  {\ForwardInv\Kl \sepimp \crb
    {\represents\locl\El \sepimp
       \represents\locr\Er \sepimp \crb
      {\isCont\kont \sepimp \crb
        {\ewpnomask
           {(\uniquecoderef{add-body})}
           {\protR}
           {\_.\;\BackwardInv\Kl}
        }
      }
    }
  }
\end{array}\]
\end{goal}

The manner in which~$\Kl$ is shared between
the precondition~$\ForwardInv\Kl$
and the postcondition~$\BackwardInv\Kl$
expresses the fact that
the execution of this branch
begins at an arbitrary point of the forward phase,
after a sequence of arithmetic
operations described by~$\Kl$ has been performed,
and ends at the corresponding point in the backward phase,
when the sequence of pending bindings is~$\Kl$.

The precondition also includes
the assertions~$\represents\locl\El$
and~$\represents\locr\Er$,
which mean that
the values~$\locl$ and~$\locr$
represent two expressions~$\El$ and~$\Er$.
The effect handler can make these assumptions about~$\locl$
and~$\locr$ thanks to the protocol~$\protad$
(\defref{def:protad}),
where these assertions appear as a requirement
(a precondition) from the point of view of the
code that performs an effect.

The last part of the precondition
is the assertion~$\isCont\kont$,
whose definition is as follows.
%
%
%
%
%
For the sake of simplicity,
we
present a definition of $\isContname$ where two universal quantifiers have
been instantiated in a wise manner. This makes the hypothesis $\isCont\kont$
weaker than it could be, but still sufficient for our purposes. This
allows us to use the abbreviation $\isHandlername$ in the definition
of $\isContname$ and thereby to present a simplified definition of
$\isContname$.
%
%
\newcommand{\Bu}{\bind\locu\Add\locl\locr}
\newcommand{\KlBu}{\listsnoc\Kl\Bu}
\newcommand{\BuKr}{\listcons\Bu\Kr}
\newcommand{\parenKlBu}{(\KlBu)}
\newcommand{\parenBuKr}{(\BuKr)}
\newcommand{\Klprime}{\Kl'}
\newcommand{\Krprime}{\Kr'}
\vspace{-\baselineskip} 
\begin{defi}
\label{def:isCont}
The assertion $\isCont\kont$ is defined as follows:
\begin{equation*}
\begin{array}{l}
\isCont\kont \eqdef \crqb{
  \forall\,\locu.\crb{
  \metalet\Klprime\parenKlBu{\crb{ 
  \later\isHandler\Klprime
  \sepimp \crb{
    \represents\locu{(\Node\Add\El\Er)} \sepimp \crb{
      \ewpnomask{(\App\kont\locu)}
        {\protR}
        {\_.\;\BackwardInv\Klprime}
      }
    }
  }}}}
\end{array}
\end{equation*}
\end{defi}

The assertion $\isCont\kont$ expresses a hypothesis about the
continuation~$\kont$. It is the \emph{specification of the
continuation}: it states under what condition one can invoke
the continuation, and what property one can expect to hold as
a result of this call.

This specification takes the form of an $\ewpname$ judgment about a function
call of the form~$\App\kont\locu$, where $\locu$ is a value.
Because an $\ewpname$ judgment is \emph{not} persistent, the assertion
$\isCont\kont$ is not persistent either. It represents a permission to invoke
the continuation~$\kont$ \emph{at most once}.

In essence,
this specification states that~$\kont$
can be applied to a vertex~$\locu$,
provided that~$\locu$
represents the expression~$\Node\Add\El\Er$.
The handler must satisfy this constraint
when invoking~$\kont$
because
of the protocol~$\protad$ (\defref{def:protad}),
where this constraint appears as a guarantee (a postcondition)
from the point of view of the
code that performs an effect.

The continuation invocation~$\App\kont\locu$
is subject to another precondition,~%
$\isHandler\Klprime$,
and its postcondition is~%
$\BackwardInv\Klprime$,
where we have chosen
to let $\Klprime$ stand for $\KlBu$,
that is, for the extension
of the sequence $\Kl$
with the new binding $\Bu$.
We have made this decision because we know
that this invocation
begins during the forward phase
after the sequence of operations~$\Klprime$ has been performed
and ends during the backward phase
when the sequence of pending bindings is $\Klprime$.
%
The precondition~%
$\isHandler\Klprime$
reflects the fact that
this is a deep handler:
a new instance of the handler is reinstalled
as the bottommost
frame
inside the continuation~$\kont$.
Thus,
in order to be allowed to invoke this continuation,
we must prove that this new instance of the handler
is also a correct handler.




Let us now attack
the proof of
\goalref{goal:add:branch}.
The code at \extracoderef{add-body}
is a sequential composition of three segments,
namely:
the allocation of a new auxiliary variable,
at \extracoderef{add-fwd},
part of the forward phase;
the invocation of the continuation,
at \extracoderef{cont};
and
two update instructions,
at \extracoderef{add-bwd},
part of the backward phase.
The sequencing rule and the frame rule of \SL let us verify
each segment independently,
as follows.


\paragraph{Segment 1: Forward Phase}


\begin{goal}
\label{goal:add-fwd}
The correctness of the instruction at label~\extracoderef{add-fwd}
is expressed as follows:
\[\begin{array}{l}
\ForwardInv\Kl\sepimp
\crb{
  \represents\locl\El \sepimp
  \represents\locr\Er \sepimp
  \crb{
    \ewpVerticalAlt{}
      {(\uniquecoderef{add-fwd})}
      {\protR}
      {\locu}
      {\metalet\Klprime\parenKlBu{\crb{ 
       \ForwardInv\Klprime \isep
       \represents\locu{(\Node\Add\El\Er)
      }}}}
  }
}
\end{array}\]
\end{goal}
%
The appearance of the forward invariant~%
$\ForwardInvname$
in the pre- and postcondition
indicates that the execution of~%
\extracoderef{add-fwd} takes place during
the forward phase.

The fact that $\ForwardInv\Kl$
is transformed into $\ForwardInv\Klprime$
reflects the fact that a new arithmetic operation
has taken place.
To perform this transformation,
one must begin with the following two steps:
\begin{itemize}
\item
unfold~$\ForwardInv\Kl$
(\defref{def:forward:inv}),
thereby revealing $\currCtx\Kl$;
\item
perform a ghost update,
by applying \RULE{NewBinding},
thereby replacing $\currCtx\Kl$ with
the conjunction of
$\currCtx\parenKlBu$,
which is necessary to reestablish the
forward invariant,
and
$\isEntry{(\Bu)}$,
which is necessary to prove
that~$\locu$ represents~$\Node\Add\El\Er$.
\end{itemize}

To establish $\ForwardInv\Klprime$,
one must also check that,
for every auxiliary variable
$\x\in\{\tx\}\cup\defs\Klprime$,
the assertion
$\isVar
  {\x}
  {\evalbasic{\interp\Klprime{\x}}}
  {0}
$ holds.
This is easy; we say no more.
%

\paragraph{Segment 2: Continuation Invocation}

From this point on, the variable~\oc|u| (bound at \lineref{handle:Add:mk}) is
  in scope, and (in our proof) so is the value~$\locu$ that this variable
  denotes.
Therefore, we may introduce the following abbreviations:
we let $\Klprime$ stand for $\parenKlBu$,
and
we let $\Krprime$ stand for $\parenBuKr$.

\begin{goal}
\label{goal:cont}
The correctness of the continuation invocation
is expressed as follows:
\[\begin{array}{l}
\isCont\kont\sepimp\crb
  {\ForwardInv\Klprime \sepimp\crb
    {\represents\locu{(\Node\Add\El\Er)}\sepimp\crb
      {\ewpnomask
        {(\uniquecoderef{cont})}
        {\protR}
        {\_.\;\BackwardInv\Klprime}
      }
    }
  }
\end{array}\]
\end{goal}

This goal amounts to checking that invoking the continuation is permitted.
By unfolding the definition of~$\isContname$~(\defref{def:isCont}),
it is easy to see that this assertion
allows such an invocation.
%
%
%
\begin{goal}
\label{goal:handler:stepped}
\goalref{goal:cont} reduces to the following goal:
\[
  \ForwardInv\Klprime \sepimp
  \later\isHandler\Klprime
\]
\end{goal}
In other words, we must check that, after the forward phase advances
from~$\Kl$ to~$\Klprime$, the handler continues to behave correctly.

\goalref{goal:handler:stepped} is a direct consequence from
our induction hypothesis
(\hypref{hypo:lob}),
which is applicable
because the conclusion of this goal
is guarded by a later modality.


\paragraph{Segment 3: Backward Phase}

\begin{goal}
\label{goal:add-bwd}
The correctness of the instructions at label~\extracoderef{add-bwd}
is expressed as follows:
\[
\BackwardInv\Klprime\sepimp
\ewpnomask
  {(\uniquecoderef{add-bwd})}
  {\protR}
  {\_.\;\BackwardInv\Kl}
\]
\end{goal}

This goal states that
the two update instructions
identified by the label \extracoderef{add-bwd}
advance the backward invariant by one step.
More precisely,
if the backward invariant holds
of the list of pending bindings~$\Klprime$,
then,
after the execution of these two update instructions,
the backward invariant holds of the list~$\Kl$.
Hence,
these update instructions correctly
process the binding~$\Bu$:
they remove it from the list of pending bindings.

The proof of \goalref{goal:add-bwd} begins
by unfolding
the backward invariant~$\BackwardInv\Klprime$
(\defref{def:backward:inv})
and eliminating the existential
quantifiers that appear as a result.
This yields the following hypothesis and new goal:

\begin{hypo}
We assume that a context~$\Kr$
and an auxiliary variable~$\y$ are given.
\end{hypo}

\begin{goal}
\label{goal:add-bwd:unpacked}
\goalref{goal:add-bwd} is replaced with:
\[
\BackwardInvBody\Klprime\Kr\y
\sepimp
\ewpnomask
  {(\extracoderef{add-bwd})}
  {\protR}
  {\_.\;\BackwardInv\Kl}
\]
\end{goal}

The next step is to unfold $\BackwardInv\Kl$ in the postcondition and to
introduce the existential quantifiers that appear as a result.
We instantiate them with $\Krprime$ and $\y$.

\begin{goal}
\label{goal:add-bwd:unpacked:intro}
\goalref{goal:add-bwd:unpacked} is replaced with:
\[
\BackwardInvBody\Klprime\Kr\y
\sepimp
\ewpnomask
  {(\extracoderef{add-bwd})}
  {\protR}
  {\_.\;\BackwardInvBody\Kl\Krprime\y}
\]
\end{goal}

There remains to unfold the two occurrences of $\BackwardInvBodyname$
(\defref{def:backward:inv})
and prove that the two update instructions achieve the desired effect.
The first occurrence,
which describes the state before these updates have taken place,
is expanded as follows:
\begin{equation*}
\BackwardInvBody\Klprime\Kr\y
\quad\equiv\quad
  \ISEP\left\{\begin{array}{ll}
    \glabel{EA1} &
    \Equiv\Num
      {\evalmap{\deriv\E}{\parenvX\num}}
      {\Diffbasic
        {\interp{(\Klprime\listconcat\Kr)}\y}\tx
      }                                                     \\[2mm]
    \glabel{EB1} &
    \begin{array}{l}
      \forall\,\x\in\{\tx\}\cup\defs\Klprime. \\[1mm]
        \quad\isVar\x
          {\evalbasic{\interp\Klprime\x}}
          {\Diffext\Klprime
                {\interp\Kr\y}\x
          }
    \end{array}
  \end{array}\right.
\end{equation*}




The second occurrence,
which describes the state
after these updates have taken place,
is expanded as follows:
\begin{equation*}
\BackwardInvBody\Kl\Krprime\y
\quad\equiv\quad
  \ISEP\left\{\begin{array}{ll}
    \glabel{EA2} &
    \Equiv\Num
      {\evalmap{\deriv\E}{\parenvX\num}}
      {\Diffbasic
        {\interp{(\Kl\listconcat\Krprime)}\y}\tx
      }
                                                   \\[2mm]
    \glabel{EB2} &
    \begin{array}{l}
      \forall\,\x\in\{\tx\}\cup\defs\Kl.           \\[1mm]
        \quad\isVar\x
          {\evalbasic{\interp\Kl\x}}
          {\Diffext\Kl
                {\interp\Krprime\y}\x
          }
    \end{array}
  \end{array}\right.
\end{equation*}

Claims~\gref{EA1} and~\gref{EA2} coincide,
because $\Klprime\listconcat\Kr$ and
$\Kl\listconcat\Krprime$
are the same sequence.

Establishing that
Claim~\gref{EB2}
holds after the two update instructions
is the crux of this proof.
Intuitively,
this claim asserts that the \dfield of
every pending auxiliary variable is
correctly updated:
%
after the update,
the \dfield of every auxiliary variable~$\x\in\{\tx\}\cup\defs\Kl$
must hold the partial derivative of
$\interp\Krprime\y$
with respect to~$\x$.

The key mathematical tool that is used in the proof of Claim~\gref{EB2}
is the Left-End Chain Rule (\lemmaref{lemma:diff:filling}).
%
The proof begins with
a three-way case disjunction on~$\x$:
it must be the case that either
(1)~$\x \notin \{\locl,\locr\}$, or
(2)~$\locl \neq \locr$ and~$\x\in\{\locl, \locr\}$, or
(3)~$\x = \locl = \locr$.

In case (1),
\lemmaref{lemma:diff:filling}
boils down to the following simple equality:
\[
\Diffext\Kl
     {\interp\Krprime\y}\x \Equiv\Num{}{}
  \Diffext\Klprime
       {\interp\Kr\y}\x
\]
This equality means that the \dfield of the auxiliary variable~$\x$ does
not need to be updated.
In this case,
the update instructions at~\extracoderef{add-bwd}
are indeed correct,
since they do not update~$\x$:
they update only the auxiliary variables~$\locl$ and $\locr$.

In case (2),
\lemmaref{lemma:diff:filling}
yields the following equality:
\[
\Diffext\Kl
     {\interp\Krprime\y}\x \Equiv\Num{}{}
  \Diffext\Klprime
       {\interp\Kr\y}\x
    \,+\,
  \Diffext\Klprime
       {\interp\Kr\y}\locu
\]
This equality implies that,
in order to reach the final state
described by Claim~\gref{EB2},
it suffices to increment the \dfield of the auxiliary variable~$\x$
by the quantity~%
$\Diffext\Klprime
      {\interp\Kr\y}\locu$.
By Claim~\gref{EB1},
this number is stored in the \dfield
of the auxiliary variable~$\locu$.
In this case,
the update instructions at~\extracoderef{add-bwd}
are indeed correct,
since they update~$\x$
(which is either~$\locl$ or~$\locr$, but not both)
precisely in the desired way.

Finally, in case (3),
\lemmaref{lemma:diff:filling}
yields the following equality,
where~$2$ denotes~$1 + 1$: 
\[
\quad\quad
\Diffext\Kl
     {\interp\Krprime\y}\x \Equiv\Num{}{}
  \Diffext\Klprime
       {\interp\Kr\y}\x
  \,+\,
  2
  \,\times\,
  \Diffext\Klprime
       {\interp\Kr\y}\locu
\]
This equality implies that
the \dfield of~the auxiliary variable~$\locl$
must be incremented by twice the number~%
$\Diffext\Klprime
      {\interp\Kr\y}\locu$.
Again, this is precisely the behavior
of the update instructions at~\extracoderef{add-bwd},
since in this case they update~$\x$
(which is the same as~$\locl$ and~$\locr$)
twice in succession.

This concludes the proof.
We have verified that Statement~\ref{formal:spec:diff} holds.
Therefore, we have verified that the \rmAD algorithm in \fref{fig:ad}
is correct.


\section{Related Work}
\label{sec:related}
We organize our review of the related work into three categories:
sources of inspiration for the specification and the code
that are presented in this paper
(\sref{sec:related:sources});
approaches to reasoning about effect handlers
(\sref{sec:related:reasoning});
and
formal presentations and correctness arguments for
automatic differentiation algorithms
(\sref{sec:related:ad}).


\subsection{Sources of Inspiration}
\label{sec:related:sources}



Our minimalist API for define-by-run \AD (\fref{fig:sig}),
which relies on a tagless-final representation of
expressions~\cite{carette-kiselyov-shan-09,kiselyov-tagless-final-10},
appears to be new.
This API seems remarkable insofar as it is very simple:
in short,
\diff has type \oc|exp -> exp|.
It is used as the basis for an arguably
similarly simple specification in \hazel
(Statement~\ref{formal:spec:diff}).

This API can be implemented in several ways.
We provide three implementations:
a forward-mode algorithm based on dual numbers (\fref{fig:fmad}),
a reverse-mode algorithm that exploits an explicit stack (\fref{fig:stad}),
and
a reverse-mode algorithm that exploits effect handlers (\fref{fig:ad}).
One can imagine other implementations,
such as a reverse-mode algorithm
that constructs backpropagator functions
\cite{pearlmutter-siskind-08,brunel-mazza-pagani-20,smeding-vakar-23}.
These implementations may involve complex programming-language features, such
as dynamically allocated mutable state, higher-order functions, and effect
handlers, but this complexity is (to a large extent) abstracted away in our
API (\fref{fig:sig} and Statement~\ref{formal:spec:diff}), as this API
describes only the \emph{interaction} between the differentiation algorithm
and the outside world.

It is worth remarking that this outside world is in fact split in two
components, namely the \emph{subject}
(that is, the program fragment that is differentiated)
and
the \emph{customer} 
(that is,
the program fragment that requests the evaluation of the derivative
at a certain point).
This is reflected in our API by the fact that the type \oc|exp| occurs
twice in the type of \diff (\fref{fig:sig}) and, similarly, the predicate
$\isExpname$ occurs twice in the specification of \diff
(Statement~\ref{formal:spec:diff}).
One occurrence describes the interaction between
the subject
and the differentiation algorithm:
in this dialogue,
the subject plays the role of an expression,
which the differentiation algorithm is allowed to query.
The other occurrence describes the interaction between
the differentiation algorithm
and
the customer:
there,
the differentiation algorithm plays the role of an expression,
which the customer queries.



The code that we present in \fref{fig:ad} and that we verify is inspired by
Wang \etal.~\cite{wang-rompf-18,wang-19} and by Sivaramakrishnan \cite{kc-18}.
Wang \etal.'s key observation is that the use of delimited-control operators,
such as \texttt{shift} and \texttt{reset}, allows an elegant compositional
presentation of reverse-mode \AD.
Furthermore, by making clever use of staging, they are able to propose both
define-by-run and define-then-run implementations that share this common
compositional architecture.
They present an implementation in Scala and evaluate its performance.
%
%
Inspired by Wang \etal.'s ideas, Sivaramakrishnan~\cite{kc-18} proposes a
minimalist implementation in \mocaml that uses effect handlers instead of
\texttt{shift} and \texttt{reset}. We retain the essence of this
implementation and we adapt it so as to satisfy our more abstract API.

Another implementation of \rmAD, written in Frank, is documented by
Sigal~\cite{sigal-21}. It differs from ours in several aspects. First, Frank
does not have primitive mutable state, so it is simulated via effect handlers.
Second, Sigal presents other \AD algorithms, including
a reverse-mode algorithm that performs checkpointing.


\subsection{Reasoning about Effect Handlers}
\label{sec:related:reasoning}

Goodenough~\cite{goodenough-structured-75} provides a meticulous study of
resumable exceptions, a mechanism that allows the exception handler to resume
the suspended computation. This feature can be seen as a restricted form of
effect handlers where the continuation is not a first-class object and must be
invoked while the effect handler is running. Goodenough describes the
exception-handling methods of the time and compares them according to several
criteria, including implementation difficulty, efficiency, and readability.

Effects and effect handlers offer an interface to delimited control;
that is, they
offer the ability to capture a delimited continuation and to reify it as a
first-class value.
Before Plotkin and Pretnar's seminal work~\cite{plotkin-pretnar-09}, which
introduced effect handlers, a large family of delimited-control operators had
been studied already.
Filinski~\cite{filinski-96} shows that many of these operators can be
expressed in terms of \texttt{reify} and \texttt{reflect}, two novel
programming constructs that he introduces to simulate various
forms of effects in a pure language.
Intuitively, the \texttt{reflect} construct allows the programmer
to introduce an effect by giving its implementation as a term in a monad.
The \texttt{reify} construct then translates a program to a monadic
expression by transforming \kw{let} bindings into monadic binds.
Filinski shows how these constructs can be implemented using
\texttt{call/cc} and a single mutable cell.
He proves the correctness of this encoding using
a logical-relations argument.
%
%

%
Plotkin and Pretnar~\cite{plotkin-pretnar-09} introduce a denotational
semantics for a programming language equipped with effect handlers.
This semantics allows one to think of a computation as a
tree whose nodes are effectful operations,
%
%
and to think of an effect handler as a \textit{deconstructor} of such
computations: an effect handler traverses the tree and substitutes an
implementation for each effectful operation.
%
%
Plotkin and Pretnar require handlers to be correct
in the sense that
the operational behavior of the handler
must satisfy an effect theory,
that is,
a set of equations,
which is presented to an end user as a tool
to understand and reason about effects.
However, they do not investigate through what technical means one might prove
that a handler is correct.
%
%
Plotkin and Pretnar also adapt their previous equational
logic~\cite{plotkin-pretnar-08} to account for effect handlers.
This logic allows one stating and proving that two programs are equivalent.
Once such a logical judgement is proven, the soundness of the
logic implies an equality between the denotational interpretations
of the programs.

Xia \etal.~\cite{xia-20} build a Coq library, \textsc{ITrees},
which defines a coinductive data structure, \textit{interaction trees}.
An interaction tree is a possibly infinite tree-like structure whose nodes
are either effectful operations or silent reduction steps.
Handlers act on interaction trees by providing an interpretation of the
operations into a user-defined monad.
%
%
%
Xia \etal.~\cite[\S8.2]{xia-20} observe that the library currently does not
support handlers in their most general form: in particular, the handler does
not have access to the continuation.

%

In Plotkin and Pretnar's work and in interaction trees, the idea is to reason
about programs that involve effect handlers through a denotational model.
Another approach is to reason directly in terms of contextual equivalence
between programs.

In the setting of a restricted and untyped programming language
equipped with effect handlers,
Biernacki et al.~\cite{biernacki-lenglet-polesiuk-20}
show that contextual equivalence coincides with bisimilarity.
To simplify proofs of bisimilarity,
the authors propose \textit{up-to techniques},
which they illustrate through a number of simple examples.
However, it remains to see if this verification methodology scales to
the setting of a realistic programming language.


In the setting of a typed language, one can establish
contextual equivalence by means of logical relations.
Biernacki et al.~\cite{biernacki-al-18} present the first
logical-relations model of a type system with support for
effect handlers.
%
%
%
%
%
%
%
Later,
Zhang and Myers~\cite{zhang-myers-19} and
Biernacki et al.~\cite{biernacki-al-20}
propose a logical-relations model of a type system with support
for \textit{lexically scoped handlers}, a restricted use case of effect handlers
where generating a fresh effect name and installing a handler are
combined into a single operation.

Our work differs from the papers cited above in two respects. First, we
consider a programming language with both effect handlers and dynamically
allocated mutable state. Second, we propose a compositional program logic, in
the style of Hoare logic and \SL. That is, we propose a method that allows
writing a logical specification of the expected behavior of a program
component in isolation.
%
%
The proof that is presented in this paper emphasizes the benefits of
compositionality: we prove that our define-by-run \AD library works
correctly in an arbitrary context, provided, of course, that this
context respects the requirements imposed by our specification.
%

\nocite{letan-al-18,letan-al-21}

\subsection{Formal Presentations and Proofs of \AD Algorithms}
\label{sec:related:ad}

There is a huge literature on real-world implementations
of \AD and on their engineering aspects.
%
%
We cannot give a survey of this literature, and refer the reader
to Griewank and Walther's textbook~\cite{griewank-walther} for an
introduction to the field.

Focusing more specifically on the programming-language literature,
many researchers have contributed to setting \AD on a firm theoretical footing
by presenting formal definitions of \AD algorithms, by proposing formal proofs
of the correctness of these algorithms, and by making these algorithms
applicable in richer settings.
%

A large part of this work seems dedicated to define-then-run presentations
of \AD, where \AD takes the form of a program transformation, which transforms
an abstract syntax tree into an abstract syntax tree. In this style, an \AD
algorithm is essentially a compiler, and its proof is
a~\emph{compiler correctness} argument.
Our paper may be the first formal study of define-by-run \AD, where the
\AD algorithm is packaged as a library, and where its proof requires a
program verification task.
Our paper certainly offers the first proof of correctness of an \AD algorithm
that involves delimited-control operators: the similar algorithms previously
described by Wang \etal.~\cite{wang-rompf-18,wang-19},
Sivaramakrishnan~\cite{kc-18}, or Sigal~\cite{sigal-21} are not
accompanied with a proof.
Furthermore, our proof is machine-checked.

Whereas the define-then-run approach involves an inspection of the syntax
of the subject (that is, the program that one wishes to differentiate) and
therefore requires limiting the set of programming-language constructs that
the subject may use, the define-by-run approach involves executing the
subject and monitoring its execution, therefore requires limiting the set of
behaviors that the subject may exhibit.
%
%
So, these approaches may seem quite different.
This said, the dividing line can in fact be blurry.
For instance, a common approach to implementing \AD involves overloading the
arithmetic operations. In Haskell, this can be done by exploiting type
classes. Because overloading itself can be implemented either via a
compile-time program transformation or via runtime dictionary passing, it can
be difficult to decide in which category this approach falls.
As another example,
the implementation of
a \emph{differentiable programming language},
where differentiation is a primitive construct,
may well involve a marriage of
program transformation and program monitoring techniques.



In the following, we briefly list some of the recent work on \AD
that has appeared in the programming-language literature,
with a focus on its semantics and on the correctness of its
implementation.
This is not a true survey: we provide these citations merely
as starting points for the interested reader.




Karczmarczuk~\cite{karczmarczuk-98,karczmarczuk-01} implements forward-mode
and reverse-mode \AD in Haskell. He uses type classes to overload the
arithmetic operators. His forward-mode implementation computes not just the
first derivative, but the infinite lazy sequence of all higher-order
derivatives. His reverse-mode implementation uses backpropagator functions.
%
%
%
Also in Haskell,
Kmett, Pearlmutter and Siskind~\cite{haskell-ad} implement
forward-mode, reverse-mode, and mixed-mode \AD combinators,
with a common API, as a library.
Their implementation relies on mutable state,
stable names~\cite{peyton-jones-marlow-elliott-99},
and on a form of reflection.
%



%
%


Pearlmutter and Siskind~\cite{pearlmutter-siskind-08} present \VLAD,
a functional programming language
where differentiation is a primitive construct.
%
They propose a program transformation that eliminates this construct.
Because this transformation is \emph{non-compositional},
it is not necessarily an attractive implementation technique.
For this reason, in their prototype implementation of \VLAD,
Pearlmutter and Siskind
use a different technique:
they implement an interpreter,
and rely on the fact that an interpreter has access
to the source code of a function at runtime.



Elliott~\cite{elliott-09} presents an implementation of higher-dimensional,
higher-order, forward-mode \AD in the general setting of calculus on
manifolds.
Starting from the specification of \AD,
Elliott~\cite{elliott-18}
derives a general \AD algorithm,
which he then specializes in various ways
by varying the representation of derivatives.
His implementation in Haskell
relies on a compiler plugin
which performs a form of reflection
and allows him to alter the conventional meaning
of function abstraction and function application.




A number of authors present \AD as a program transformation
for calculi that do \emph{not} include differentiation as
a primitive construct.
Shaikhha \etal.~\cite{shaikhha-al-19}
implement forward mode first,
then show how reverse mode
can be reconstructed by
combining forward mode with a number of standard compiler optimizations.
Their calculus is simply-typed and
does not allow a function to return a function.
%
Barthe \etal.~\cite{barthe-crubille-dal-lago-gavazzo-20}
prove the correctness of a fragment of
Shaikhha \etal.'s transformation.
Alvarez-Picallo \etal.~\cite{alvarez-picallo-al-21}
prove the correctness of a simplified version of
Pearlmutter and Siskind's transformation~\cite{pearlmutter-siskind-08}.
They phrase the proof in terms of hierarchical string diagrams,
or \emph{hypernets}, and rewriting rules.
%
Brunel, Mazza and Pagani~\cite{brunel-mazza-pagani-20}
define a reverse-mode transformation
for simply-typed \lc
The transformation is compositional
and does not use mutable state;
backpropagator functions are used instead.
A \emph{linear-factoring} reduction rule,
which is built into the semantics of the calculus,
is required for the transformation to be cost-preserving.
Smeding and \vakar~\cite{smeding-vakar-23}
improve on Brunel, Mazza and Pagani's work 
by showing how their approach can be efficiently implemented
in a standard programming language,
whose semantics does not include a linear-factoring rule.
Mazza and Pagani~\cite{mazza-pagani-21}
prove the soundness of \AD transformations in the setting of PCF,
a typed \lc equipped with real numbers, recursion,
and conditionals.
In the presence of conditionals, \AD cannot be expected to
yield a correct result everywhere: Mazza and Pagani show
that it yields a correct result almost everywhere.
  %
With similar motivation,
Lee \etal.~\cite{lee-al-20}
isolate a class of functions for which
an \emph{intensional derivative}
always exists and coincides almost everywhere with the standard derivative.
%
Huot \etal.~\cite{huot-staton-vakar-20,huot-staton-vakar-22} present a
denotational semantics based on diffeological spaces for a simply-typed \lc
equipped with algebraic data types. They use this semantics to state and prove
the correctness of a forward-mode transformation.
%
Extending Elliott's work~\cite{elliott-18},
\vakar \etal.~\cite{vakar-21,nunes-vakar-21,vakar-smeding-22} present
forward- and reverse-mode transformations for pure $\lambda$-calculi equipped
with rich type disciplines.
Their proofs of correctness exploit logical relations.
Krawiec \etal.~\cite{krawiec-krishnaswami-22}
present forward-mode and reverse-mode transformations for a simply-typed \lc.
The transformations are cost-preserving.
The correctness proof involves logical relations.
Radul \etal.~\cite{radul-al-23} propose a modular presentation of reverse-mode
\AD in a typed calculus equipped with reals, tuples, and first-order
functions. The transformation is decomposed in three steps, namely
forward-mode \AD, \emph{unzipping}, and transposition. They argue that this
decomposition allows a better understanding and a more economical
implementation of reverse-mode \AD.



Treading in the footsteps of Pearlmutter and
Siskind~\cite{pearlmutter-siskind-08},
several authors propose
semantics and implementation techniques
for differentiable programming languages,
that is,
for calculi that \emph{do} include differentiation
as a primitive construct.
Vytiniotis \etal.~\cite{vytiniotis-curry-19}
sketch a compilation scheme
for a simply-typed, higher-order language.
Abadi and Plotkin \cite{abadi-plotkin-20}
give operational and denotational semantics for
a first-order language.
They prove that these semantics coincide, thereby establishing
the correctness of the \AD algorithm that is built into the
operational semantics.
Cockett \etal.~\cite{cockett-al-20}
propose ``a~starting point to build categorical semantics of
differentia[ble] programming languages''.
%
Mak and Ong~\cite{mak-ong-20}
design a higher-order differentiable programming language,
where the reduction strategy simulates reverse-mode \AD.
The correctness of this reduction strategy is proved
via a categorical interpretation.
  %
%
Sherman \etal.~\cite{sherman-michel-carbin-21}
describe $\lambda_S$, a higher-order programming language that includes
higher-order derivatives as well as constructs for
integration, root-finding and optimization.
Its denotational semantics is
based on Clarke derivatives.
They describe an implementation of $\lambda_S$
as an embedded language inside Haskell.

\section{Conclusion}
\label{sec:conclusion}



In this paper, we have verified a very small implementation
of reverse-mode \AD,
packaged as a library.
The code exploits
dynamically allocated mutable state,
higher-order functions,
and effect handlers.
%
%
We have proposed an original API for this library (\fref{fig:sig})
as a transformation of expressions in tagless-final style.
We have used Hazel \cite{de-vilhena-pottier-21},
a~variant of higher-order \SL,
to write a specification of the library
(\sref{sec:verif:spec})
and to construct a proof of its correctness.
(\sref{sec:verif:proof}).
This proof is machine-checked \cite{repo}.
We view this as a nontrivial exercise in modular program verification and an
illustration of the power of \SL, in the presence of mutable state,
higher-order functions, and effect handlers.
%
%
To the best of our knowledge, this is a first: outside of our own previous
work~\cite{de-vilhena-pottier-21}, no correctness proofs for programs that
involve primitive mutable state and effect handlers have appeared in
the literature.




Our specification of \diff illustrates how effects are described in Hazel.
According to Statement~\ref{formal:spec:diff},
\diff itself performs no effects: it
is a function from expressions to expressions. According to
\defref{formal:isExp}, an expression in tagless-final style is a
computation that is parameterized with a dictionary of arithmetic operations.
These arithmetic operations may perform unknown effects, represented by an
abstract protocol~$\prot$: the expression is not allowed to handle these
effects or to perform any effects of its own. It can perform effects
internally, but if it does so, then it must handle them, so they are not
observable.


By virtue of working in \SL, we naturally reap the benefits of modular
reasoning. The verified function \diff can be safely combined with other
software components, provided they respect the expectations expressed by the
specification of \diff. The use of mutable state and effect handlers inside
\diff cannot interact in unexpected ways with these foreign components. As an
example of particular interest, the specification of \diff allows composing
\diff with itself: it is clear (and one can easily verify) that \oc|fun e ->
diff (diff e)| computes a second-order derivative. This is a nontrivial
result: it would be very difficult to reason operationally about how
mutable state and effect handlers are used when the expression
\oc|diff (diff e)| is evaluated.



By now, many programmers are familiar with the fundamental principles of Hoare
logic and \SL: they know (at least informally) that one reasons about a loop
with the help of a loop invariant, and that one reasons about a function
through a~precondition and a postcondition. We would like this paper to
popularize the idea that one reasons about an effect through a protocol,
a~pair of a precondition and a postcondition, which represents a contract
between the handlee (which performs an effect) and the handler (which handles
this effect).
%
%
Our proof illustrates both handlee-side and handler-side reasoning.
On the handlee side, performing an effect is akin to calling a function: the
protocol describes the pre- and postcondition of the effect. On the handler
side, naturally, the postcondition of the effect becomes the precondition of
the captured continuation. Less obviously, because a~deep handler is sugar for
a shallow handler wrapped in a recursive
function~\cite{hillerstrom-lindley-18}, verifying a deep handler requires
spelling out the pre- and postcondition of this recursive function,
as well as any universal quantifiers that are shared between
the pre- and postcondition.
In our proof, the forward invariant (\defref{def:forward:inv}) and the
backward invariant (\defref{def:backward:inv}) serve as pre- and
postcondition for the handler,
and they are combined in a universally quantified statement
that expresses the correctness
of an arbitrary instance of the handler (\goalref{goal:handler:general}).
Finally, when writing down the protocol that the handlee and handler obey, it
is often necessary to introduce custom \SL assertions, whose definition may
involve ghost state. In our proof, the predicate $\representsname$
(\defref{def:represents}), which appears in the protocol that
describes the effects \oc|Add| and \oc|Mul| (\defref{def:protad}),
is defined in terms of a
ghost history~$\K$ of the past effects.
This ghost history appears in the forward invariant as well.
Ghost history variables are common in proofs of concurrent and distributed
algorithms~\cite{abadi-lamport-88,lamport-merz-17};
here, although the code is sequential, the handlee and the handler form two
distinct logical threads, so it should not be surprising that the need
for a history variable appears.
%



Our work is limited in several ways. Because our emphasis is on program-%
verification techniques, as opposed to automatic differentiation techniques,
the code that we verify has been distilled to the simplest possible form.
It is limited to expressions
that involve one variable, two constants (zero and one) and two primitive
arithmetic operations (addition and multiplication). It is not at all
optimized for efficiency. We see no obstacle in principle to supporting
multiple variables and a richer set of primitive arithmetic operations. In
future work, it would be interesting to investigate which real-world \AD
libraries rely on effect handlers and what obstacles remain before these
libraries can be verified.

Another caveat about our work is that there exists a gap between the code that
we present in the paper and the code that we verify. While the code that we
present (\fref{fig:ad}) is written in Multicore OCaml 4.12.0, the code that we
verify is expressed in \lang, a $\lambda$-calculus with mutable state and
effect handlers, whose syntax and operational semantics are defined in Coq.
The main difference between them is that the declarations \oc|effect Add| and
\oc|effect Mul| in Multicore OCaml generate fresh effect names at runtime,
whereas \lang does not
have fresh name generation nor effect names, so we encode \oc|Add| and
\oc|Mul| in \lang using left and right injections into a binary sum.
In other words, in our encoding, \oc|Add| and \oc|Mul| are essentially
global names.
We believe that this difference should not fundamentally influence the manner
in which one reasons about effect handlers.
Nevertheless, in the future, it would be desirable to propose a \SL that
allows reasoning about effect handlers in the presence of multiple effect
names and dynamic generation of fresh effect names.
This is currently an active area of research~\cite{%
biernacki-al-18,%
biernacki-al-19,%
biernacki-al-20,%
zhang-myers-19,%
effekt-20,%
de-vilhena-pottier-23%
}, where the final word has not yet been said.
%
%

\section*{Acknowledgments}

The authors wish to thank the anonymous reviewers, whose comments have greatly
helped improve the presentation of the paper.

\hbadness=10000
\bibliographystyle{alphaurl}
\bibliography{english,local}

\end{document}